\documentclass[10pt,journal,compsoc]{IEEEtran}
\ifCLASSOPTIONcompsoc
  \usepackage[nocompress]{cite}
\else
  \usepackage{cite}
\fi
\ifCLASSINFOpdf
  \usepackage[pdftex]{graphicx}
\else
\fi

\usepackage{multicol,graphicx,amssymb,mathrsfs,amsmath,pifont,amscd,latexsym,amsthm,color,fancyhdr,CJK,url,hyperref,longtable, pifont, subfigure,epstopdf,setspace,multirow,colortbl,tabularx,threeparttable, booktabs, verbatim, bbm, balance}
\definecolor{mygray}{gray}{.7}
\begin{document}
%
\title{Mining Behavioral Patterns from Millions of Android Users}
%
%
%
%

\author{Xuanzhe~Liu,~\IEEEmembership{Member,~IEEE,}
        Huoran~Li, Xuan~Lu, Tao~Xie~\IEEEmembership{Senior Member,~IEEE}, Qiaozhu~Mei, 
        ~Feng~Feng, and Hong~Mei~\IEEEmembership{Fellow, ~IEEE}

\IEEEcompsocitemizethanks{\IEEEcompsocthanksitem
 Xuanzhe~Liu, Huoran~Li, and Xuan~Lu are with the Key
Laboratory of High Confidence Software Technologies (Peking
University), Ministry of Education, Beijing, China, 100871. Email: \{liuxuanzhe, lihuoran, luxuan\}@pku.edu.cn }%
\IEEEcompsocitemizethanks{\IEEEcompsocthanksitem
Tao~Xie is with the University of Illinois Urbana-Champaign, USA. Email: taoxie@illinois.edu }%
\IEEEcompsocitemizethanks{\IEEEcompsocthanksitem
Qiaozhu~Mei is with the University of Michigan. Email: qmei@umich.edu }%
\IEEEcompsocitemizethanks{\IEEEcompsocthanksitem
Feng~Feng is the co-founder and Chief-Technology Officer of Wandoujia. Beijing, China, 100084. Email: jackfeng@wandoujia.com }

\IEEEcompsocitemizethanks{\IEEEcompsocthanksitem
Hong~Mei is with both Beijing Institute of Technology and  the Key
Laboratory of High Confidence Software Technologies (Peking
University), Ministry of Education, Beijing, China. Email: meih@pku.edu.cn. }

\thanks{Manuscript received Oct. 29, 2016; accepted Feb. 27, 2017. Recommended for acceptance by S. Malek. }

}

%
%

\markboth{IEEE TRANSACTIONS ON SOFTWARE ENGINEERING ,~Vol.~XX, No.~X, Feb~2017}%
{Liu{\textit{et al.}}: Understanding Diverse Usage Patterns from Large-Scale Appstore-Service Profiles}
%



\IEEEtitleabstractindextext{%
\begin{abstract}

The prevalence of smart mobile devices has promoted the popularity of mobile applications (a.k.a. apps). Supporting mobility has become a promising trend in software engineering research. This article presents an empirical study of behavioral service profiles collected from millions of users whose devices are deployed with Wandoujia, a leading Android app-store service in China. The dataset of Wandoujia service profiles consists of two kinds of user behavioral data from using 0.28 million free Android apps, including (1) app management activities (i.e., downloading, updating, and uninstalling apps) from over 17 million unique users and (2) app network usage from over 6 million unique users. We explore multiple aspects of such behavioral data and present patterns of app usage. Based on the findings as well as derived knowledge, we also suggest some new open opportunities and challenges that can be explored by the research community, including app development, deployment, delivery, revenue, etc. 

\end{abstract}

}

\maketitle

\IEEEdisplaynontitleabstractindextext

%
\IEEEpeerreviewmaketitle


%
%
%
%
\section{Introduction}

\IEEEPARstart{T}{he} release of iPhone in 2007 has opened a new era of mobile computing. Currently, smart devices such as iPhones, iPads, and Android devices have played an indispensable role in our daily lives. The increasing popularity of mobile devices and apps has induced an evolution of software industry. One of the currently inspiring trends is the emergence of online app stores (e.g., the Apple App Store, Google Play, and Microsoft Marketplace) for distributing mobile software applications (a.k.a., apps)~\cite{Picco:FOSE14}. For the first time in the history of software development, individual developers and small companies can access distribution infrastructures that allow them to sell (mobile) apps to millions of potential customers at the tap of a finger.

The emergence of mobile apps and online app stores has been a game changer to breed a new ``\textit{mobile app ecosystem}"~\cite{Petsas:IMC13}  constituting stakeholders such as app developers, marketplace operators, end-users, network service providers, and advertisers. Such an ecosystem also creates new opportunities and challenges for software engineering research. Recently, a few efforts have been proposed, covering aspects including requirement analysis~\cite{lim2014investigating,Falaki:IMC10}, code/library analysis~\cite{Xie:USENIXSecurity13, Huang:ICSE2014, Zeller:ICSE14,Thomas:ICSE14,Gui:ICSE2015}, version evolution~\cite{Geoffrey:ASE2015}, and system/tool supports~\cite{Chen:MobiCom2015,Crussell:MobiSys2014}.

Other than the preceding efforts, in software engineering research, understanding user behaviors is a natural and effective channel to align software development activities with user requirements. However, in practice, app developers have quite limited communications with their users, and thus have difficulties to comprehensively identify target users and understand their needs. Although developers can receive user ratings and reviews towards their apps, the reviews and ratings can be quite sparse and even low-quality for some apps~\cite{Liu:KDD13}, and only very few successful apps can receive useful user feedback~\cite{lim2014investigating}. Previous in-field user studies have made efforts to understand user behaviors towards using apps~\cite{Zhong:TMC13, Zhong:CHI2012, Falaki:MobiSys10, Apaolaza:W4A13,lim2014investigating}, but most of these studies were conducted using rather limited or biased datasets, typically based on subjects such as college students and questionnaire volunteers. Other behavioral signals were collected through a monitoring app voluntarily installed on the subjects' devices. Such a study cannot be widely applied by a variety of crowds because of security and privacy concerns. Certainly, app stores and network-service providers can have a lot of data on app usage, but no evidence shows that such data has been accessed and studied by external researchers. \textbf{Due to the absence of user behavioral data, it is currently difficult for the research community to directly extrapolate existing results and make representative understandings of how, where, and when the apps are \textit{actually} used at scale, and thus explore what knowledge can be derived for apps development, maintenance, revenue, etc., accordingly}.

The main goal of the work presented in this article is to bridge this knowledge gap. We are fortunate to collect a variety of user behavioral data from millions of users with a leading Android app-store service provider. We then conduct our study from three main folds. First, we make an empirical study to characterize the diverse app-usage behaviors. To this end, we propose various research questions in terms of app popularity, management activities, and network usage. We conduct a series of \textbf{measurements} over the dataset covering 17 million users, 0.28 million apps, with various metrics. Second, we exploit \textbf{the derived knowledge} from our measurement study and provide some \textbf{helpful implications} to current app-centric software engineering research, ranging from better understanding user requirements and needs, to improving the workload and ranking system of app stores, and to affecting the app's development activities and revenue strategies, etc. Last, although this article does provide some primitive answers and implications, researchers can go much deeper into each of these research questions and conduct a much more in-depth study. In addition, we expect that our dataset can establish \textbf{a valuable resource}\footnote{Part of our dataset has been released along with the publication of our IMC 2015 paper~\cite{Li:IMC15}, and one can find the brief description from \url{http://sei.pku.edu.cn/~liuxzh/appdata/}. The dataset can be requested by the researchers who have the IRB approval.} for the research community to explore more potential research topics and opportunities.

Our dataset comes from a leading Android app store in China, called \textbf{Wandoujia}\footnote{http://www.wandoujia.com}. Similar to Google Play and Apple AppStore, Wandoujia provides its own services with a native management app that facilitates users to search, browse, download, install, update, and uninstall apps. Additionally, compared to Google Play and AppStore, Wandoujia can provide advanced monitoring services and features that can be optionally enabled by its users. Once these features are enabled, the Wandoujia management app runs as a background system-wide daemon service but without requiring the ``root" privilege, and is able to collect network-activity information per app, e.g., data traffic and access time under Wi-Fi and cellular, respectively. 

Our dataset covers millions of active users who frequently use Wandoujia. The dataset contains two types of service profiles reflecting user behaviors: (1) app-management activities (i.e., installation, updating, and uninstallation) from over 17 million (17,303,122) unique users; (2) app-network usage from 6 million (6,632,303) unique users. Based on this extensive dataset, we conduct a systematic analysis that includes the distributions of app popularity, possible reasons of app selection, life cycles of abandoned apps, fine-grained network analysis of apps, and the impact of device-specific factors on app usage. In total, our dataset covers over 0.28 million (283,922) Android apps. Although the apps provided by Wandoujia are all free, such a large-scale dataset can provide useful knowledge on app usage patterns.

Part of this work was previously presented in our IMC 2015 conference paper~\cite{Li:IMC15}. The main extension presented in this article is that we employ a larger dataset spanning five months (May 1, 2014 to September 30, 2014) instead of the previous one-month data. The two datasets have the same kinds of information, but the new dataset contains more users (17 million instead of 8 million) and more apps (0.28 million instead of 0.26 million) and thus can enable more comprehensive analysis. In this way, our new dataset can overcome some limitations of the previous one-month data such as impacts by release time and update frequency of apps. We conduct the same measurements proposed in our IMC 2015 conference paper~\cite{Li:IMC15} over our new dataset, i.e., the app popularity, app management, and network usage. Additionally, we extend some entirely new statistical measurements, i.e., \textit{how the user reviews can correlate to app popularity} (Section~\ref{rating}), and \textit{how app usage is affected by the choice of device models} (Section~\ref{device}). Most of results are quite consistent with those made over the old dataset. In addition, the new longer-timespan dataset also enables us to explore more insights.

This article includes new suggestions on how our findings and implications can help explore open opportunities and problems for software engineering research (Section~\ref{implication}). We also provide more discussions (Section~\ref{threats}), such as limitations of our dataset and threats to validity of our study. 

Based on the unique dataset, we conduct a systematic empirical study from various perspectives. We not only confirm some results derived from previous studies that were conducted over relatively small datasets or limited users, but also derive some new knowledge and implications from diverse user behavioral patterns on app usage. More specifically, this article makes the following main contributions:
 


\begin{itemize}
\item We characterize the popularity of apps with various ranking metrics including the number of downloads, the number of unique users, the volume of data traffic, and the length of network-access time. We validate the Pareto-like principle and further explore the power-law of apps' popularity distribution. Additionally, we also find some ``clustered'' apps that are frequently to be requested together, indicating their locality on the servers. These findings suggest the significant improvement for cache placement where the copy of some apps can be placed on the app store's servers. We then simulate the request traces on various typical cache mechanisms and devise an adaptive mechanism for optimizing the app store's workload.

\item We describe how users perform app-management activities such as downloading, updating, uninstalling, and rating their apps. We reveal the diurnal regularities of app-management activities and the lifecycle of those abandoned apps. We demonstrate how to find the apps having possible ``fake'' downloads, e.g., some apps have an abnormal number of downloads. In particular, we surprisingly find that the user ratings of an app are not always consistent with the numbers of downloads and unique users of this app, especially for apps with very few ratings. These findings suggest that  app-store operators should incorporate new ranking mechanisms to predict the adoption of an app.

\item We investigate the network usage including the volume of data traffic and the length of access time under cellular and Wi-Fi as well as at foreground and background, respectively. We find that apps from specific categories (e.g., \textit{VIDEO}) can account for substantial traffic under different networks. We are especially surprised to find that numerous apps keep ``long-live" TCP connections and produce data drain at background after they are launched but without user interaction. These findings suggest that both developers and end-users need to justify whether such dynamic behaviors (especially those at background) and potentially extra cost are \textit{really} reasonable with respect to the regular functionality of the apps. 

\item We reveal that the choice of device models can lead to significantly diverse usage patterns of apps. We find that the device models are heavily ``fragmented'', i.e., the number of users per device model varies significantly. Additionally, the user behaviors of app download \& uninstall and network-access time are affected by the choice of device models. These findings suggest that Android developers need to carefully prioritize device models. Furthermore, the users holding different device models have quite distinct preferences of selecting ``competing apps". For instance, lower-end users prefer the \texttt{Opera Mini} browser while higher-end uses prefer \texttt{Chrome}, since the latter claims  to reduce data traffic with advanced compression services. These findings suggest that app developers should take into account the device-specific features in releasing their apps to gain more users and potential revenues. 


\end{itemize}

We approach our study from perspectives with the intention that interested readers could focus on parts relevant to their research. In addition, various open opportunities can be explored over our dataset and findings. Although most users studied in our work are from China, the measurement approach and derived knowledge from such an extensive dataset can be generalizable to the populations from other app stores.  

The remainder of this article is organized as follows. Section~\ref{dataset} describes the dataset. Sections~\ref{methodology} presents the measurement approach by proposing some research questions. Sections~\ref{popularity}, \ref{management}, \ref{network}, and \ref{device} describe the inferred diverse app-usage patterns in four aspects: app popularity patterns, management patterns, network patterns, and device-sensitive patterns, respectively. Section~\ref{implication} summarizes the findings and implications to different stakeholders in the app-centric ecosystem. Section~\ref{threats} discusses some possible limitations of our study. Section~\ref{related} makes comparisons with related work, and Section~\ref{conclusion} concludes the article.  
\section{Dataset}\label{dataset}

In this section, we briefly introduce the Wandoujia app store and describe the information covered by our dataset. To protect the user privacy and assure the academic ethics of our research, we also discuss how the data is processed with a rigorous workflow.

\subsection{Wandoujia}

Wandoujia\footnote{Visit its official site via \url{http://www.wandoujia.com}.} is a free Android app store in China. Wandoujia was founded in 2009 and has grown to be a leading Android app store~\cite{Web:Wandoujialeading}. Similar to other app stores, third-party app developers can upload their apps to Wandoujia and get them published by passing Wandoujia's authentication system\footnote{The authentication system of Wandoujia checks whether uploaded apps contain illegal contents and performs basic anti-virus tasks.}. Wandoujia also provides a categorization system, in which each app is annotated with a category tag, such as \textit{COMMUNICATION}, \textit{GAME}, \textit{MEDIA}, \textit{MUSIC}. Developers can choose the category by themselves or Wandoujia can annotate the app's category information\footnote{Apps not belonging to any category are annotated as ``MISCS".}. Compared to other app stores such as Google Play, apps on Wandoujia are all free, but apps are fully allowed to have ``\textit{in-app purchase}". 

Our dataset comes from Wandoujia's management app. The Wandoujia management app is a native Android app that provides various services, by which people can manage their apps, e.g., searching, downloading, updating, and uninstalling apps. The logs of these management activities are all automatically recorded. 

\begin{figure}
\centering
	\begin{center}
	\subfigure[]
	{
		\includegraphics[width=0.13\textwidth]{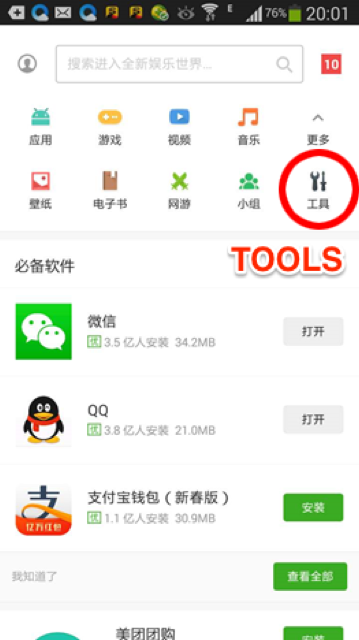}
	}
	\subfigure[]
	{
		\includegraphics[width=0.13\textwidth]{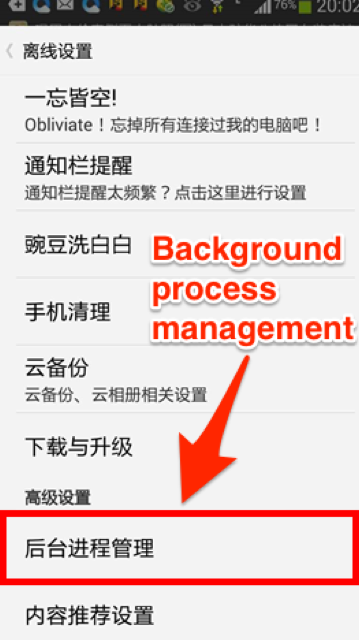}
	}
	\subfigure[]
	{
		\includegraphics[width=0.13\textwidth]{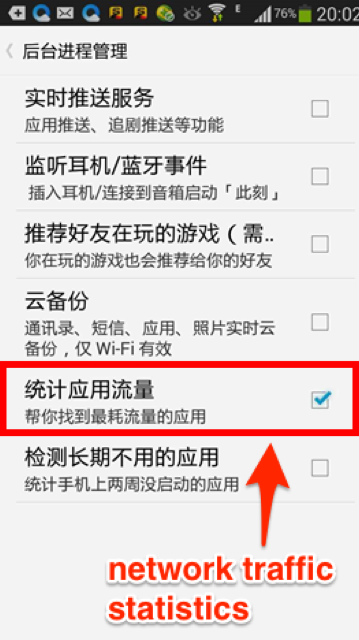}
	}
	\caption[7.5pt]{Screenshots of advanced settings in the Chinese version of the Wandoujia management app (the advanced settings is not supported in the current English version). \textit{(a) is the homepage of the Wandoujia management app, where users can navigate to ``\textit{settings}" by clicking the text circled by red; (b) refers to the background management service setting, which is highlighted by the red rectangle; (c) is to toggle whether to allow Wandoujia to collect the data of network activities}.}\label{wdj}
	\end{center}
\end{figure}

Besides these basic features, the Wandoujia management app is developed with some advanced but optional services that can monitor and optimize a device. These services include network activity statistics, permission monitoring, content recommendation, etc. All services are developed upon standard Android system APIs and do not require the ``root" privilege. Users can decide whether to enable these features, as shown in Figure~\ref{wdj}. However, these services are supported \textbf{only} in the Chinese version.

When installed, the Wandoujia management app is automatically launched and it works as a system-wide service after the device starts up. The data collected by such a service per device are uploaded to the Wandoujia server when Wi-Fi is available. 

\begin{table*}[htbp]\scriptsize
\newcommand{\tabincell}[2]{\begin{tabular}{@{}#1@{}}#2\end{tabular}}
  \centering
  \caption{Chosen apps by category.}
  \label{topapps}
  \begin{threeparttable}
    \begin{tabular}{l|r|r|r|r|r|r|r|r|r}
    \hline
    \textbf{App Category} & \textbf{Apps} &\tabincell{c} {\textbf{Users}\\($10^6$ devices)}&\tabincell{c} {\textbf{Downloads}\\($10^6$ times)}&\tabincell{c} { \textbf{Traffic}\\(GB)} &\tabincell{c} {\textbf{Access Time}\\($10^7$ hours)}&\tabincell{c}{\textbf{\textit{C-}}\\\textbf{Traffic}} & \tabincell{c}{\textbf{\textit{C-}}\\\textbf{Time}}&\tabincell{c}{\textbf{\textit{W-}}\\\textbf{Traffic}}& \tabincell{c}{\textbf{\textit{W-}}\\\textbf{Time}}\\
\hline

    \bfseries BEAUTIFY & 38,072 & 2.71 & 5.82 & 43,060.85 & 4.24 & 0.54\% & 11.96\% & 0.69\% & 11.33\% \\
    \bfseries COMMUNICATION & 1,745 & 8.13 & 18.76 & 863,020.84 & 17.21 & 42.18\% & 48.32\% & 10.12\% & 46.08\% \\
    \bfseries EDUCATION & 22,849 & 3.41 & 7.19 & 41,186.12 & 1.51 & 1.16\% & 4.87\% & 0.58\% & 3.48\% \\
    \bfseries FINANCE & 3,634 & 0.90 & 1.76 & 3,143.70 & 0.08 & 0.26\% & 0.27\% & 0.02\% & 0.20\% \\
    \bfseries GAME & 80,762 & 7.61 & 34.24 & 51,854.80 & 2.00 & 3.01\% & 5.88\% & 0.55\% & 5.14\% \\
    \bfseries IMAGE & 2,738 & 0.25 & 0.35 & 2,309.98 & 0.00 & 0.04\% & 0.01\% & 0.04\% & 0.01\% \\
    \bfseries LIFESTYLE & 30,623 & 2.97 & 6.67 & 27,542.35 & 0.29 & 1.74\% & 0.83\% & 0.28\% & 0.78\% \\
    \bfseries MOTHER\_AND\_BABY & 743 & 0.22 & 0.38 & 4,797.12 & 0.04 & 0.16\% & 0.06\% & 0.07\% & 0.13\% \\
    \bfseries MUSIC & 1,002 & 5.22 & 8.86 & 190,438.70 & 0.67 & 4.23\% & 2.15\% & 2.84\% & 1.55\% \\
    \bfseries NEWS\_AND\_READING & 18,583 & 2.73 & 5.80 & 120,835.76 & 1.40 & 4.48\% & 3.48\% & 1.59\% & 4.13\% \\
   
    \bfseries PRODUCTIVITY & 6,155 & 1.78 & 3.43 & 18,439.92 & 0.07 & 0.52\% & 0.18\% & 0.26\% & 0.21\% \\
    \bfseries SHOPPING & 4,664 & 3.90 & 10.03 & 194,146.89 & 0.72 & 5.52\% & 1.17\% & 2.75\% & 2.69\% \\
    \bfseries SOCIAL & 6,474 & 5.13 & 11.25 & 235,489.22 & 1.33 & 6.69\% & 3.01\% & 3.34\% & 4.17\% \\
    \bfseries SPORTS & 1,852 & 0.50 & 0.65 & 883.37 & 0.01 & 0.04\% & 0.03\% & 0.01\% & 0.03\% \\
    \bfseries SYSTEM\_TOOL & 9,818 & 7.30 & 18.86 & 231,470.89 & 1.12 & 3.92\% & 3.01\% & 3.60\% & 3.11\% \\
    \bfseries TOOL & 29,808 & 8.20 & 25.44 & 692,746.52 & 4.22 & 11.39\% & 12.16\% & 10.80\% & 11.05\% \\
    \bfseries TRAFFIC & 930 & 0.19 & 0.27 & 637.16 & 0.01 & 0.05\% & 0.02\% & 0.01\% & 0.01\% \\
    \bfseries TRAVEL & 3,381 & 3.56 & 7.43 & 59,791.09 & 0.17 & 1.92\% & 0.75\% & 0.82\% & 0.21\% \\
    \bfseries VIDEO & 15,321 & 6.50 & 17.47 & 3,588,326.52 & 1.41 & 12.02\% & 1.82\% & 61.56\% & 5.66\% \\
 \bfseries MISCS & 4,768 & 0.12 & 0.14 & 5,884.40 & 0.01 & 0.13\% & 0.02\% & 0.09\% & 0.02\% \\
    \hline  
    \end{tabular}
    \begin{tablenotes}
    \footnotesize
    \item \textbf{The users, downloads, traffic, and access time are all computed by aggregating the data of each app in the category}
    \item \textbf{The percentage of \textit{W}-Traffic (\textit{C}-Traffic) and \textit{W}-Time (\textit{C}-Time) refer to the data traffic and foreground access time over Wi-Fi (W) and cellular (C) network, respectively. For each category, the percentage value is computed based on the apps from this category relative to all apps in our dataset.} 
    \end{tablenotes}
    \end{threeparttable}
\end{table*}

%
%
%
%
%
%

\subsection{Data Collection}
As of 2014, Wandoujia has over \textbf{350} million users\footnote{\url{http://www.chinatechnews.com/2014/05/07/20496-chinese-android-app-store-inks-deal-with-japanese-messaging-app}}. Each user is actually identified by a unique Android device, which could be either a smartphone or tablet computer. In the study described in this article, we collected five-month usage data from May 1, 2014 to September 30, 2014. 

To avoid ``zombie'' users who contribute little to the analysis, our five-month dataset chooses only users who are actively using Wandoujia. To this end, the users should launch and use the Wandoujia app for more than 120 days, according to the service profile logs generated by the Wandoujia management app. The data reflecting user behaviors consist of two types of service profiles: (1) the \textbf{\textit{profile of app-management activities}} (i.e., installation, update, and uninstallation)  (2) the \textbf{\textit{profile of app-network usage}} (i.e., the traffic and access time per app). Finally, we obtain the behavioral data from more than 0.28 million (283,922) Android apps. The overall statistical information of our dataset is described in Table~\ref{topapps}, which contains aggregated information including the number of apps, users, traffic, and network access time per category. Such a dataset occupies about 5.6 TB disk space. 

As the network activity monitoring is an optional service of the Wandoujia management app, we distinguish the two kinds of service profiles as ``App-Management Activities" and ``App-Network Usage Activities". In addition, we collect the user rating (against an app) and device model information.  

\subsubsection{App-Management Activities}

App-management activities consist of downloading, updating\footnote{In the Wandoujia management app, a pop-up of installation wizard is presented to users when an app is successfully downloaded or updated. So we simply treat ``download" and ``installation" equally, if not specifically distinguished.}, and uninstalling apps. The monitoring of app-management activities is always enabled when the Wandoujia management app is installed on device. When an app is installed or updated via the Wandoujia management app, its installation counter is automatically incremented by one and a log entry of this activity is created. The logs of uninstallation via the Wandoujia management app are processed similarly, and the count of uninstallations is incremented automatically. 

Unlike the Apple App Store, the Android platform allows users to install various app stores other than Google Play\footnote{In fact, the access to Google Play is currently banned in China.}; hence, our dataset focuses on only the apps that are operated by the Wandoujia management app. To this end, we maintain a list of popular app stores in China such as 360, Baidu, Tencent, and Xiaomi, and filter out all users who install these app stores out of our dataset\footnote{Essentially, these app stores are also native apps that can be monitored by Wandoujia.}. Finally, we collect the management activity logs from \textbf{17,303,122 unique users} (unique devices in fact). We denote the dataset as ``\textbf{Universal User Set}." The logs of management activities are used to explore an app's popularity, and can implicitly reflect the app's quality.

\subsubsection{App-Network Usage Activities}

When the advanced features are enabled, the Wandoujia management app collects daily network statistics of each app, when the app is connected to network either from Wi-Fi or cellular (2G/3G/LTE). If an app is never launched or generates no network connections, the app is not recorded in the network statistic logs. 

\indent To reduce the overhead of runtime monitoring, the Wandoujia management app does not record network details of each session of an app. Instead, it summarizes the total daily traffic drain and access time of an app by examining flows at the TCP level. The traffic drain and access time are accounted for Wi-Fi and cellular, respectively. In particular, the traffic drain and access time generated from \textbf{foreground} and \textbf{background} are accounted separately. To this end, the Wandoujia management app determines whether an app is running at ``foreground'' by probing the Android system stack for every 2 seconds. In this way, the ``foreground'' access time can imply how long the user interacts with an app. The Wandoujia management app checks whether an app running at ``background'' every minute. If any network activity is detected during this interval, this app is regarded to be ``online'' and its ``background'' access time is increased by a minute. Such a time interval is reasonable to initiate and release a TCP connection. 


In summary, the statistic of network activities provides 8 dimensions of information, i.e., 2 \textbf{\textit{metrics}} (access time and traffic) * 2 \textbf{\textit{networks}} (Wi-Fi and Cellular) * 2 \textbf{\textit{states}} (foreground and background). 

As the statistic of network activities is an optional feature for end-users, the covered users are a subset of the ``\textbf{Universal User Set}." We take into account only the users who successively contributed the statistics for more than three weeks. In our five-month dataset, the network activities cover  \textbf{6,632,303} unique users from our 0.28 million apps. We denote such a dataset as ``\textbf{Networked User Set}." 
\subsubsection{User Ratings}
Most app stores allow users to commit their user reviews and make ratings to an app. Compared to Google Play where an app is ranked by a 5-star model, Wandoujia's users can simply rank an app with a binary-metric voting model of ``\textit{like}" or ``\textit{dislike}", i.e. a user who installs an app can vote ``like" if he/she is in favor of the app, or ``dislike" otherwise.  To encourage user participation, Wandoujia allows multi ratings of an app per user. But, to prevent possible ``fake" ratings, Wandoujia now restricts that each user can rate only once every 3 days. Note that currently Wandoujia \textbf{does not} associate the rating towards a specific version of the app, but just the app generally. Hence, it is possible that users installing an older version of an app can still rate this app even after the app has released updates. In practice, on most app stores such as Apple App Store and Google Play, users can have the information of ratings of an app from its profile page, but it is observed that the ratings are usually provided as an overall score, not specific to versions. In this way, we aggregate the ratings given to an app in our five-month dataset.
 
In addition, users can optionally commit textual reviews. In this article, we collect only the number of ``like" (and ``dislike") that can be publicly collected from the profile page of an app on Wandoujia. 

\subsubsection{Device-Model Information and Price}

The Wandoujia management app also records the manufactural information of each device, e.g., \texttt{Samsung Galaxy Note 2, Google Nexus}.  We employ the information of device models to classify users. There are 19,147 different device models in total. Such a result immediately implies the heavy fragmentation of Android device models. To better organize these models, we collect their on-sale price information when they were firstly put onto market. We provide details of clustering the device models against their subscribers in Section~\ref{device}.

\subsection{Ethical Considerations}

Undoubtedly, collecting user behavioral data always needs to be very careful. As an app-store service provider, Wandoujia explicitly declares what and why the preceding data should be collected in its privacy policy statement. We take a series of steps to preserve the research ethics and user privacy of involved users in our dataset. First, all raw data collected for this study are kept within the Wandoujia's data warehouse servers, which are placed behind the company firewall. Second, our data-collection logic and analysis pipelines are completely governed by three Wandoujia employees\footnote{One co-author serves as a co-founder and the current CTO of Wandoujia. He supervises the process of data collection and de-identification.} to ensure compliance with the commitments of Wandoujia privacy policy in the \textit{Term-of-Use} statements. Finally, the Wandoujia employees take charge of anonymizing the user identifiers. The dataset includes only the aggregated statistics for the users covered by our study period. No actual users can be traced at all.

\subsection{Limitations}
Some limitations should be also addressed before we perform our measurement, as they can have potential impacts on the analysis and may narrow the generalization of results. First, the collected data come from only a single app store in China, and the covered users are mainly Chinese. As a result, the demographical differences can occur in other app stores or other countries. Second, the dataset currently cannot trace an app's versions where the data come from, and thus cannot capture the impact of app release, which is proposed to be a factor to affect the app's success~\cite{Nayebi:SANER2016}. Third, apps published on the Wandoujia app store are free, and hence we cannot infer the users' payment behaviors on the paid apps. Indeed, it is reported that user behaviors can be a bit different on free and paid apps~\cite{Petsas:IMC13}. 

However, our dataset is quite unique in terms of both the scale of users and behavior dimensions. As presented in the subsequent sections, our study not only validates or contradicts some results derived from a small scale of users, but also provides new findings. In addition, we discuss how to alleviate the preceding limitations in Section~\ref{threats}.

\section{Research Questions}\label{methodology}
Starting from this section, we demonstrate how our dataset can be useful for general research on the diverse user-behavior patterns of Android apps. In particular, we propose a series of research questions, which are concerned with the popularity, management activities, and network characteristics of using apps. We show that these research questions are rather interesting and can inspire various potential research projects; if answered, these research questions can provide many insights on app-usage behaviors. Although in this article we do provide preliminary answers, researchers can go much deeper into each of these research questions and conduct a much more in-depth study. Our goal is to provoke the interest of the research community to study these questions (based on such a dataset). 

\subsection{RQ1: How can we identify the multi-dimensional popularity distribution of apps?}

For app-store operators, a fundamental task is to identify and determine which apps are actually popular or unpopular, so that the operators can improve their ranking and recommendation system, allocate server-side resources, and place ads for the most popular apps. Currently, the popularity of an app is usually measured by the number of downloads of this app. Such a metric should not be always sufficient or accurate enough in some cases. For example, it is quite possible that some apps can be purposely re-downloaded by automatic programs. In addition, downloading an app does not mean that the app could be actually used. To this end, we aim to make a comprehensive study of app popularity from multiple aspects, i.e., the download times, the unique subscribers, and the network usage. \textbf{Indeed, all these aspects are meaningful, but no previous study can synthesize them comprehensively}. Hence, we aim to understand the distribution of app popularity from various indicators. Additionally, we can further examine the consistency of these indicators, and thus provide a representative genre of the apps that can substantially account for all indicators. Such a genre can be released to researchers who can explore further topics. 

\subsection{RQ2: How do the users manage their apps?}

The app-management activities include downloading, updating, and uninstalling apps. Exploring app-management activities is motivated for various reasons. Essentially, the management activities can reflect the user attitudes towards an app, i.e., downloading and updating can mean that the user needs this app, while uninstalling can imply that the user does not need or even dislikes this app. In addition, we can associate the app-management activities with the publicly available user ratings, and infer users' attitudes towards the app in a more comprehensive way. On the other hand, as the app-management activities are usually made manually, they can implicitly reflect the density and frequency of user interactions with the app stores. This information can help app-store operators better understand when a large number of concurrent user requests arrive, so that the operators can optimize their servers for faster app delivery and more reliable network bandwidth. Therefore, we plan to break down RQ2 to the following questions.\\

\noindent \textbf{RQ 2.1}: How do the users perform their diurnal management activities of apps?\\

\noindent \textbf{RQ 2.2}: What apps are more likely to be selected and liked by users?\\

\noindent \textbf{RQ 2.3}: How can we identify an app that is more likely to be disliked by users?\\

\noindent \textbf{RQ 2.4}: Are the user ratings of an app consistent with the app-management activities, with respect to the user attitude towards this app?

\subsection{RQ3: How do apps perform in terms of access time and traffic drain over network?}

Understanding network activities of apps is always a highly interesting topic. Undoubtedly, most of current smartphone apps heavily rely on the network to function and provide features. End-users would like to know how their data plan is spent by an app, and whether some potentially unnecessary ``hidden" or even ``stealthy" network behaviors may occur. To avoid the low ratings or even user loss, the developers should carefully check the design/implementation such as improperly granting permissions, or fix some possible bugs. As our dataset contains the detailed information of access time and traffic volume  generated at foreground/background under cellular/Wi-Fi, respectively, we decompose the network characteristics of apps by exploring the following questions. \\

\noindent \textbf{RQ 3.1}: Which apps are the users likely to interact with, when these apps are under Wi-Fi and cellular networks, respectively?\\

\noindent \textbf{RQ 3.2}: Which apps are more ``traffic-intensive'' and how much traffic is generated by these apps?\\

\noindent \textbf{RQ 3.3}: How much ``hidden'' traffic is consumed when using an app?

\subsection{RQ4: How does the choice of device models affect the app usage?}

Intuitively, the answers to \textbf{RQs 2-3} can reflect the overall knowledge of app usage from millions of users. To break down whether the classification of users can have impact on the app usage, we categorize the users according to the device models that they hold, i.e., high-end, medium-end, and low-end, ranked by the on-sale price. Hence, we revisit some important aspects of \textbf{RQ2} and \textbf{RQ3}, such as  app selection, app abandonment, and network usage.\\

As an empirical study, we employ \textbf{descriptive and statistical} analysis of our dataset to answer the preceding research questions. We apply some well-established statistical metrics such as the Spearman correlation coefficient and linear regression model. The large-scale dataset enables our findings to be comprehensive. 

\begin{figure*}[!t]
  \centering
  \begin{center}
    \subfigure[Percentage of downloads against rank\label{fig:percentage_ndownload}]{\includegraphics[width=0.32\textwidth]{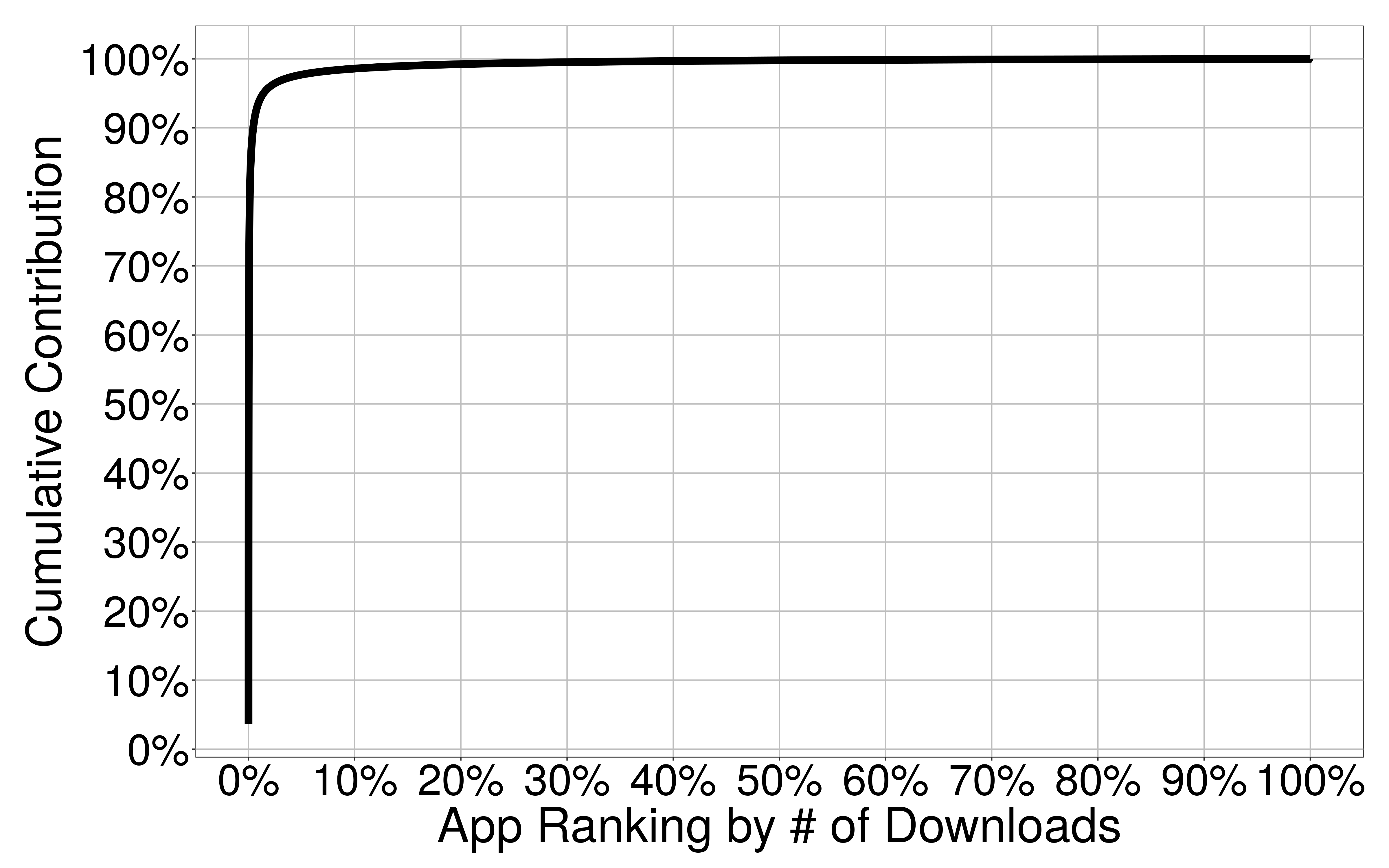}}
        \subfigure[Power-law of app downloads\label{downloadpower}]{\includegraphics[width=0.32\textwidth]{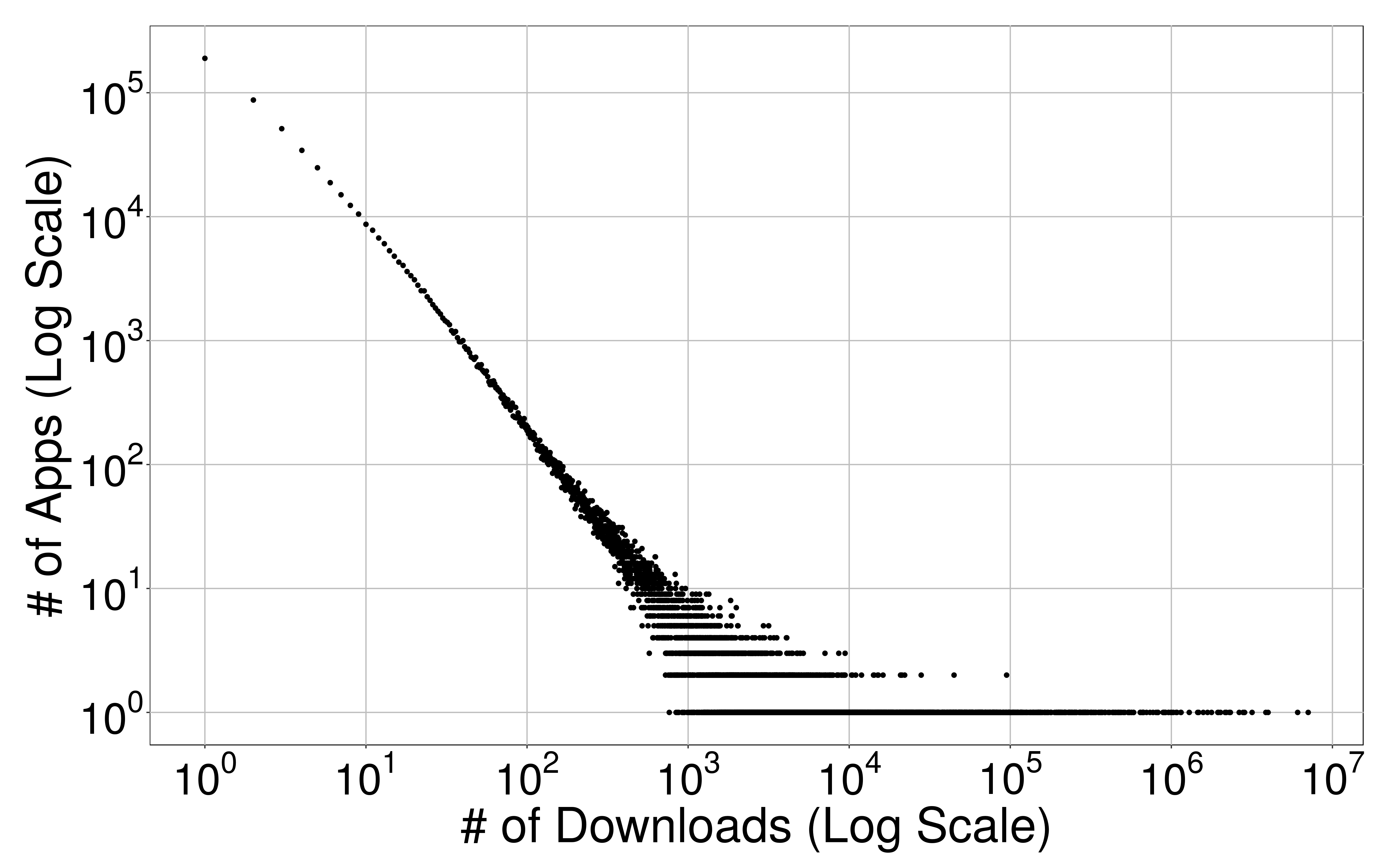}}
       \subfigure[Correlation between numbers of downloads and users\label{fig:percentage_nuser}]{
    \includegraphics[width=0.32\textwidth]{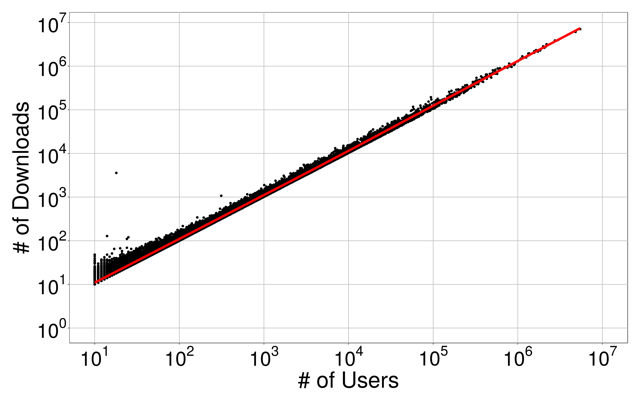}}
     \subfigure[Power-law of apps per user\label{userpower}]{
    \includegraphics[width=0.32\textwidth]{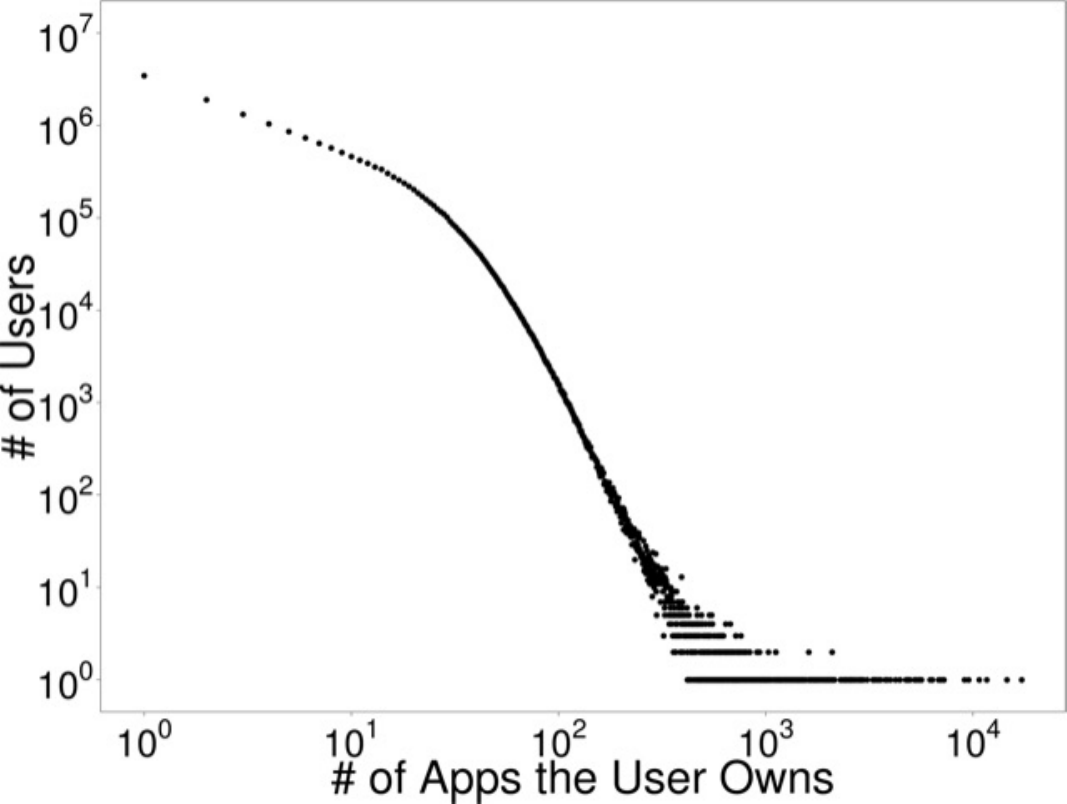}}
          \subfigure[Average numbers of downloads and updates per app \label{dpdistribution}]{\includegraphics[width=0.32\textwidth]{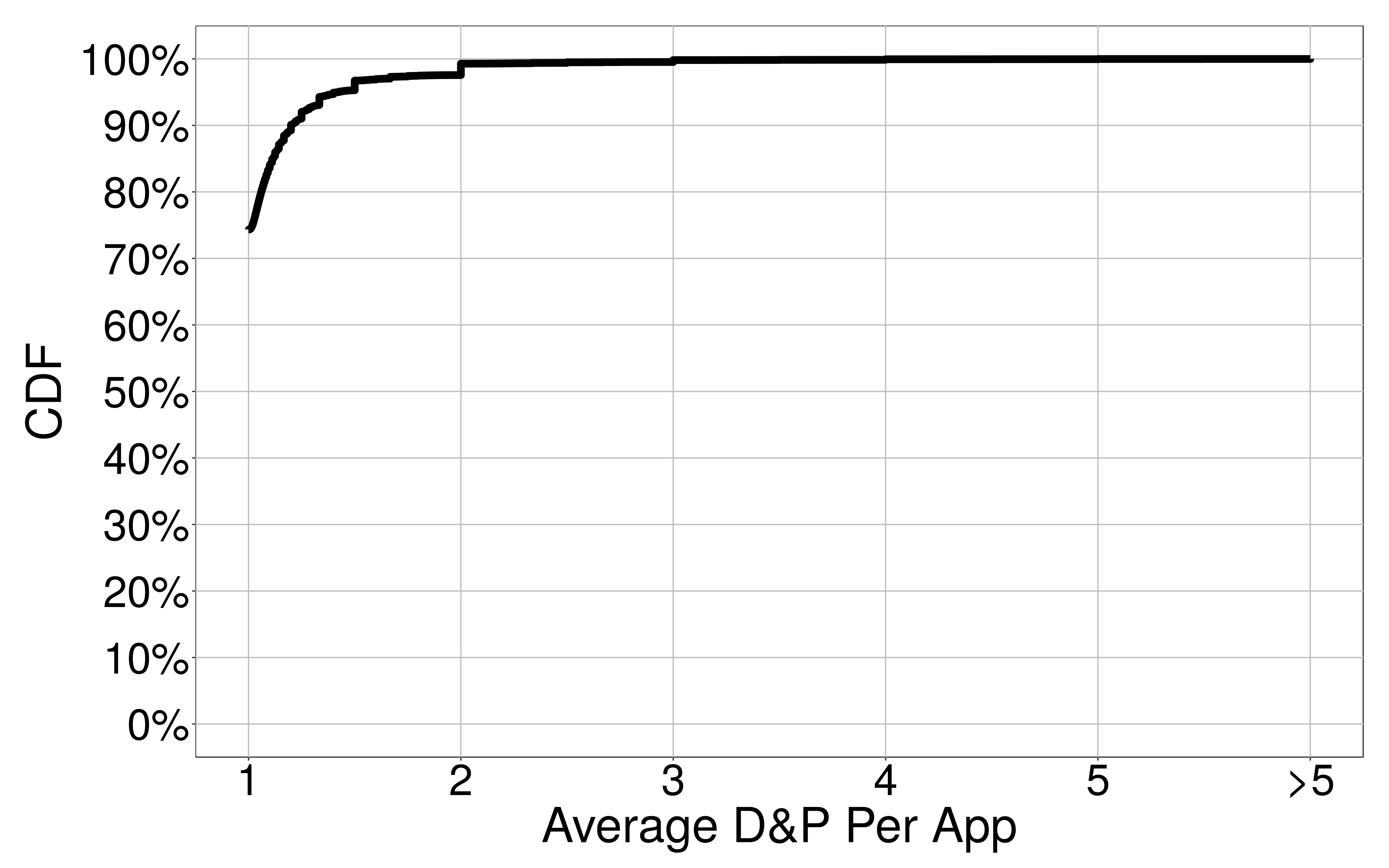}}
      \caption[7.5pt]{\textbf{App popularity by unique users and downloads.} Figure~\ref{fig:percentage_ndownload} to Figure~\ref{userpower} demonstrate that the number of downloads and the number of unique users for a single app are observed to follow the ``Pareto-like" principle and power law. Figure~\ref{dpdistribution} indicates that users do not very frequently update their apps. 
      }
      \label{devices}
  \end{center}
\end{figure*}
\section{App Popularity Patterns with Different Metrics}\label{popularity}

In this section, we first analyze the apps' popularity distribution. Essentially, the app store can be viewed as a special kind of system for sharing web contents and resources. Indeed, it is demonstrated that analyzing the exact form of popularity distribution not only helps understand the underlying mechanisms (i.e., cache, bandwidth, etc.), but also helps improve important design solutions in other systems of sharing web contents and resources such as search engines, online video systems, e-commerce systems~\cite{Wiki:Pareto}. For instance, the scale-free nature of web requests has been used to improve search engines, advertising policies, and recommendation systems. 

In a previous study~\cite{Petsas:IMC13}, app popularity is usually measured by only the number of app downloads on the app store. To make a comprehensive analysis, we employ four metrics for an app in this article: \textbf{(1) the number of unique devices that ever used the app; (2) the number of downloads of the app; (3) the aggregated data traffic generated by the app; (4) the aggregated access time that users interacted with the app.} The former two metrics can indicate how widely an app is owned by users, and the latter two can indicate how much an app is really used.
\subsection{Popular Apps by Downloads}\label{download}
We investigate the most intuitive metric of app popularity, i.e., the number of downloads of an app. Many app stores take the number of downloads (i.e., accumulatively total, monthly, or weekly) as the key indicator to rank app popularity. We then investigate app downloads from the management activities of the  \textbf{Universal User Set}.

Various data points show that the Pareto principle exists in networked application domains such as web content, audio, and video downloads~\cite{Wiki:Pareto}, i.e., 20\% of objects account for about 80\% number of downloads. In practice, the Pareto principle can be extended as ``\textit{a small proportion of the objects account for a substantial proportion of downloads.}" Figure~\ref{fig:percentage_ndownload} demonstrates the cumulative distribution function (CDF) of the percentage of app downloads against app ranking by downloads. It indicates that the distribution of app downloads exactly follows the ``Pareto-like'' principle, or more specifically, 3\% of apps account for about 95\% downloads of all apps. 

Other than the Pareto principle, the power-law distribution was discovered to be one basic law of the networked systems~\cite{Adamic:Science2000, Barabasi:sciecne99}, and has been increasingly used to explain various statistics appearing in computer science and networking applications, such as Youtube~\cite{Cha:IMC2007} and search engines~\cite{Wiki:Pareto}. Therefore, we next explore whether the number of app downloads can follow the power-law. A distinguished feature of power law is a straight line in the log-log plot of views versus frequency. However, there are other distributions (e.g.,  log-normal) with a very similar shape. In the real world, the shape of the natural distribution can be affected for various reasons. In fact, it has been found that many distributions whose underlying mechanism is power law fail to show clear power-law patterns, especially at the two ends of the distribution: the most popular and the least popular items~\cite{Barabasi:sciecne99}. Hence, the distribution of app downloads is yet another typical presence of power law in collective behaviors. 

The easiest way to spot a power law is observing a straight line on a log-log plot: the power-law exponent is essentially the slope of the straight line. As illustrated in Figure~\ref{downloadpower}, the apps are ranked by the number of their downloads (in X-Axis). The result indicates that the main trunk follows a quite linear slope, and is truncated at both ends.  While there are many techniques to estimate the slope, the best way of estimating the power-law exponent is using a maximum likelihood estimator~\cite{clauset:SAIM09}. In practice, such an exponent (\texttt{r} = 1.699) can be obtained by  \texttt{Python} \footnote{More specifically, we use the \texttt{Fit} function provided by the \texttt{powerlaw} package. More details can be found at~\url{https://pypi.python.org/pypi/powerlaw}.}. At the curve's head, some extraordinarily popular apps, such as \texttt{WeChat} (having over 6 million downloads) and \texttt{QQ} (having about 7 million downloads), gain substantial proportion of downloads than other apps. In contrast, about 80\% of the apps are downloaded fewer than 10 times, and 30\% of the apps are even downloaded only once. We can then validate that the ``\textit{fetch-only-once}'' principle~\cite{Cha:IMC2007} still applies for app stores. 

In a sense, the power law can be a guiding indicator for content-service providers to allocate the server resources and bandwidth~\cite{Flavio:TOIT2014}. Therefore, a straightforward implication that can be immediately taken away from the popularity of apps is that app-store operators can optimize the resource allocation (e.g., more bandwidth or servers) to the apps that are more frequently downloaded and updated. Indeed, it is quite possible that most app stores have already found and applied such knowledge. However, in Section~\ref{implication}, we  synthesize the power law and the app-management patterns to further help the design and optimization of an app store's workload in terms of cache mechanisms.  

\subsection{Popular Apps by Unique Users}\label{subscriber}
Usually, many app stores take the number of an app's downloads to indicate the app's popularity. However, it is a common observation that people can update their installed apps, and the updates can be affected by the app's release time~\cite{Nayebi:SANER2016}. In addition, some apps can be possibly downloaded or updated with ``\textbf{faking}'' behaviors. For example, apps can be downloaded and updated by automated programs to increase their ranks on an app store. 

In our opinion, the number of an app's unique users is an  intuitive and straightforward indicator that cannot be affected by the preceding factors. We then aggregate the \textbf{unique users (devices)} that ever downloaded, updated, or uninstalled an app in our dataset, and each user is counted once and only once. Such processing can involve more users who ever used this app, because some apps were downloaded in advance to the starting time of our dataset but the updates and uninstallations can still be captured. Figure~\ref{fig:percentage_nuser} shows the correlation between the number of downloads \& updates and the number of unique users per app; the correlation is linearly positive. In other words, apps having more users can gain more downloads \& updates. Note that there are some outliers at the left top.  These outliers actually refer to the apps that have substantially more downloads against the number of the unique users that the apps have.

Figure~\ref{userpower} plots the distribution of the number of apps that a user installs on the device. Intuitively, many users use only a few apps, while a few others try out a large number of apps. The distribution obeys the power law in its tail distribution. We need to mention that the number of apps installed on a device is likely to be underestimated as a lot of devices have pre-loaded apps, and users can install apps directly from the app developers' websites other than Wandoujia. As a result, the distribution somewhat does not strictly follow the power law.  


We then investigate how frequently an app is downloaded and updated by its users. Computed by Formula~\ref{dpperusercount}, we can observe the user preferences of an app and the possibility of adopting its released new versions. As shown in Figure~\ref{dpdistribution}, we are surprised to find that more than 95\% apps received only one download \& update in our five-month dataset. Such a finding indicates that users do not tend to update apps very frequently\footnote{On the Android system, some apps can notify the users the release of updates and navigate them to directly download the updates from their websites rather than an app store. Such a behavior cannot be captured by our dataset.}.

\begin{equation}\label{dpperusercount}
Avg. D\&P per app = \frac{\mbox{Number of Downloads \& Updates}}{\mbox{Number of Unique Users}}
\end{equation}


This simple metric can help identify  some apps that receive an extraordinary number of downloads \& updates. We find that 408 apps receive more than 5 downloads \& updates per user in our dataset, i.e., at least one operation per month. These apps include some popular apps such as \texttt{QQ}. Such a finding confirms that more-popular apps usually have more updates~\cite{Nayebi:SANER2016}. However, we are surprised to observe that some apps can have extremely abnormal behaviors. For example, an app has only 18 unique users, but receives 3,581 downloads \& updates, and 3,563 downloads \& updates come from only one user. In addition, we find that some apps receiving an extraordinary number of downloads \& updates per user can share quite similar behaviors: (1) the management activities are mostly ``downloading" but very few ``updating"; (2) the user reviews are quite sparse, but most user ratings are marked as ``like". Such a finding indicates that some app developers may purposely increase the number of downloads in possibly ``faking" ways, e.g., by automatic programs. We plan to release the detailed information of these apps, including the \texttt{apk} name and the exact number of downloads \& updates per unique user. Some of these apps can still be accessed on Wandoujia, as we cannot make sure that they are problematic ones. However, these apps are moved to the watchlist, and their rankings are tuned down accordingly.

\begin{figure*}[!t]
  \centering
  \begin{center}
        \subfigure[Aggregated data traffic of apps\label{percentage_traffic}]{\includegraphics[width=0.32\textwidth]{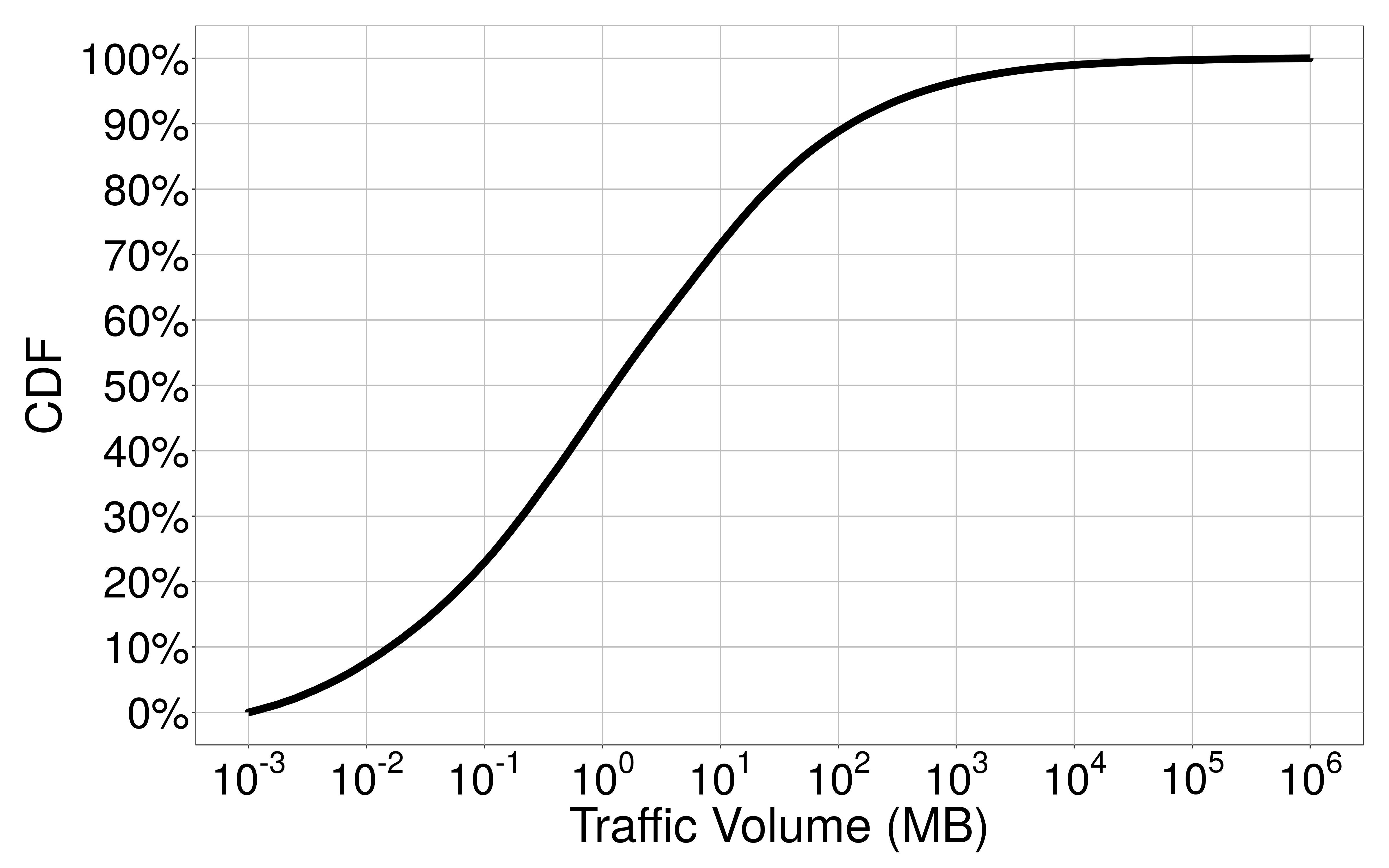}}
        \subfigure[Correlation between number of owned users and traffic drain \label{usertraffic}]{\includegraphics[width=0.32\textwidth]{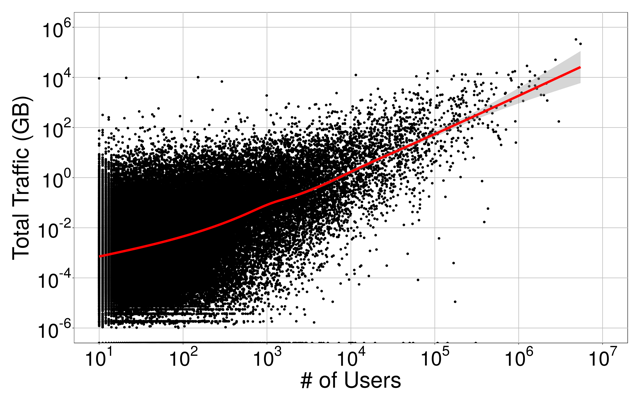}}
         \subfigure[Contributions of the top X apps to total data traffic\label{apptraffic}]{\includegraphics[width=0.32\textwidth]{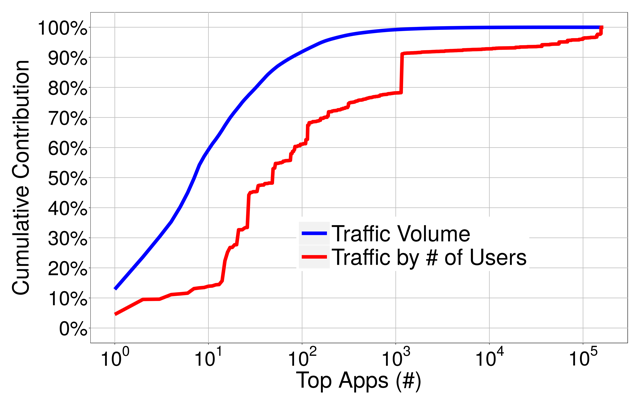}}
            \caption[7.5pt]{App popularity by traffic drain}
    \label{trafficusage}
  \end{center}
\end{figure*}
\begin{figure*}[!t]
  \centering
  \begin{center}
    \subfigure[Aggregated access time of apps\label{percentage_time}]{\includegraphics[width=0.32\textwidth]{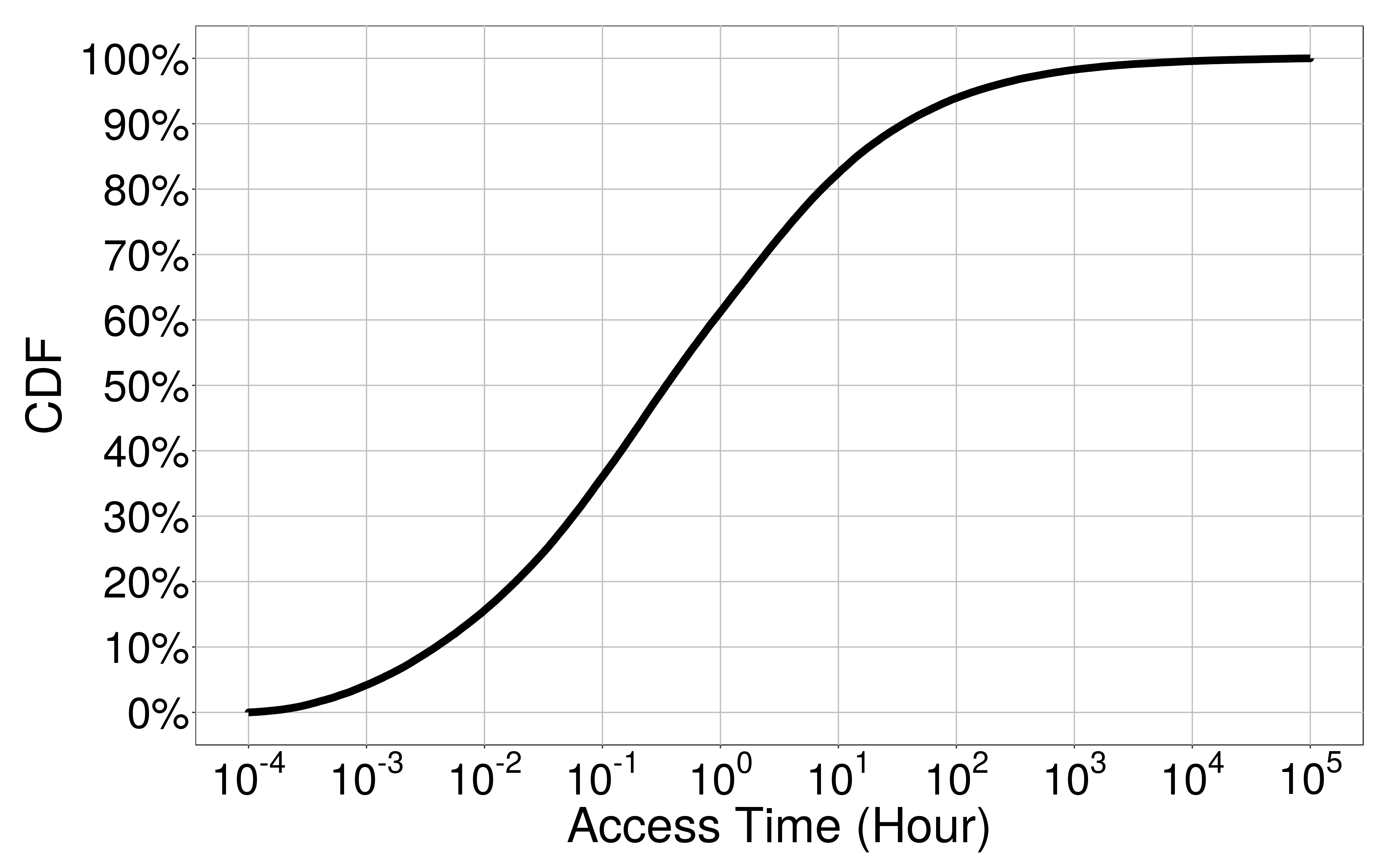}}
    \subfigure[\# of users and access time\label{usertime}]{\includegraphics[width=0.32\textwidth]{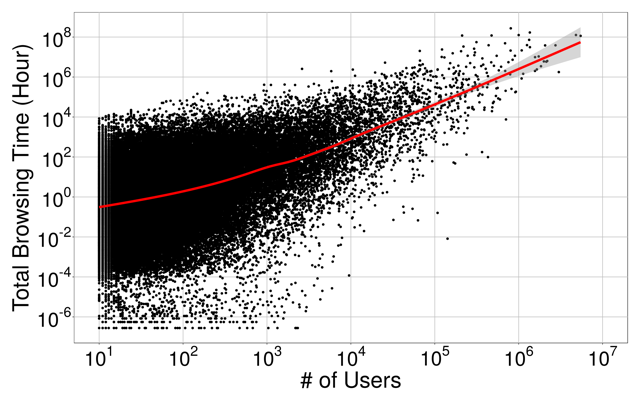}}
  \subfigure[Contributions of the top X apps to total access time\label{apptime}]{\includegraphics[width=0.32\textwidth]{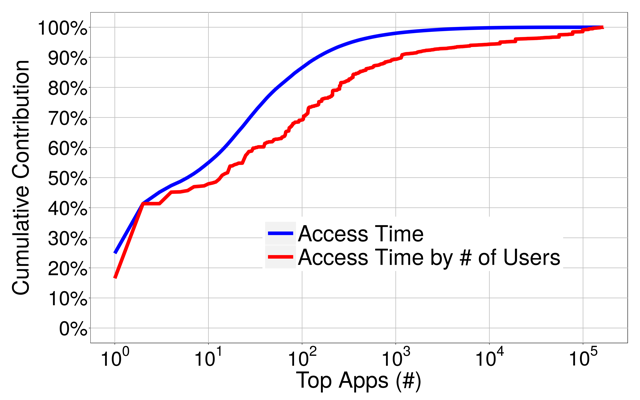}}
         \caption[7.5pt]{App popularity by access time}
    \label{accesstime}
  \end{center}
\end{figure*}

\subsection{Popular Apps by Network Usage}\label{networkpop}

The preceding analysis can identify popular apps based on their numbers of downloads and their unique users. But at the same time, we do not want to discriminate against those apps with few users but with a significant impact on the network, i.e., generating a lot of traffic or accessing the network for long time periods. Indeed, either the number of downloads or that of unique devices of an app can indicate only that this app is downloaded and installed, but we cannot judge whether the app is ``really'' used by users. From the logs of the \textbf{Networked User Set}, an app cannot generate network logs if it is never launched by users. Hence, we use the network activities to examine whether the app is really used. Although we may miss some apps that are usually used offline, e.g., PDF readers or dictionaries, it is a common sense that most of current smartphone apps heavily rely on network.

To illustrate the usage of network, we distinguish the \textbf{aggregated} data traffic and the network-access time from \textbf{all users} that an app owns, respectively. The data traffic comes from both foreground and background. In contrast, we take into account \textbf{only} the access time from foreground, because such a metric indicates how long users really interact with the app when they are connected to the network.

\indent Figures~\ref{percentage_traffic} and~\ref{percentage_time} illustrate the distribution of aggregated traffic/access time of apps, respectively. We can find that the Pareto-like principle still holds for the network activities of investigated apps. We can observe that about 90\% apps consume less than 100 MB traffic volume in five months, and about 94\% apps are used less than 100 hours. Considering that our dataset comes from five months, we can regard that most of apps are not quite active over network. 

Intuitively, the more users an app has, the more traffic and access time the app accounts for. Such an intuition can be reflected in Figures~\ref{usertraffic} and~\ref{usertime}, respectively. Furthermore, if the number of unique users is a good metric for filtering, the top X apps based on the number of unique users should contribute similar amount of data traffic and access time as the top X apps based on the data traffic and access time. We compare the contribution of the top X apps based on these two metrics. Figure~\ref{apptraffic} compares the contribution of the top X apps based on the number of unique users against the top X apps based on the data traffic. We can observe that the cumulative contributions of the top X apps based on traffic and the top X apps based on number of unique users are quite close, by comparing the ``data traffic from all apps'' and the ``data traffic from top apps.'' Likewise, the contributions of the top X apps based on the network-access time and the number of unique users are still  close in Figure~\ref{apptime} although a little difference does exist. We note that over 90\% of the total data traffic and access time is accounted for around the top 2,500-3,000 apps based on the number of unique users. 

The preceding analysis studies the app popularity from various metrics including the number of downloads, the number of unique users, data traffic, and access time. The immediate finding is summarized as follows.\\

\noindent \fbox{%
\parbox{0.485\textwidth}{
\textbf{Finding (F1)}: The popularity of apps can typically follow the Pareto principle. Furthermore, the distributions of the numbers of downloads and unique users even follow the power law. By exploring the average number of  downloads \& updates of an app per user, some possibly faking behaviors can be detected.}} 
\\
\subsection{Released Popular Apps}
\label{sec:released}

Indeed, for researchers who are interested in our dataset, it is not realistic or necessary to make the data for all 0.28 million apps released, as a substantial percentage of them have a very limited number of downloads or unique users. We plan to release the information of some representative apps. Hence, we should define a reasonable threshold instead of releasing all the apps. We choose the intersection of apps having at least 50 downloads, 50 unique users, aggregated 100-hour foreground network access time, and 100-MB traffic from users, as the genre of ``\textbf{Popular App Set}." In total, we have around 3,500 apps in this set. Indeed, it is known that app sampling can have selection bias~\cite{Harman:MSR15}. Nevertheless, the  released data of the chosen apps have sufficient information that can help explore more research topics.



\section{App Management Patterns}\label{management}

In this section, we study how users manage their apps, i.e., which apps are frequently installed, and when they are installed, which apps are more likely to be uninstalled, etc. In addition, we also explore whether the users' app-management activities are consistent with the ratings of the apps.

\begin{figure}
	\centering
	\begin{center}
		\includegraphics[width=0.48\textwidth]{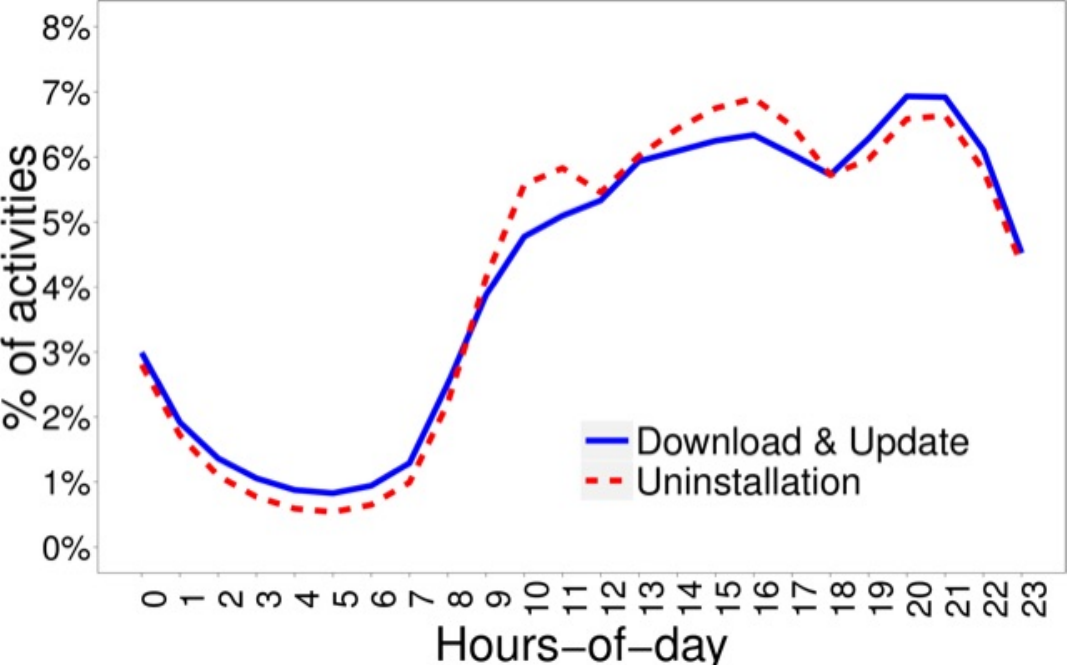}
		\caption[7.5pt]{\textbf{Distribution of diurnal app-management activities.} Each point on the curve represents the percentage of activities performed during the one-hour time interval against the total activities, during the whole day. \textit{For example, activities during 10:00 am-11:00 am account for about 6\% of all activities.}}\label{periodicaldiurnal}
	\end{center}
\end{figure}

\subsection{Diurnal Patterns of App Management}\label{diurnal}

We first investigate \textbf{RQ 2.1}, i.e., \textit{how do the users perform their diurnal management activities of apps?} To this end, we investigate the diurnal downloading, updating, and uninstallation distribution. We aggregate the activities of downloading and updating per app, because these two kinds of activities both reflect the users' interest towards this app and access to the app store. Each entry of activity log is associated with the smartphone's local timestamp to indicate when the activity is performed. We align the timestamp to avoid the inconsistencies caused by different time zones.

As shown in Figure~\ref{periodicaldiurnal}, \textbf{the app-management activities are ``\textit{periodically and regularly}"  performed during a day}. The extent of app downloading and updating activities keeps growing from 6:00 am and reaches the first peak around 11:00 am. The downloading and updating activities decline slightly between 11:00 am to 12:00 pm. It is not very surprising because users may take lunch at this time. The same observation can be found between 4:00 pm to 6:00 pm, i.e., the time on the way back home or at dinner. We can also find that about 32\% of downloading and updating activities are performed from 7:00 pm to 11:00 pm, where they reach the maximum around 8:00 pm-9:00 pm. Such a distribution is quite consistent with human regularity. After 9:00 pm, the downloading and updating activities decline quite sharply, and reach the minimum around 5:00 am. However, at  midnight, downloading and updating activities occupy about 7\% in total, implying that there are still a considerable number of active users at this time. 

The preceding results indicate the temporal patterns when users access the app store. Hence, the app-store operators  should reserve sufficient bandwidth at the peak to reduce user-perceived latency. In addition, app developers can leverage this finding in their release planning, e.g., pushing the update notifications to their users at the right time.

Activities of uninstalling apps present a similar distribution to those of downloading/updating apps. However, knowing when users uninstall apps may be less useful, because the uninstallation activities do not have interactions with app-store operators or app providers.\\

\noindent \fbox{%
\parbox{0.485\textwidth}{
 \textbf{Finding (F2)}: The app-management activities can reach fixed peaks and are performed quite periodically during a day.} }

\subsection{App Selection Patterns}\label{selection}

Then we explore \textbf{RQ 2.2}, i.e., \textit{what apps are more likely to be selected and liked by users?} Such activities can imply the user interests and needs towards apps. For app-store operators, such information can help improve the recommendation systems.

\begin{figure}
	\centering
	\begin{center}
		\includegraphics[width=0.45\textwidth]{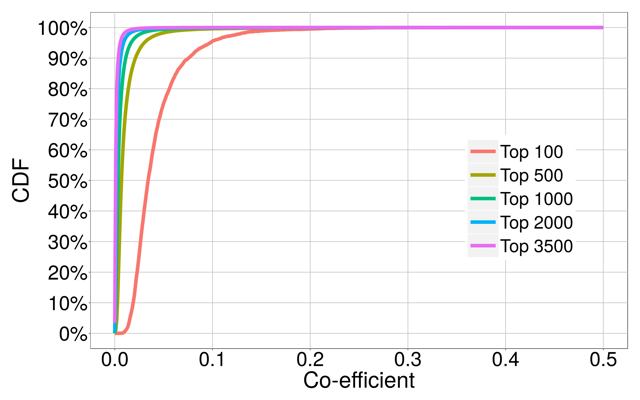}
		\caption[7.5pt]{Jaccard coefficient of co-installed apps}\label{coefficient}
	\end{center}
\end{figure}

In fact, some previous studies investigated how users select apps~\cite{MoreleyMao:IMC11, Liao:CIKM13, Zhong:TMC13, Petsas:IMC13}, and some findings were reported. A straightforward metric is to check the ``\textbf{cluster effect}": which apps are more likely to be selected together. We adopt a similar metric, but explore the study at two levels: the \textbf{micro-level} of co-installed apps, and the \textbf{meso-level} of correlated app categories. 

From our previous analysis, a substantial percentage of apps are rarely downloaded and updated. Therefore, we choose only the top 3,500 apps mentioned in Section~\ref{sec:released}. 


\subsubsection{Clustering Co-Installed Apps}

We study the frequently co-installed apps. Given two apps $app_{m}$ and $app_{n}$, we employ the  \textbf{\emph{Jaccard Similarity Coefficient}} (denoted as $\lambda$) to measure the possibility of how they are installed together by users. We denote the number of unique devices that install either $app_{m}$ or $app_{n}$ as $\mathbb{D}$ $(app_{m} \cup app_{n})$, and the number of unique devices that install both $app_{m}$ and $app_{n}$ as $\mathbb{D}$ $(app_{m} \cap app_{n})$. We compute $\lambda$ as $\frac{\mathbb{D}(app_{m} \cap app_{n})}{\mathbb{D} (app_{m} \cup app_{n})}$.

Figure~\ref{coefficient} shows the \textbf{Jaccard Similarity Coefficient} of the top-$N$ apps, where $N$ varies from 100 to 3,500. With the increasing number of $N$, the value of $\lambda$ decreases significantly. The CDF  indicates that the $\lambda$ value of over 95\% of app pairs is lower than 0.1. In other words, there are a very small fraction of apps that are frequently co-installed together. 

To better demonstrate which apps are frequently co-installed, we employ the metric of Point-wise Mutual Information (PMI), which is widely used in information retrieval to identify co-occurrence of objects. Formally, let us assume that $n_i$ represents the number of downloading and updating activities that contain $app_{i}$, $n_{ij}$ represents the number of activities that contain both $app_{i}$ and $app_{j}$, and $N$ denotes the total number of activities, and thus the PMI is computed as follows:

\begin{align}
PMI(i,j) = log(\frac{p_{ij}}{p_i * p_j}) &= log(\frac{n_{ij} / N}{(n_i / N) * (n_j / N)}) \\
&=log(\frac{n_{ij} * N}{n_i * n_j}) \\
&=log(N) + log(\frac{n_{ij}}{n_i * n_j})
\end{align}

\begin{figure}[htbp]
    \centering
    \includegraphics[width=0.5\textwidth]{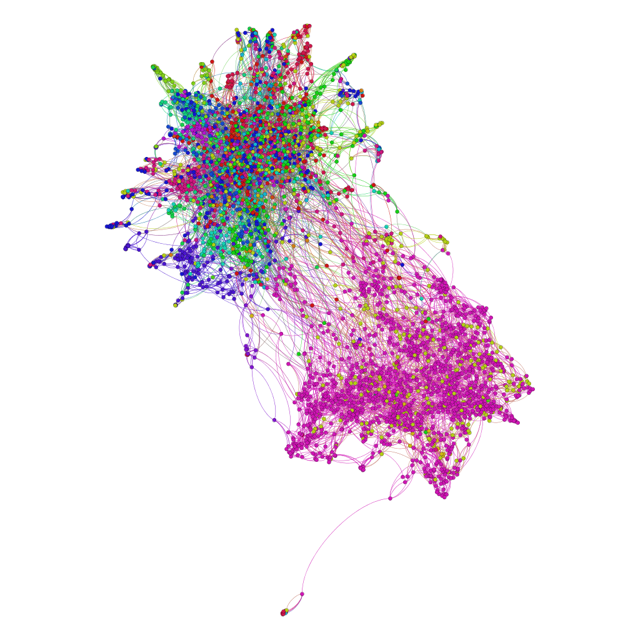}
    \caption{Network structure of co-installed apps}
    \label{toppmi}
\end{figure}

The larger PMI that two apps hold, the more probably they are co-installed. We visualize the network structure of co-installed apps based on the PMI, and employ a force vector algorithm \cite{jacomy:plosone2014} to detect the community structure. As shown in Figure~\ref{toppmi}, there exist some significant clusters, i.e., the apps  \texttt{WeChat, QQ, QQMusic, Dianping} are more likely to be co-installed. To make further exploration, we find that apps from some big clusters share some characteristics, i.e., developed by the same \textit{vendor} or from the same \textit{category}.

\begin{figure*}[t]
	\centering
	\begin{center}
		\includegraphics[width=0.92\textwidth]{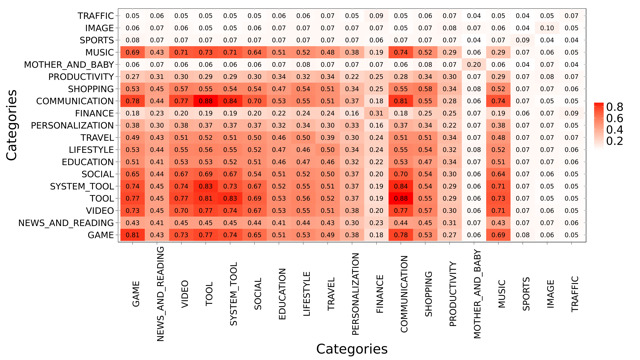}
        \caption[7.5pt]{Heatmap of the category-level relationships of co-installed apps.}
        \label{coinstallation}
	\end{center}
\end{figure*}

The \textit{vendor} information of an app can reflect who develops this app. Usually, the naming rules of an Android app can reflect the vendor information. For example, the package \texttt{com.tencent.mm} can be processed by removing the general stopword ``\texttt{com}" and the app's feature word ``\texttt{mm}", and the vendor information ``\texttt{tencent}" is extracted. We find that \textbf{a number of co-installed apps with high $\lambda$ values come from the same vendor}. For example, the pair of $<$ \texttt{Huawei Backup} (used by 555,332 devices), \texttt{Huawei Account Manager} (used by 151,541 devices) $>$ has the $\lambda$ value of 0.274, and these two apps are both provided by Huawei. Furthermore, \textbf{the $\lambda$ value can be much higher, if two apps developed by the same vendor also belong to the same category}. For example, the pair of $<$ \texttt{WeChat} (used by 3,048,557 devices), \texttt{QQ} (used by 7,225,074 devices)$>$ holds the $\lambda$ value of 0.43, and the two apps are provided by Tencent.

There are many possible reasons why apps from the same vendor are often co-installed. A vendor usually focuses on a specific application domain, e.g., Tencent is the largest messaging service provider in China. Tencent \texttt{QQ} is the most popular instant messaging app in China;  \texttt{WeChat} not only supports instant messaging, but also provides social communication features such as content sharing. Another reason is that there might be ``\textbf{in-app bundled installation}" in some apps. For example, when users install an app, the app's vendor may implicitly or explicitly recommend the users to install their other apps. For simple validation, we make field studies by selecting 50 apps from well-known app developers such as Qihoo, Baidu, and Tencent, and install them manually. 14 apps out of the 50 apps recommend installing other apps in their installation wizard, and 8 apps of these 14 ``bundled" installations are provided by the same vendor. 
 
\subsubsection{Correlation of App Categories}

The \textit{category} information of an app indicates the functionality and application domains of the app. We can infer the users' needs and interests according to their selected app's category. Given two app categories \textit{M} and \textit{N}, we denote the number of unique users who install an app either from $M$ or $N$ as $\mathbb{D}$ $({M} \cup {N})$, and the number of unique devices that install apps from both $M$ and $N$ as $\mathbb{D}$ $({M} \cap {N})$. We then compute $\frac{\mathbb{D}({M} \cup {N})}{\mathbb{D} ({M} \cap {N}}$ to indicate how likely that the apps in $M$ and $N$ are installed together. We also take into account the special case where \textit{$M$=$N$}, indicating how many users install more than one app in the same category.

Figure~\ref{coinstallation} shows the probability distribution that the apps from different categories are selected together. The categories are sorted by the descending order of the number of apps (in X-axis). Apps providing related functionalities are more likely to be selected together. For example, users may want to share a video from a video-player app to friends in a communication app (e.g., correlation between \textit{COMMUNICATION} and \textit{VIDEO} is 0.77), or use a viewer app to open a document that is received by an instant messenger app (e.g., correlation between \textit{TOOL} and \textit{COMMUNICATION is 0.88}). 

It is not surprising that users may install more than one app in the same  category. For example, in \textit{GAME} and \textit{COMMUNICATION}, the correlations are both more than 0.8. The result suggests that users have more interests and needs in these categories.\\

\noindent \fbox{%
\parbox{0.485\textwidth}
{
\textbf{Finding (F3)}: In terms of app installation, apps from some app categories are frequently installed together such as \textit{COMMUNICATION} and \textit{TOOL}. Additionally, apps that come from the same vendor or category are more likely to be installed together.
}
}

\subsection{Uninstallation Patterns}\label{uninstallation}

\begin{figure*}
\centering
\begin{center}
\subfigure[Possibility of app abandonment\label{duratio}]
{\includegraphics[width=0.32\textwidth]{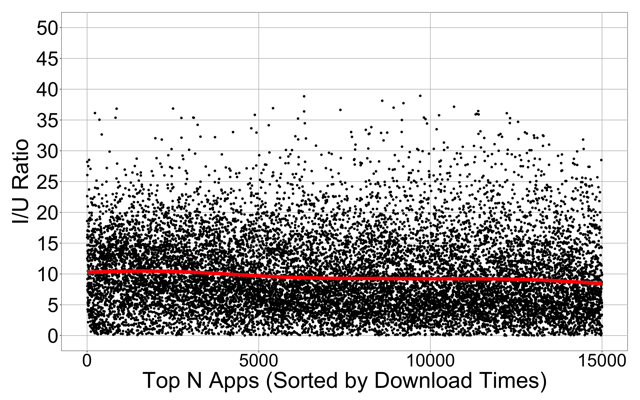}}
\subfigure[Lifecycle of abandoned apps\label{lifecyledistribution}]{\includegraphics[width=0.32\textwidth]{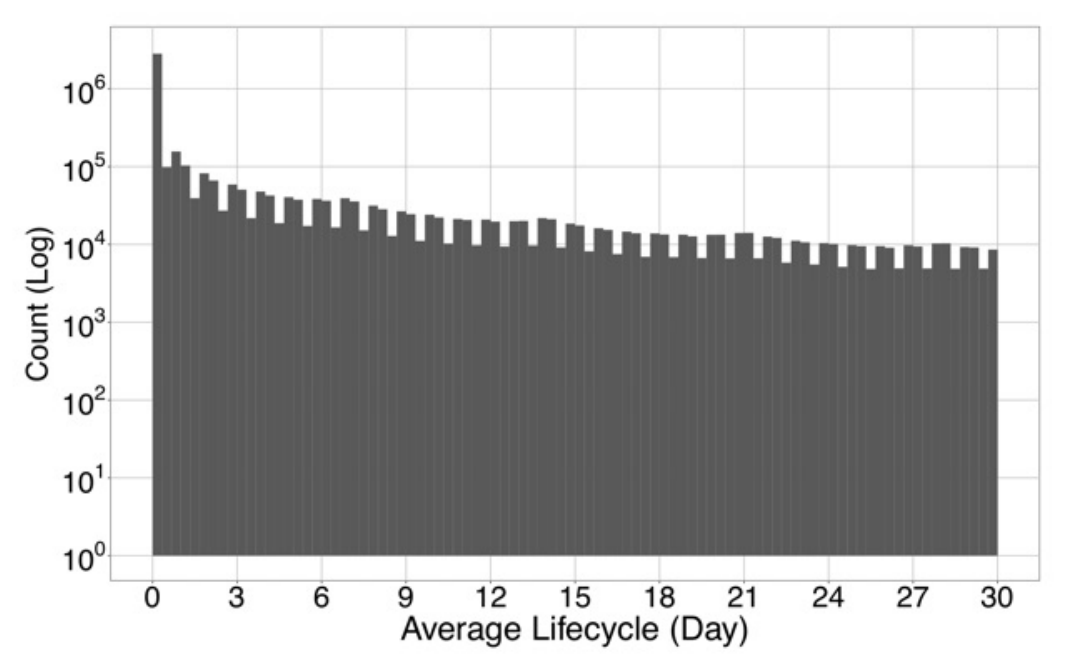}}
\subfigure[Lifecycle of frequently abandoned apps \label{lifecyclewithIU}]{
\includegraphics[width=0.32\textwidth]{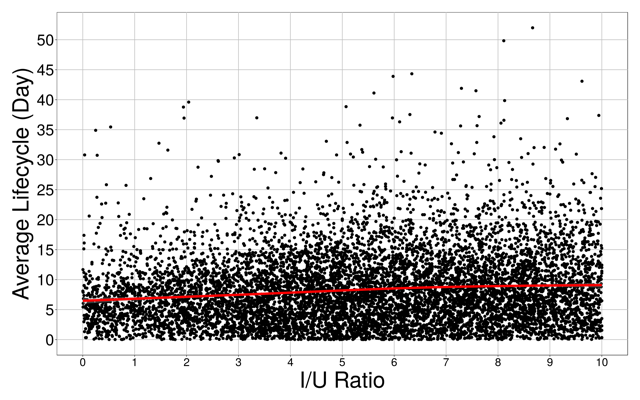}}
\caption[7.5pt]{Lifecycle of abandoned app.}\label{lifecycle}
\end{center}
\end{figure*}

We next explore \textbf{RQ 2.3}, i.e., \textit{how can we identify an app that is more likely to be disliked by users?} Such a question is quite crucial to app developers and app-store operators. App developers can know whether their apps are appreciated by users or not, so that they can examine their apps in time to avoid losing users. App-store operators can improve their recommendation systems to filter unpopular, low-quality, or even malware apps.

However, to answer the question, only the absolute number of unistallations of an app may not be a good indicator. For example, apps with a high number of uninstallations may also have a high number of downloads. So we compute the metric of \textit{installation/uninstallation ratio}, or \textbf{``I/U ratio"} for short, denoted as $\Omega(app_{i})$, to indicate how likely an app could be abandoned. Given an app $app_{i}$, we compute $\Omega(app_{i})$ as $\frac{\sum\mathbb{I}_{device_{i}}}{\sum\mathbb{U}_{device{i}}}$, where $\sum\mathbb{I}_{device_{i}}$ and $\sum\mathbb{U}_{device_{i}}$ represent the number of devices that install and uninstall $app_{i}$, respectively. We extract all devices that appear in both installation and uninstallation activity logs of $app_{i}$ from the \textbf{Universal User Set}. 

Note that there may be some biases when computing $\Omega(app_{i})$, i.e., we cannot capture how many installations have already existed for an app before the starting time of our dataset, nor can we know how many uninstallations would be performed for this app after the ending time of our dataset. However, our analysis aims to derive the overall trend that an app would be abandoned by users during our dataset's time span, we still simply rely on the preceding computation.    

\indent The lower value of $\Omega(app_{i})$ an app holds, the higher likelihood that this app could be abandoned. \textbf{$\Omega$ can tell how much an app is actually abandoned by users.} For better illustration, Figure~\ref{duratio} shows the scattered distribution of $\Omega(app_{i})$. The mean and median values of $\Omega$ are 7.89 and 5.875, respectively.  However, the value of $\Omega(app_{i})$ seems to be irrelevant to the number of downloads, indicating that $\Omega(app_{i})$ is at least  not a good signal to comprehensively reflect how much an app is disliked by users, because users may not always uninstall an app even if they do not need the app any longer. 

To further infer the users' attitude towards apps, we evaluate the lifecycle of abandoned apps by combining the temporal information with $\Omega$. Such an evaluation is motivated by an intuition that an app is likely to be a disliked one if it is uninstalled shortly after being installed. To this end, we compute the app's lifecycle by the timestamps of installation and uninstallation. We have two immediate observations. First, from Figure~\ref{lifecyledistribution}, if an app is uninstalled, its lifecycle can be identified. \textbf{About 60\% of abandoned apps can ``survive" for only less than 1.5 days, and about 80\% of abandoned apps can ``survive" for less than a week}. Such results are largely consistent with the ones derived from the one-month data in our previous work~\cite{Li:IMC15}, i.e., \textit{60\% abandoned apps can ``survive'' for only less than two days, and about 93\% abandoned apps can ``survive'' for less than a week}. Second, from Figure~\ref{lifecyclewithIU}, we can find \textbf{a quite weak positive correlation between $\Omega$ and the lifecycle of abandoned apps}. In other words, apps with a lower $\Omega$ seem to be a bit more probably uninstalled within a shorter interval. However, such a finding implies that we should find more meaningful indicators.  In practice, we have devised some new signals to more accurately predict how an app could  be adopted by users~\cite{Li:WWW2016}.  \\

\noindent \fbox{%
\parbox{0.485\textwidth}{
     \textbf{Finding (F4)}: An app's installation/uninstallation ratio exhibits a weakly positive correlation to its lifecycle. Most ``abandoned" apps are often uninstalled within 1.5 days after they are installed. } }

\subsection{User Rating Patterns}\label{rating}
Usually, on most app stores such as Apple App Store and Google Play, the users' attitudes towards an app can be somewhat reflected by the ratings of the app. End-users may be simply attracted by the overall ratings of an app at their first sight, before scrolling down the page to see textual user reviews. To some extent, the ratings of an app can imply the quality of the app. For example, Google Play allows users to rate apps with a 5-star model, where 1-star refers to the lowest rating and 5-star refers to the highest rating. In contrast, Wandoujia allows users to simply tag an app with a binary metric, i.e., ``\textit{like}'' or ``\textit{dislike}.'' 

Intuitively, an app is considered to be of higher quality if it receives a higher average rating from its users. Previous study reported that the score of ratings can have positive correlation with the app rank by downloads~\cite{Harman:MSR2012}. In practice, there are many issues of directly using this simple and straightforward metric. Online ratings can suffer from the sparseness of some apps. To this end, we argue that the management activities may be more objective, e.g., downloading and updating an app can reflect positive attitudes of this app, while uninstalling the app can reflect negative attitudes. In this way, we are interested in investigating whether user ratings of an app are typically consistent with management activities of the app.

To this end, we then move to \textbf{RQ 2.4}, i.e., \textit{are the user ratings of an app consistent with the app-management activities, with respect to the user attitude towards this app?}

We compute the average rating of an app on Wandoujia, namely \emph{likerate}, as denoted in Equation~\ref{likerate}. In other words, the higher likerate an app holds, the more possibly it is preferred by the users. We then correlate the number of downloading and updating activities (actually the installations) to the likerate of apps. Indeed, the likerate metric reflects only the general attitude of an app by its users, but suffers from the absence of the attitude towards a specific version of the app.
\begin{equation}\label{likerate}
\mbox{likerate}=\frac{\mbox{number of likes}}{\mbox{number of likes} + \mbox{number of dislikes}}
\end{equation}

\subsubsection{Correlation between Rating and Selection}
\begin{figure}[hbtp]
    \centering
    \includegraphics[width = 0.42\textwidth]{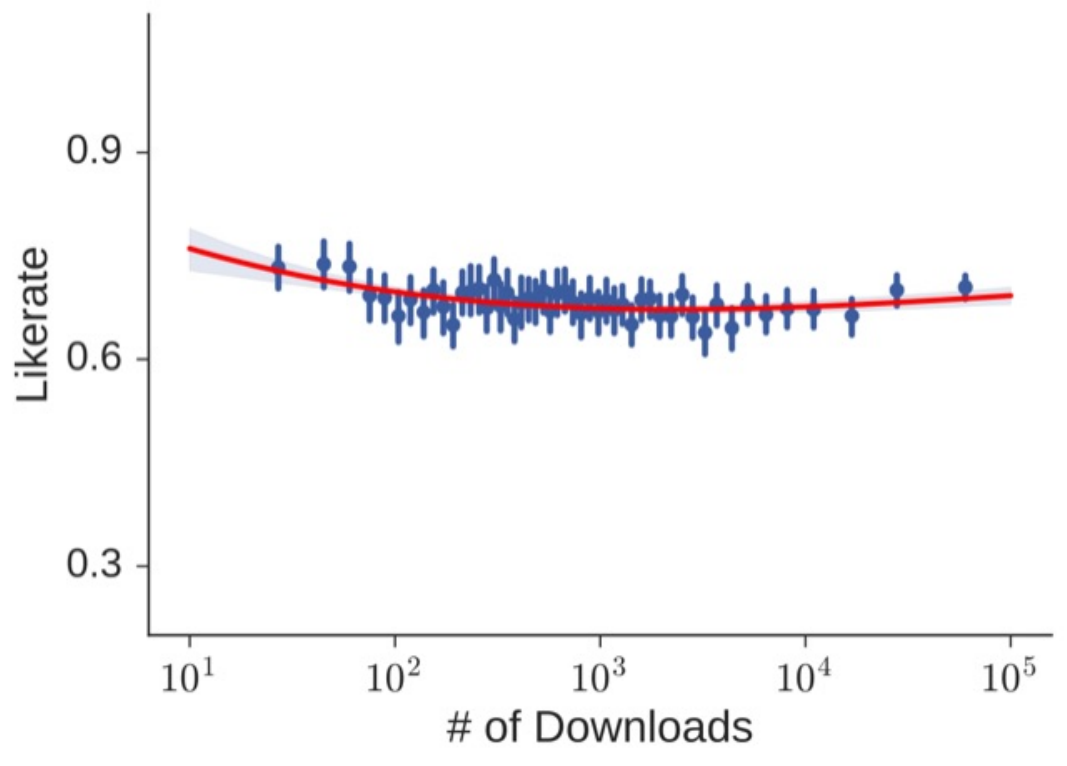}
    \caption{\textbf{The number of downloads against likerate}. The number of downloads is weakly correlated with likerate for popular apps and negatively correlated for unpopular apps. }
    \label{fig:d-likerate}
\end{figure}

In Figure~\ref{fig:d-likerate}, we rank all apps that have received at least 5 ratings during our five-month period, split them into equally-sized bins\footnote{We employ the \texttt{regplot} function provided by Python's \texttt{seaborn} package, where one can control the size of bins by the parameter \texttt{x\_bins}.}, and demonstrate the mean and standard deviations of their likerates.  
As the number of downloads follows the power law, we plot the results at the \textit{log} scale (X-axis). We then run a regression process to derive the correlation between the likerates and the number of downloads, and use the \texttt{seaborn} package of \texttt{Python} to plot the results, as shown in Figure~\ref{fig:d-likerate}.

Some immediate observations can be reached. Surprisingly, when the number of downloads is less than 1,000, it is negatively correlated with the likerate. In other words, the more times an app is downloaded, the more likely it is \textit{disliked} by users. Such an observation is rather counter-intuitive. One explanation is that the apps not frequently downloaded may be sensitive to fake ``\textit{like}" ratings, while some frequently downloaded apps may be maliciously rated down by their competitors. It reminds that app-store operators should pay attention to address such an issue in their ranking and recommendation system. When the number of downloads exceeds 10,000, the correlation becomes quite weakly positive. \textbf{In either case, the result indicates that only the number of downloads is weak as a guiding indicator for ranking apps. In particular, it may not be valid for the apps that are not popular, with few ratings, or newly published on an app store.} 

\begin{figure}[hbt]
    \centering
    \includegraphics[width = 0.42\textwidth]{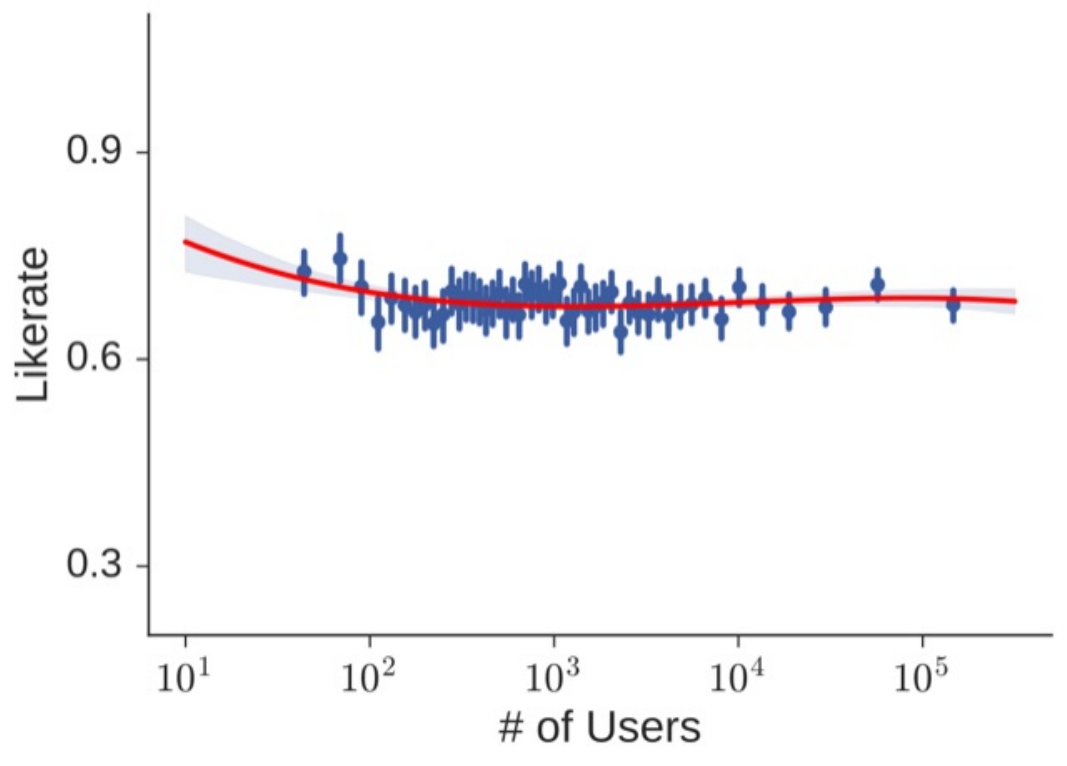}
    \caption{\textbf{The number of unique users against likerate of apps}. The number of unique users seems not to be a positive indicator of likerate.}
    \label{fig:user-likerate}
\end{figure}

One may argue that the ratings are given to all versions of an app rather than a specific version. To alleviate the bias, we explore the likerate and the number of users (devices) that ever used the app in our five-month dataset. From Figure~\ref{fig:user-likerate}, we can observe that the number of users (devices) installing an app is even negatively correlated with its likerate, although the correlation is quite weak either. From a macro perspective, it indicates that the app's versions do not  have very significant impact on its ratings. 
\subsubsection{Correlation between Rating and Abandonment} 

\begin{figure}[hbt]
    \centering
    \includegraphics[width = 0.42\textwidth]{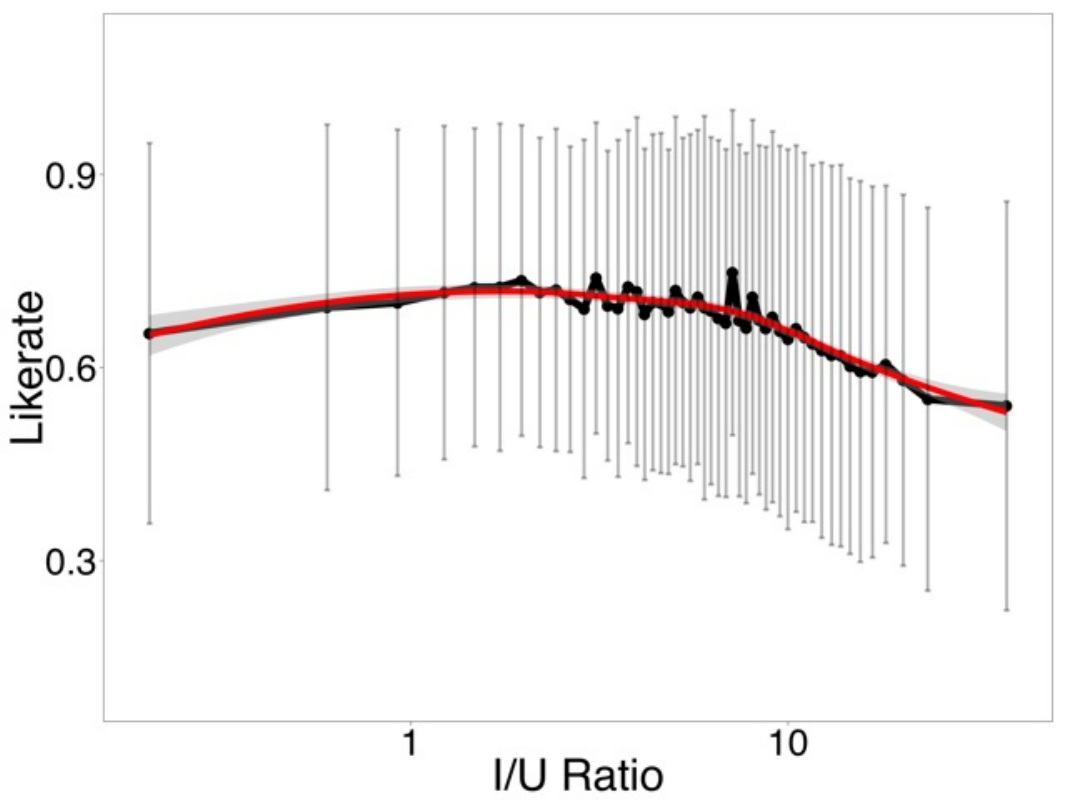}
    \caption{\textbf{I/U ratio against likerate of apps}. I/U ratio is not a promising indicator of likerate.}
    \label{fig:du-likerate}
\end{figure}

Another intuition is that an uninstallation may indicate that a user dislikes an app. Although most app stores usually do not report this statistic, we can compare the number of uninstallations of an app with its user ratings. Instead of using the raw number of uninstallations, we still use the metric of $\Omega$, which is computed as the total number of downloads divided by the total number of uninstallations. Intuitively, the lower $\Omega$ an app holds, the more likely the app is disliked. We plot the correlation of $\Omega$ of an app and the corresponding likerate in Figure~\ref{fig:du-likerate}. Again, we rank all apps with at least 5 ratings by $\Omega$ and split them into equally-sized bins. 

Interestingly, when $\Omega$ is below 1, a quite weakly positive correlation is observed between $\Omega$ and the likerate. However, when $\Omega$ is over 10, the correlation becomes more negative. In either case, the correlation is not significant and presents a long error bar. Apparently, the I/U ratio is also at best a weak indicator of user preferences. To infer user preferences from activities, new signals need to be conducted. In practice, our recent work~\cite{Li:WWW2016} explored how app-management sequences can be a promising indicator with machine learning algorithms. Due to page limit, the details are not included in this article. \\

\noindent \fbox{%
\parbox{0.485\textwidth}{
     \textbf{Finding (F5)}: Neither the number of unique users nor the users' attitudes towards an app (installing and uninstalling) can exhibit significant correlation of users' ratings towards this app. Such a finding is somewhat inconsistent with a previous study~\cite{Harman:MSR2012}, which reported that the score of user ratings can be positively correlated with the app ranking determined by the number of downloads of the app. For apps that have few or spare reviews, new indicators are required to judge the user attitudes.} }

\section{Network Activity Patterns}\label{network}

\begin{table*}[htbp]\small
\newcommand{\tabincell}[2]{\begin{tabular}{@{}#1@{}}#2\end{tabular}}
	\centering
	\caption{Network summary of all app categories}
	\begin{threeparttable}
	\begin{tabular}{l|r|r|r|r|r|r|r|r}
	\hline
	App Category & \tabincell{c}{\textit{C}-Time\\
	(B)}& \tabincell{c}{\textit{W}-Time\\
	(B)} & \tabincell{c}{\textit{C}-Time\\(F)}& \tabincell{c}{\textit{W}-Time\\(F)}& \tabincell{c}{\textit{C}-Traffic\\(B)} & \tabincell{c}{\textit{W}-Traffic\\(B)} & \tabincell{c}{\textit{C}-Traffic\\(F)}& \tabincell{c}{\textit{W}-Traffic\\(F)}  \\
	\hline

	\bfseries BEAUTIFY & 43.49\% & 43.28\% & 6.33\% & 6.90\% & 4.56\% & 71.48\% & 4.03\% & 19.94\% \\
	\bfseries COMMUNICATION & 42.56\% & 51.39\% & 2.88\% & 3.17\% & 10.88\% & 18.71\% & 22.33\% & 48.08\% \\
	\bfseries EDUCATION & 43.26\% & 53.76\% & 1.63\% & 1.34\% & 9.78\% & 34.36\% & 9.38\% & 46.48\% \\
	\bfseries FINANCE & 42.26\% & 57.07\% & 0.37\% & 0.31\% & 29.13\% & 20.80\% & 26.68\% & 23.39\% \\
	\bfseries GAME & 47.23\% & 50.52\% & 1.12\% & 1.13\% & 12.95\% & 20.90\% & 26.44\% & 39.71\% \\
	\bfseries IMAGE & 44.15\% & 55.64\% & 0.09\% & 0.12\% & 7.11\% & 64.61\% & 4.42\% & 23.85\% \\
	\bfseries LIFESTYLE & 41.36\% & 58.38\% & 0.13\% & 0.14\% & 30.25\% & 31.42\% & 12.61\% & 25.72\% \\
	\bfseries MOTHER\_AND\_BABY & 34.25\% & 65.21\% & 0.15\% & 0.39\% & 11.26\% & 23.59\% & 11.23\% & 53.91\% \\
	\bfseries MUSIC & 46.14\% & 52.83\% & 0.56\% & 0.47\% & 6.89\% & 34.21\% & 8.20\% & 50.70\% \\
	\bfseries NEWS\_AND\_READING & 42.24\% & 56.07\% & 0.71\% & 0.97\% & 7.85\% & 15.85\% & 17.36\% & 58.94\% \\
	\bfseries PRODUCTIVITY & 41.55\% & 58.14\% & 0.13\% & 0.18\% & 15.42\% & 63.69\% & 3.64\% & 17.25\% \\
	\bfseries SHOPPING & 38.05\% & 61.50\% & 0.12\% & 0.33\% & 8.29\% & 15.17\% & 11.04\% & 65.50\% \\
	\bfseries SOCIAL & 42.94\% & 55.97\% & 0.42\% & 0.67\% & 6.67\% & 20.06\% & 12.62\% & 60.65\% \\
	\bfseries SPORTS & 41.49\% & 57.83\% & 0.33\% & 0.36\% & 17.72\% & 23.06\% & 14.93\% & 44.29\% \\
	\bfseries SYSTEM\_TOOL & 46.51\% & 53.18\% & 0.14\% & 0.17\% & 9.56\% & 77.83\% & 1.95\% & 10.67\% \\
	\bfseries TOOL & 46.19\% & 52.50\% & 0.64\% & 0.67\% & 5.11\% & 49.30\% & 6.06\% & 39.53\% \\
	\bfseries TRAFFIC & 40.91\% & 58.94\% & 0.10\% & 0.05\% & 31.17\% & 29.41\% & 17.41\% & 22.01\% \\
	\bfseries TRAVEL & 43.57\% & 56.28\% & 0.11\% & 0.04\% & 14.15\% & 46.65\% & 7.69\% & 31.51\% \\
	\bfseries VIDEO & 40.84\% & 58.38\% & 0.17\% & 0.61\% & 0.93\% & 35.69\% & 1.34\% & 62.04\% \\
	\bfseries MISCS & 48.19\% & 51.74\% & 0.03\% & 0.04\% & 11.11\% & 55.40\% & 3.86\% & 29.63\% \\
	\hline
	\end{tabular}
	\begin{tablenotes}
    \footnotesize
    \item \textbf{\textit{W}} and \textbf{\textit{C} refer to Wi-Fi and Cellular, respectively.}
    \item \textbf{\textit{B}} refers to background and \textbf{\textit{F}} refers to foreground. 
    \end{tablenotes}
	\end{threeparttable}
	\label{networksummary}
\end{table*}

Previous studies have already revealed some observations on network usage of apps, e.g., by the TCP flows on tier-1 network~\cite{MoreleyMao:IMC11}, or usage logs by field studies~\cite{Zhong:SIGMobileComm13, Zhong:CHI2012}. In contrast, our study is performed at a much finer granularity. First, we distinguish the daily data traffic and access time from Wi-Fi and cellular network, respectively. Second, we distinguish the daily data traffic and access time from foreground and background, respectively. 

Based on the granularity of network activities, our study aims to explore some issues that are not covered by previous efforts. End-users can know which apps are network-intensive, and thus result in more data traffic and battery consumption. In this way, end-users can identify the apps that generate ``\textit{undesirable}" traffic, pay attention to granted network permissions, or even uninstall these apps. App-store operators can identify some potentially problematic apps. App developers can fix possible bugs, and OS vendors can patch their frameworks to avoid potential threats. 

\subsection{Access Time Patterns}
First, we aim to answer \textbf{RQ 3.1}, i.e., \textit{which apps are the users likely to interact with, when these apps are under Wi-Fi and cellular networks, respectively?} 
We investigate the access time of the network activity log. Intuitively, access time may reflect two important insights. First, the foreground access time of an app indicates how long a user stays in this app when he/she is connected to the Internet. Therefore, such a metric can somewhat imply how much the user likes or needs the app. Second, similar to the background data traffic, the background access time indicates how long an app connects to network when users do not interact with it. Therefore, the background access time can imply the ``liveness" of the app after it is launched.

We illustrate the access time distribution among app categories, as shown in Table~\ref{topapps} (in Section~\ref{dataset}). When the foreground access time is explored, it is not surprising that the \textit{COMMUNICATION} apps account for 48.32\% of cellular time and 46.08\% of  Wi-Fi time against all apps. It is also interesting to find that users spend a lot of time on \textit{BEAUTIFY} (11.96\% under cellular and 11.33\% under Wi-Fi) to personalize their smartphones (typical \textit{BEAUTIFY} apps include choosing themes, background, icon types, and rings), and \textit{TOOL} (12.16\% under cellular and 11.06\% under Wi-Fi) to optimize the smartphones (typical \textit{TOOL} apps can include battery manager, third-party input method, and  weather).  

We then break down the access time spent at foreground and background under cellular and Wi-Fi of all apps of a category, respectively. From Table~\ref{networksummary} (Columns 2-5), we are surprised to observe that foreground time accounts for only less than 2\% (by aggregating \textit{W}-Time(F) and \textit{C}-Time(F)) in most categories, but the background time occupies more than 98\% (by aggregating \textit{W}-Time(B) and \textit{C}-Time(B)). Even for the \textit{COMMUNICATION} apps, the background time accounts for more than 94\% of all network time. In other words, \textbf{most apps still keep ``long-and-live" TCP connection at background after being launched, even users do not interact with them}. The background time may be reasonable for the apps that heavily rely on network, e.g., \textit{COMMUNICATION} and \textit{SOCIAL}. Most of these apps require auto synchronization or notification. However, it is hard to confirm whether many apps from other categories should have ``reasonable" continuous network connection at background. 

Indeed, the background TCP connection does not always indicate data traffic loss, as our dataset can monitor the actual data transmission. In other words, the live TCP connection can probably generate no data traffic. This finding also indicates that currently most Android apps could stay silently in memory even when they are switched to background. Such a mechanism can be useful, as the app can be quickly ``waken up'' and the network connection can be fast restored. However, it can also lead to large memory occupation and have side effect of system performance. In fact, many Android users complain that their devices become too sluggish to respond to user interaction, and the user experiences really are unsatisfying. It would be interesting to explore whether keeping ``long-live'' network connection at background could be a potential factor of such unsatisfactory user experiences.\\
 
\noindent \fbox{%
\parbox{0.485\textwidth}{
     \textbf{Finding (F6)}: A large number of apps keep long-lived TCP connection when they are not currently ``used" by users. It is not quite sure whether the background connection is always reasonable.} } 
\subsection{Data Traffic Patterns}

We then move to \textbf{RQ 3.2}, i.e., \textit{which apps are more ``traffic-intensive'' and how much traffic is generated by these apps?}
We identify the apps that consume substantial data traffic. We aggregate apps by their categories and summarize the total traffic consumption (in GB) from Wi-Fi and cellular, respectively. As shown in Table~\ref{topapps}, \textit{VIDEO} apps are the most ``\textit{traffic-intensive}". Apps from \textit{VIDEO} category consume 61.56\% of Wi-Fi traffic and 12.02\% of cellular traffic against all apps. Interestingly, apps from \textit{TOOL} and \textit{SYSTEM\_TOOL} consume a lot of data traffic. The apps in these two categories include input method, anti-virus, and app management. It indicates that Android users heavily rely on these apps to manage, optimize, and personalize their devices. 

We then classify data traffic into two dimensions: (1) Wi-Fi and cellular; (2) foreground and background. Such classifications can provide more details of traffic consumption.

\subsubsection{Traffic from Wi-Fi and Cellular}

As shown in Table~\ref{networksummary} (Column 6-9), in most categories, it is not surprising that the data drain generated under Wi-Fi accounts for more than \textbf{70\%} of total traffic. In the categories of \textit{TOOL}, \textit{MUSIC}, \textit{SYSTEM\_TOOL}, \textit{SHOPPING}, and \textit{EDUCATION}, more than 75\% of data traffic is from Wi-Fi. For \textit{VIDEO}, almost 97\% of traffic drain comes from Wi-Fi. A possible reason is that most of these apps are usually used in places with stable Wi-Fi connection, e.g., at home or cafe. The situation is a bit different in \textit{COMMUNICATION, GAME, LIFESTYLE, FINANCE}, and \textit{TRAFFIC}, where the traffic from cellular network accounts for more than 30\%. Such results can be consistent to the purposes of the apps. For example, users may use \textit{COMMUNICATION} apps such as instant messaging and \textit{LIFESTYLE} apps such as searching restaurant whenever a network connection is available, and need to synchronize to the servers for latest stock information when using \textit{FINANCE} apps.

\subsubsection{Traffic from Foreground and Background}
We then move to \textbf{RQ 3.3}, i.e., \textit{how much ``hidden'' traffic is consumed when using an app?}. We distinguish the foreground and background traffic of an app, respectively. Often, the foreground traffic is generated when users interact with the app. In the Android OS, a foreground app can be identified if the app is currently at the top of the activity stack. In contrast, the background traffic implies that the app is still connecting to network even when users do not interact with it. From Table~\ref{networksummary}, the foreground traffic accounts for more than 60\% in many categories. Foreground traffic accounts for less than 50\% in some categories, i.e., \textit{SYSTEM\_TOOL} (12.62\%), \textit{TRAVEL} (39.2\%), \textit{LIFESTYLE} (38.33\%),  \textit{TRAFFIC} (39.42\%), \textit{BEAUTIFY} (23.97\%), \textit{PRODUCTIVITY} (20.89\%), and \textit{IMAGE} (28.27\%). \textbf{It indicates that some apps in these categories keep consuming a large amount of traffic, when users switch to use other apps, or the screen-off traffic occurs with device sleeping}. Hence, the background traffic of these apps could be necessary. Some apps can reasonably have background network activities. For example, the \textit{SYSTEM\_TOOL} management apps such as anti-virus apps often need downloading or updating activities at background. 

\begin{figure}
	\centering
	\begin{center}
		\includegraphics[width=0.48\textwidth]{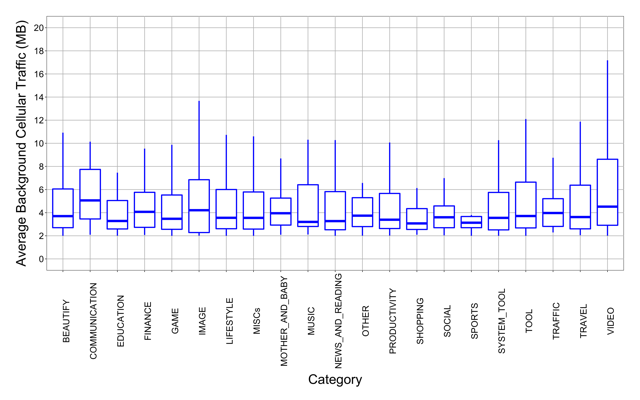}
		\caption[7.5pt]{The background data traffic generated under cellular}\label{CBD}
	\end{center}
\end{figure}

Compared to the data generated at background under Wi-Fi (abbreviated as \textbf{WBD}), the data generated at background under cellular (abbreviated as \textbf{CBD}) can bring potential loss of data plan for end-users. We demonstrate the \textbf{CBD} according to the app's category in Figure~\ref{CBD}. The median values of \textbf{CBD} of \textit{COMMUNICATION} and \textit{VIDEO} apps are relatively higher than the ones of apps from other categories. The results provide strong evidence of ``hidden'' data drain generated at background. Hence, we focus on the average daily \textbf{CBD} of an app from all users that use this app. From our dataset, we have 2,697 apps that produce at least 2 MB daily \textbf{CBD} per user. Given the average daily \textbf{CBD} of 2 MB, the monthly data drain can reach up to 60 MB, which cannot be ignored for those who have rather limited data plan. 

The immediate take-away message of this study reminds end-users to alert or kill background network activities after launching an app. Indeed, there are various possible reasons why an app produces data drain at background. One reason is that ad-libraries are widely used in a lot of apps, and may download and update advertisements according to user contexts~\cite{Li:MobiSys15}. Another reason is that the background data drain is required by the apps' features, e.g., \textit{VIDEO} and \textit{NEWS\_AND\_READING} apps may download contents and cache them locally. Such behaviors are reasonable under Wi-Fi, but are not desired under cellular, especially for users who have limited data plan. Last but not the least, the misuses or even malicious granting of network permissions cannot be neglected. For example, as found in our conference paper~\cite{Li:IMC15}, some \textit{TOOL} apps such as \texttt{flashlight} and \texttt{namecard scanning} were detected to collect the location information of their users. 

However, it is really very difficult to justify whether the background data drain generated by an app is really necessary with respect to the functionalities of this app. To the best of our knowledge, some preceding efforts such as \texttt{CHABADA}~\cite{Zeller:ICSE14} and \texttt{WHYPER}~\cite{Xie:USENIXSecurity13} can help check whether the app's granted permissions or behaviors are abnormal against their descriptions. However, existing efforts are not adequate to validate whether the background data drain is reasonable for apps, as background traffic is a dynamic behavior that can be monitored only at runtime. In addition, background data drain from in-app ads are needed for developers' revenue, but end-users may be annoyed by such a loss and thus annotate low rating of the app or even uninstall it.

Another immediate outcome of this study is that we can find some apps consuming exceptionally high cellular data at background. For example, on average, the \textit{GAME} app ``\texttt{DJMax Ray}'' with 134 users generates the daily  \textbf{CBD} of 58 MB per user, and the \textit{TRAFFIC} app ``\texttt{N5 Navigator}'' with 72 users generates the daily \textbf{CBD} of 43 MB. Undoubtedly, such a large volume cannot be ignored under cellular network, as end-users have to pay for the data plan. We believe that these apps can be an interesting genre for the research community to explore possible reasons why such substantial \textbf{CBD} occurs and  assess whether the background data drain is ``really'' necessary. \\



\noindent \fbox{%
\parbox{0.485\textwidth}{
     \textbf{Finding (F7)}: Some apps can consume a considerable amount of traffic at background, but it is challenging to determine whether such dynamic behaviors are really reasonable or necessary.}} 
\newcommand{\tabincell}[2]{\begin{tabular}{@{}#1@{}}#2\end{tabular}}

\begin{table}[t]
\scriptsize
	\centering
	\caption{Categorization of device models}
	\begin{tabular}{c|r|r|c}
		\hline
		\textbf{Group} & \tabincell{c}{\textbf{Price}\\ \textbf{Interval}}& \tabincell{c}{\textbf{Device}  \\ \textbf{Count}} 
		& \textbf{Representative Devices} \\
		\hline
		\textbf{High-End} & $\geqslant$ 3,000 RMB 
		& 177 
		& \texttt{Samsung N7100, Samsung S4} \\
		\textbf{Medium-End} & 1,000-3,000 RMB  & 239 
		 & \texttt{XIAOMI 3, Google NEXUS 4} \\
		\textbf{Low-End} & $\leqslant$ 1,000 RMB & 84 
		 & \texttt{COOLPAD 7231, LENOVO A278T}  \\
		\hline
	\end{tabular}
	\label{tab:user_division}
\end{table}

\section{Device-Specific Patterns}\label{device}

The preceding measurements provide  information on some patterns of app selection and network usage. However, the patterns are derived from the global distribution instead of the classification of users. It would be interesting to explore the diverse preferences of different users. In practice, the users can be categorized from different aspects, i.e., gender, age, country, and economic background. In this article, we select a different signal, i.e., the price of the device model that a user holds. Such a signal is motivated by two aspects. First, the price of a device model can generally reflect the hardware specifications of a device model when it is released onto market. Second, such a metric can somewhat imply a user's economic background, which can influence the user behaviors at the demographic level. Indeed, such an assumption cannot be always reliable, as it cannot identify the users who buy second-hand ``high-end'' devices, or those who use two (or more) devices, e.g., one is low-end while the other is high-end. However, from a general trend, we can rely on the device models' price as an indicator to classify users in our dataset. 

We then categorize the device models according to their on-market price, and revisit the research questions \textbf{RQ 2} and \textbf{RQ 3}, respectively.


\begin{figure}
\centering
\begin{center}
\subfigure[Treemap of the number of unique users per device model\label{block}]{
\includegraphics[width=0.42\textwidth]{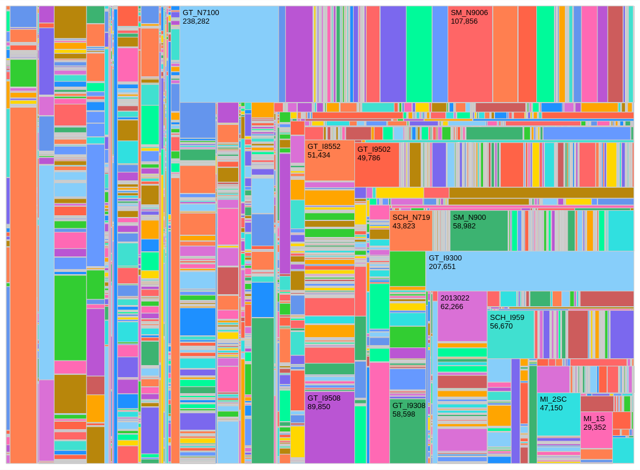}}
\subfigure[Distribution of the number of users against device models \label{devicedistribution}]{
\includegraphics[width=0.42\textwidth]{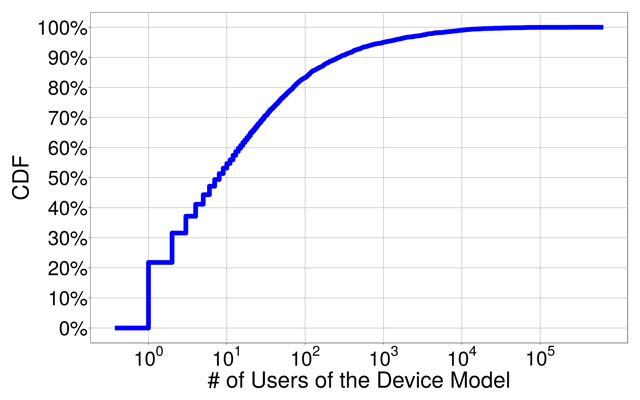}}
\caption[7.5pt]{Heavy fragmentation of Android devices}\label{devicemodel}
\end{center}
\end{figure}
\subsection{Device Model Fragmentation}
First, we compute the distribution of the number of unique users that a device model has. The result is shown in Figure~\ref{devicemodel}. In the treemap of Figure~\ref{block}, each colored block corresponds to a specific device model, and its area's size depends on the number of users. Given over 19,000 distinct device models in our dataset, such a visualized result indicates the heavy ``fragmentation'' of Android devices. In addition, for the distribution of users per device model in Figure~\ref{devicedistribution}, we can find that more than 83\% device models have only fewer than 100 unique users, while about 1\% device models own more than 10,000 unique users.  \\

\noindent \fbox{%
\parbox{0.485\textwidth}{
     \textbf{Finding (F8)}: The on-market Android device models present a heavy fragmentation, i.e., more than 19,000 distinct device models in our dataset.} }
\subsection{Device Model Clustering}

We then study the first-release price of a device model so as to classify users. In China, on the price systems of popular e-commerce web sites such as \texttt{Jd}, \texttt{Amazon}, and \texttt{Taobao}, the price of device models is usually segmented at every 1,000 RMB level, i.e., $\leqslant$ 1,000 RMB, 1,000 RMB-2,000 RMB, 2,000 RMB-3,000 RMB, 3,000 RMB-4,000 RMB, and $\geqslant$4,000 RMB. Hence, we roughly categorize the device models into three groups according to their on-sale price information that is published on  \texttt{Jd}, i.e., the \textbf{High-End} ($\geqslant$3,000 RMB, about 500 USD), the \textbf{Medium-End} (1,000 RMB-3,000 RMB, about 150-500 USD), and the \textbf{Low-End} ($\leqslant$ 1,000 RMB, about 150 USD). We choose the top 500 device models according to their number of unique users, and manually check their price-history evolution on \texttt{Jd} as well as looking up some third-party data sources such as \texttt{Dong-Dong}\footnote{\url{https://itunes.apple.com/us/app/dong-dong-gou-wu-zhu-shou/id868597002?mt=8}, is an app for inquiring history price of products on \texttt{Jd}.} and \texttt{Xitie}\footnote{\url{http://www.xitie.com}, is a website for inquiring price history of products on popular e-commerce sites.}. Most of the device models were first released to market after 2012, and can still fall into the preceding coarse-grained groups as of May 1, 2014 (the starting time of our data set). Very few device models cannot meet this criterion, e.g., the first-release price of \texttt{Galaxy S2} was 4,399 RMB, but the price fell down to about 2,700 RMB as of July 2014. For this case, we still categorize the device models by the first-release price. Luckily, only 19 exceptions out of 500 device models occur in our dataset. We list the categorization results in Table~\ref{tab:user_division}.

We then revisit \textbf{RQ2} and \textbf{RQ3}, by considering the choice of device models. We adopt the Spearman correlation coefficient\footnote{Spearman Correlation Coefficient.  \url{https://en.wikipedia.org/wiki/Spearman\%27s_rank_correlation_coefficient}}. Statistically, the Spearman correlation coefficient is a non-parametric metric of statistical dependence between the ranking of two variables, \textit{X} and \textit{Y}. Such a metric assesses how well the relationship between two variables can be described, and it does not make any assumption on the distribution of the data. Commonly, the  Spearman correlation coefficient is represented by the Greek letter $\rho$. For a sample of size $n$, the $n$ raw scores $X_i$, $Y_i$ are converted to ranks $rg X_i$, $rg Y_i$. Then, $\rho$ is computed from the following equation:

\[
\rho=\frac{cov(rg_X, rg_Y)}{\sigma_{rg_X}\sigma_{rg_Y}}
\]

where $cov$ is the covariance, and $\sigma$ is the standard deviation. In our measurement, \textit{X} represents the price of device models, and \textit{Y} represents the usage patterns including the numbers of downloads, updates, and uninstallations, the traffic volume, and the access time, respectively. For each pattern, we compute the Spearman correlation coefficient with packages provided by Python\footnote{\url{https://docs.scipy.org/doc/scipy-0.14.0/reference/generated/scipy.stats.spearmanr.html}}.  The results are shown in Table~\ref{tab:overallcorrelation}.

\subsection{Apps Selection against Device Models}

We investigate whether the choice of device models can impact the app selection. From our previous study of the global distribution of apps~\cite{Li:IMC15}, we find that users can have quite high overlap in selecting the popular apps, such as \texttt{WeChat}, \texttt{QQ}, and  \texttt{Map}, by counting the number of unique users of these apps relative to the total number of users in the dataset. 
Hence, we explore the \textbf{diverse requirements} of apps. For simplicity, we cluster the apps according to their  category information provided by Wandoujia, e.g., \textit{Game}, \textit{NEWS\_AND\_READING}. We compute the contributions of downloads and updates from every single device model relative to a specific app category. For example, if there are 1,000,000 downloads of \textit{GAME} apps and 50,000 of these downloads and updates come from the device model \texttt{Samsung S4}, we assign the contributions made by this device model as 5\%. Then we make the correlation analysis of app selection and the price of device, by means of the Spearman correlation co-efficient. We find that as the price of device models increases, the users are more likely to choose apps from the categories of \textit{TRAFFIC} (\textit{r}=0.542, \textit{p} = .000) , \textit{LIFESTYLE}  (\textit{r} = 0.565, \textit{p} = .000), \textit{NEWS\_AND\_READING}  (\textit{r} = 0.552, \textit{p} = .000), \textit{SHOPPING}  (\textit{r} = 0.659, \textit{p} = .000), \textit{FINANCE}  (\textit{r} = 0.655, \textit{p} = .000), and \textit{TRAVEL}  (\textit{r} = 0.719, \textit{p} = .000). In contrast, the correlation analysis indicates that as the price of device models increases, the users are less likely to choose the apps from \textit{GAME} (\textit{r} = -0.707, \textit{p} = .000) and \textit{MUSIC} (\textit{r} = -0.477, \textit{p} = .000).  Such observations imply that the choice of device models could significantly influence the app selections, and infer the characteristics and requirements of the users. For example, users with high-end smartphones are more likely to care about the apps from \textit{NEWS\_AND\_READING}, \textit{FINANCE}, \textit{TRAVEL}, and \textit{SHOPPING}. Users holding lower-end device models care more about the entertainment apps such as \textit{GAME} and \textit{MUSIC}.\\

\noindent \fbox{%
\parbox{0.485\textwidth}{
     \textbf{Finding (F9)}: The selection of device models has significant correlations with the selection of apps, implying the various user needs and requirements.  
     } }
\subsection{App Abandonment against Device Models}
The uninstallation can indicate the users' negative attitudes towards an app, i.e., the users do not like or require the app any longer. We then perform the correlation analysis in a similar way of downloads and updates. In most categories, the correlation is not quite significant. 

Although the uninstallation does not take significant correlation to the choice of device models at the level of app category, investigating the individual apps that are possibly abandoned by a specific device model is still meaningful, as such investigation can help developers identify some device-specific problems. To this end, we explore the apps that have been uninstalled for more than 500 times in our dataset, and obtain 6,736 apps. We then examine the distribution of uninstallations according to the device model. An interesting finding is that the manufacturer-customized or preloaded apps are more possibly uninstalled on the lower-end device models. For example, the app \texttt{Huawei News Reader} is a preloaded app on almost all device models produced by \texttt{Huawei}. This app has received 20,985 uninstallations, while 17,461 uninstallations come from the lower-end devices of \texttt{Huawei}. The similar findings can be found in other device models produced by \texttt{Samsung, Lenovo}, and \texttt{ZTE}. Such an observation implies that the lower-end users are less likely to adopt these customized or preloaded apps. Besides the preloaded apps, some apps are also more likely to be uninstalled by a specific device model. For example, two device models \texttt{Samsung Galaxy S5} and \texttt{Motorola Defy} account for 72\% of the uninstallations of an HD-Video calling app called \texttt{CIPSimple (com.hh.csipsimple)}. Such a finding implies that these apps can probably suffer from device-specific incompatibilities or bugs. Although our current finding cannot tell the root causes for such abandonments, it can help the app developers better locate some ``\textit{attention-needing}'' device models where their apps may encounter possible loss of users.\\
 
\begin{figure*}
\centering
\subfigure[WiFi time\label{WiFi-time}]{
\includegraphics[width=0.23\textwidth]{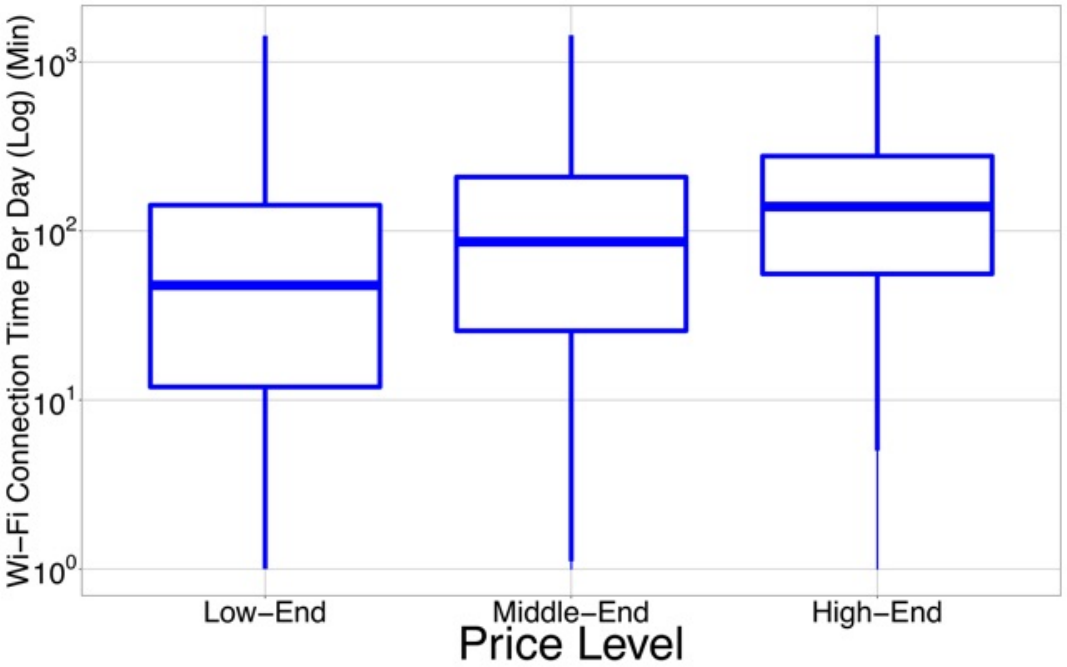}}
\subfigure[Cellular time\label{Cellular-time}]{
\includegraphics[width=0.23\textwidth]{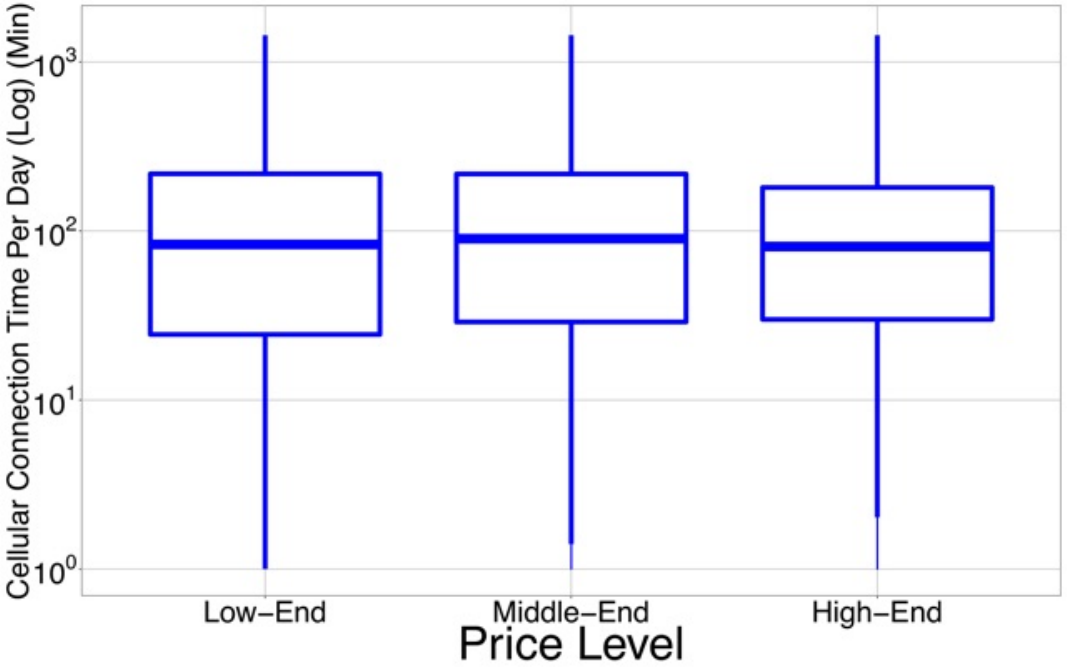}}
 \subfigure[Wi-Fi traffic volume\label{WiFi-traffic}]{
\includegraphics[width=0.23\textwidth]{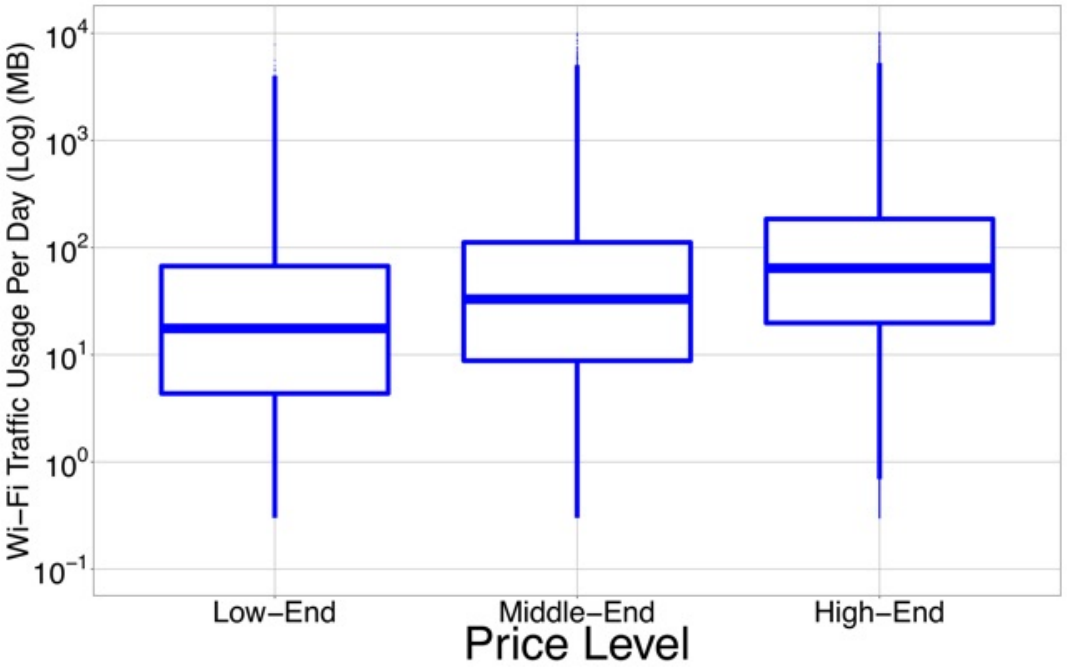}}
\subfigure[Cellular traffic volume\label{Cellular-traffic}]{
\includegraphics[width=0.23\textwidth]{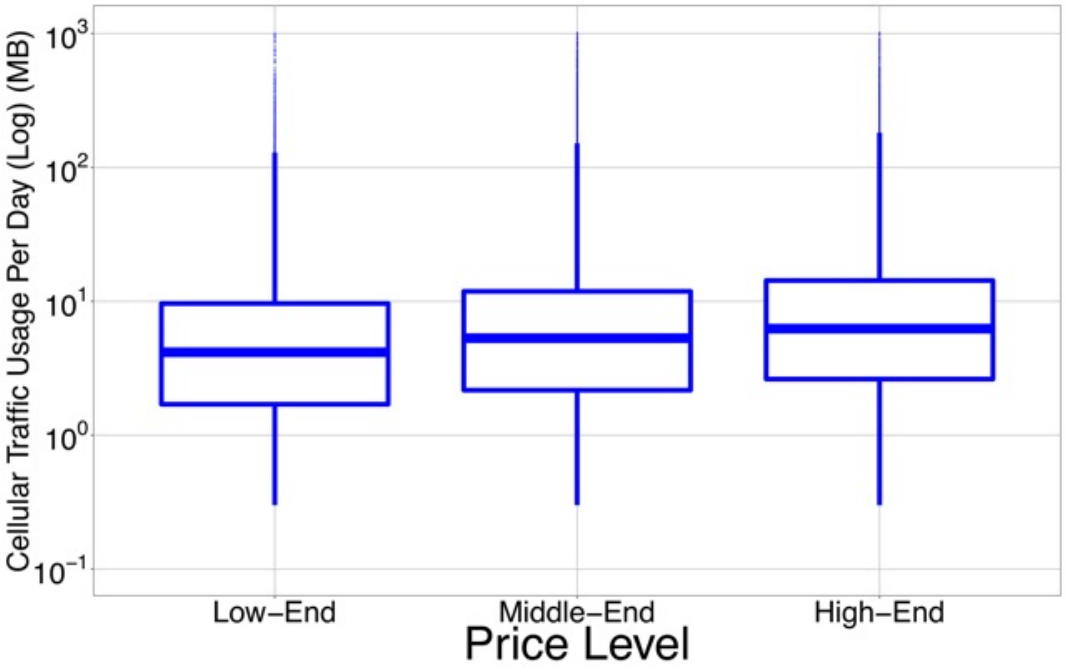}}

\caption[7.5pt]{Daily network activity comparison among user groups}\label{socio-economic}
\end{figure*}

\noindent \fbox{%
\parbox{0.485\textwidth}{
     \textbf{Finding (F10)}: Some apps are more likely to be uninstalled on specific device models, indicating users' negative attitudes towards preloaded or vendor-customized apps, or some potential device-specific problems. Such findings can help developers accurately locate some device models with more care.} }

\begin{table*}[t]
\centering
\small
\caption{The Spearman correlation co-efficient of every singe app category. The format of each cell is ``coefficient/p-value''. Cells denoted with ``*'' indicate statistical significance.}
\label{tab:overallcorrelation}
\begin{tabular}{|l|c|c|c|c|c|c|}
\hline
\textbf{Category} & \tabincell{c}{\textbf{Download \&} \\\textbf{Update}} & \textbf{Uninstallation}   & \tabincell{r}{\textbf{Cellular} \\\textbf{Time}} & \tabincell{r}{\textbf{Wi-Fi} \\\textbf{Time}} & \tabincell{r}{\textbf{Cellular} \\\textbf{Traffic}} & \tabincell{r}{\textbf{Wi-Fi} \\\textbf{Traffic}} \\
\hline
BEAUTIFY & \textbf{-0.471/0.000*} & 0.000/0.997 & -0.085/0.056 & -0.054/0.230 & -0.191/0.000 & \textbf{0.363/0.000*} \\
COMMUNICATION & -0.204/0.000 & -0.132/0.003 & 0.141/0.002 & -0.001/0.977 & -0.158/0.000 & \textbf{0.372/0.000*} \\
EDUCATION & -0.102/0.022 & \textbf{-0.315/0.000*} & \textbf{-0.382/0.000*} & 0.013/0.774 & \textbf{-0.309/0.000*} & 0.243/0.000 \\
FINANCE & \textbf{0.655/0.000*} & \textbf{0.426/0.000*} & \textbf{0.325/0.000*} & 0.274/0.000 & \textbf{0.366/0.000*} & \textbf{0.398/0.000*} \\
GAME & \textbf{-0.707/0.000*} & -0.163/0.000 & -0.016/0.714 & 0.084/0.062 & \textbf{-0.429/0.000*} & -0.091/0.043 \\
IMAGE & \textbf{0.304/0.000*} & \textbf{0.307/0.000*} & \textbf{0.424/0.000*} & \textbf{0.381/0.000*} & \textbf{0.394/0.000*} & \textbf{0.496/0.000*} \\
LIFESTYLE & \textbf{0.565/0.000*} & 0.299/0.000 & \textbf{0.365/0.000*} & \textbf{0.374/0.000*} & \textbf{0.443/0.000*} & \textbf{0.580/0.000*} \\
MOTHER\_AND\_BABY & \textbf{0.333/0.000*} & 0.228/0.000 & \textbf{0.301/0.000*} & 0.292/0.000 & 0.243/0.000 & \textbf{0.306/0.000*} \\
MUSIC & \textbf{-0.477/0.000*} & \textbf{-0.361/0.000*} & 0.159/0.000 & 0.097/0.031 & \textbf{-0.411/0.000*} & 0.220/0.000 \\
NEWS\_AND\_READING & \textbf{0.552/0.000*} & 0.249/0.000 & 0.220/0.000 & \textbf{0.318/0.000*} & \textbf{0.372/0.000*} & \textbf{0.477/0.000*} \\
PRODUCTIVITY & \textbf{0.369/0.000*} & 0.187/0.000 & \textbf{0.508/0.000*} & \textbf{0.424/0.000*} & \textbf{0.576/0.000*} & \textbf{0.553/0.000*} \\
SHOPPING & \textbf{0.659/0.000*} & 0.241/0.000 & \textbf{0.640/0.000*} & \textbf{0.590/0.000*} & \textbf{0.601/0.000*} & \textbf{0.658/0.000*} \\
SOCIAL & 0.219/0.000 & 0.240/0.000 & \textbf{0.407/0.000*} & \textbf{0.422/0.000*} & \textbf{0.424/0.000*} & \textbf{0.451/0.000*} \\
SPORTS & -0.087/0.052 & 0.136/0.002 & 0.293/0.000 & \textbf{0.354/0.000*} & 0.261/0.000 & \textbf{0.408/0.000*} \\
SYSTEM\_TOOL & -0.030/0.501 & -0.127/0.004 & 0.076/0.090 & \textbf{0.302/0.000*} & 0.127/0.004 & \textbf{0.406/0.000*} \\
TOOL & 0.111/0.013 & 0.047/0.297 & -0.103/0.021 & 0.006/0.890 & -0.257/0.000 & -0.042/0.350 \\
TRAFFIC & \textbf{0.542/0.000*} & \textbf{0.320/0.000*} & \textbf{0.459/0.000*} & \textbf{0.436/0.000*} & \textbf{0.514/0.000*} & \textbf{0.558/0.000*} \\
TRAVEL & \textbf{0.719/0.000*} & \textbf{0.382/0.000*} & \textbf{0.562/0.000*} & \textbf{0.446/0.000*} & \textbf{0.589/0.000*} & \textbf{0.516/0.000*} \\
VIDEO & \textbf{0.413/0.000*} & 0.001/0.981 & 0.285/0.000 & \textbf{-0.313/0.000*} & 0.034/0.448 & \textbf{-0.383/0.000*} \\
MISCs & -0.163/0.000 & 0.026/0.555 & 0.091/0.043 & 0.203/0.000 & 0.099/0.027 & \textbf{0.350/0.000*} \\
\hline
\end{tabular}
\end{table*}

\subsection{Access Time against Device Models}

Figures~\ref{WiFi-time} and \ref{Cellular-time} describe the distribution of daily access time at foreground under Wi-Fi and cellular, respectively. \textbf{For access time at foreground, we are surprised to find that users rely less on the cellular network when the price of device model increases. } In other words, the higher-end users typically spend less time under cellular network. For the average daily access time at foreground, the low-end users ($\leqslant$ 1,000-RMB device models) spend about 20 minutes more than the high-end users ($\geqslant$ 3,000-RMB device models) under cellular, while the high-end users spend 2 hours more than the low-end users under Wi-Fi. Immediately, we can infer that the network conditions could probably vary a lot among different users, i.e., the lower-end users are less likely to stay in the places with stable Wi-Fi connections. In contrast, the higher-end users tend to have better Wi-Fi connections. 

We then investigate whether the choice of device models can affect the usage of ``network-intensive" apps. Similar to the preceding analysis of the management activities, we compare the distribution of access time at foreground against the  device models over each app category, under cellular and Wi-Fi, respectively. As shown in Table~\ref{tab:overallcorrelation}, the cellular time of apps has no significant correlation with the price of device models, except the categories of \textit{PRODUCTIVITY} (\textit{r} = 0.508, \textit{p} = .000) and \textit{SHOPPING} (\textit{r} = 0.640, \textit{p} = .000). It is interesting to see that users holding lower-end smartphones are more likely to use \textit{EDUCATION} (\textit{r} = -0.382, \textit{p} = .000) apps under cellular network. Such a finding suggests that a considerable proportion of lower-end users may be in-school students.\\

\noindent \fbox{%
\parbox{0.485\textwidth}{
     \textbf{Finding (F11)}: The selection of device models can have significant correlations with the spent access time under different networks. For example, higher-end device users heavily rely on Wi-Fi. In contrast, lower-end device users are likely to use more cellular than lower-end device users do.} }

The correlation between the choice of device models and the access time at foreground under Wi-Fi does not seem be quite significant, either. Only in the category of \textit{SHOPPING} (\textit{r} = 0.590, \textit{p} = .000), the choice of device models seems to take positive correlation with the price of device models. Such an observation can be expected, as higher-end users are supposed to have better economic background and spend more time on shopping.\\

\subsection{Traffic Volume against Device Models}
The distribution of daily Wi-Fi and cellular traffic consumption among device models is shown in Figures~\ref{WiFi-traffic} and~\ref{Cellular-traffic}, respectively. Interestingly, although the higher-end users are observed to spend the least time under cellular network, they spend the most traffic. In other words, we can infer that the higher-end users are more likely to use those ``traffic-intensive" apps. On average, a high-end user can spend 100 MB data plan more than a low-end user in a month. For carries such as  China Mobile, such a difference of data plan can lead to 15-RMB extra data plan fee. 

It is quite meaningful to identify which apps consume more traffic on specific device models. Similar to the preceding analysis, we compute the correlation coefficients between the choice of device models from every single user and the apps on which the traffic is consumed. The cellular data traffic consumed over the apps from \textit{SHOPPING} (\textit{r} = 0.601, \textit{p} = .000) and \textit{TRAVEL} (\textit{r} = 0.589, \textit{p} = .000) presents a quite significantly positive correlation to the choice of device models. In contrast, the correlations seem to be significantly negative in \textit{GAME} (\textit{r} = -0.429, \textit{p} = .000) and \textit{MUSIC} (\textit{r} = -0.411, \textit{p} = .000) apps. In these app categories, users with lower-end smartphones tend to spend more cellular traffic. 

The traffic generated under Wi-Fi presents significant correlations with the device models in some categories. The lower-end users tend to spend a large number of Wi-Fi traffic on the \textit{VIDEO} apps. In contrast, the higher-end users are more likely to rely on the apps of \textit{COMMUNICATION}, \textit{PRODUCTIVITY}, \textit{SYSTEM\_TOOL}, \textit{TRAVEL}, \textit{BEAUTIFY}, \textit{NEWS\_AND\_READING}, \textit{LIFESTYLE}, and \textit{SHOPPING} under Wi-Fi. Such a difference in the Wi-Fi traffic usage can indicate the requirements and preferences of users holding different device models.

\subsection{Competing Apps against Device Models}

Finally, we study how the choice of device models impacts the selection of ``competing" apps with the same or similar functionalities. We choose three typical apps: \textit{News Reader}, \textit{Video Player}, and \textit{Browser}, as these apps are observed to be commonly used in daily life. For each app, we select the top  apps according to the access time at foreground that the users spend on them. The reason why we employ the access time at foreground instead of the number of downloads is that access time at foreground can be computed only when the users interact with the app. The selected competing apps are as follows. The \textbf{News Reader} contains \texttt{Phoenix News, Sohu News, Netease News, Today's Top News}, and \texttt{Tencent News}; the \textbf{Video Player} contains \texttt{QVOD, Lenovo Video, Baidu Video, Sohu Video}, and \texttt{iQiyi Video}; the \textbf{Browser} contains more apps, i.e., \texttt{Chrome, UC Web, Jinshan, Baidu, Opera Mini, Sogou, Aoyou, FireFox, Tencent}, and \texttt{360}.

\begin{figure*}
\centering
\begin{center}
\subfigure[Newsreader apps\label{newsreader}]{\includegraphics[width=0.3\textwidth]{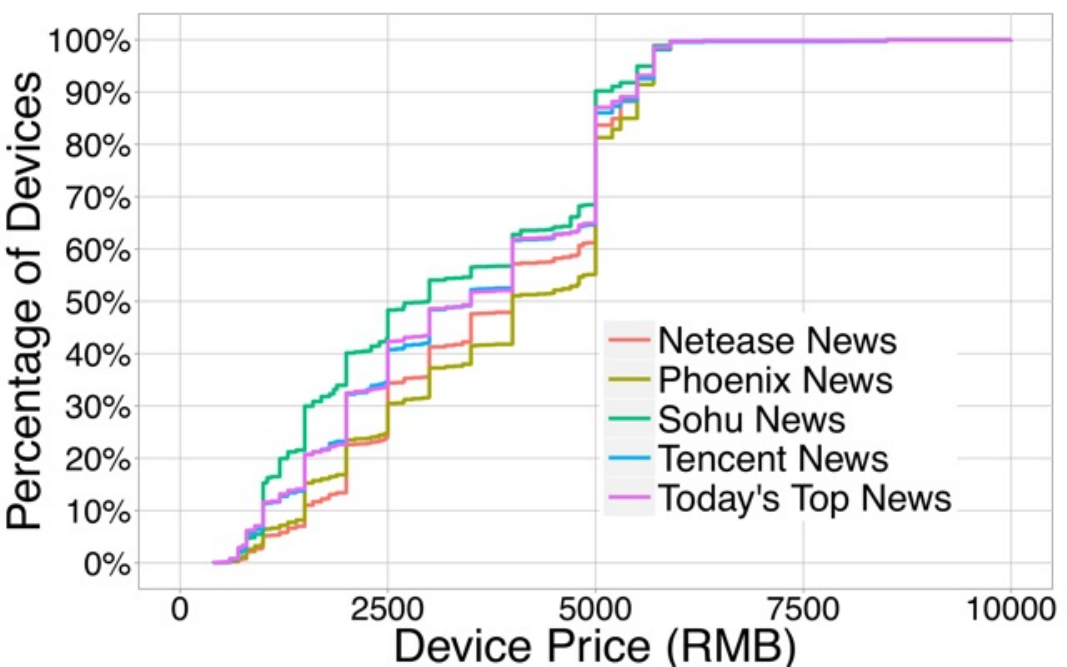}}
\subfigure[Video apps\label{video}]{\includegraphics[width=0.3\textwidth]{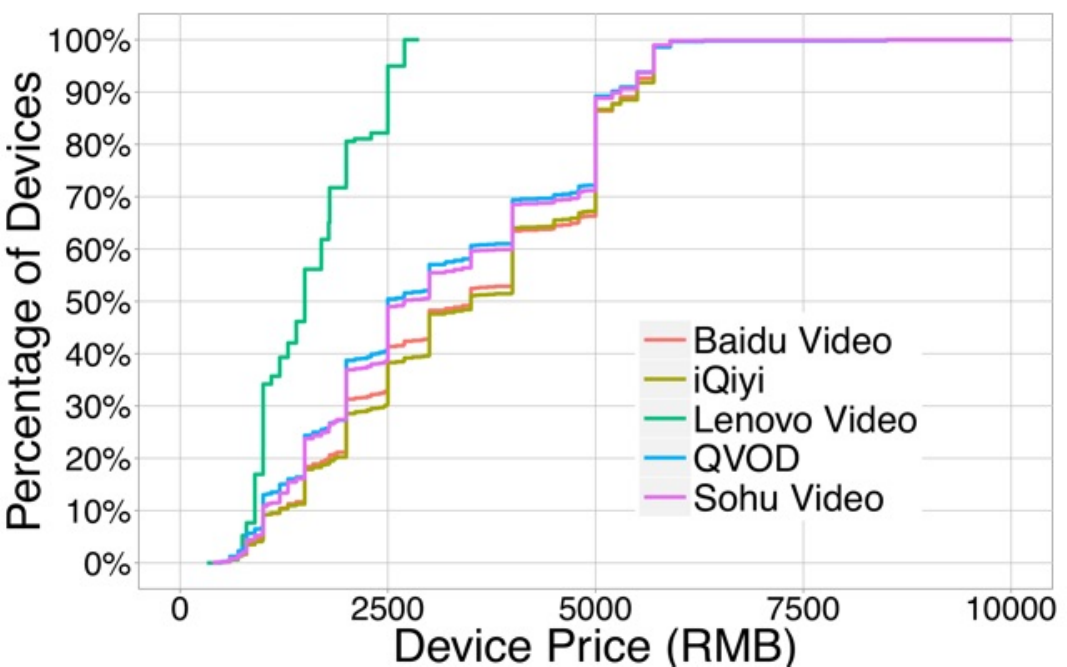}}
\subfigure[Browser apps\label{browsers}]{
\includegraphics[width=0.33\textwidth]{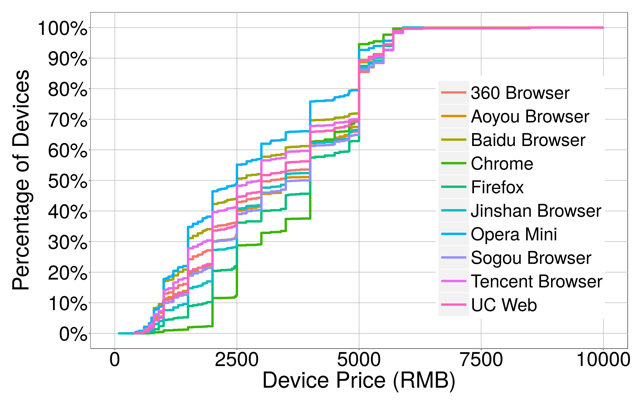}}
\caption[7.5pt]{Similar app preferences among user groups}
\label{competingapp}
\end{center}
\end{figure*}

First, we want to figure out the distribution of the user preferences against the app according to the device model. We employ the cumulative distribution function (CDF) to demonstrate such distributions, as shown in Figure~\ref{competingapp}. For each app, the X-Axis represents the price of device models that are sorted in the ascending order, and the Y-Axis refers to the percentage of the app's users holding such a device model. An app tends to be used by more higher-end users if the curve is close to the right bottom. 

We can observe that the choice of device models significantly impacts the selection of competing apps. For the \texttt{News Readers}, we can see that the \texttt{Phoenix News} and \texttt{Netease News} are more likely to be adopted by higher-end users. In contrast, the \texttt{Sohu News} tends to be more preferred by the lower-end users, possibly because \texttt{Sohu} is famous for its entertainment channels in China. The impact of device models is even more significant for the \textit{Video} players. The \texttt{Lenono Video} takes a very significant difference compared to other 4 apps, indicating most of its users are lower-end. One possible reason is that the \texttt{Lenovo Video} is a preloaded app that is used mainly on smartphones manufactured by \texttt{Lenovo}, and most of these smartphones are categorized into medium-end and low-end groups. 

Finally, in the \texttt{Browser} group, the similar findings can be observed. The most preferred browser of higher-end users is the \texttt{Chrome} browser, followed by the \texttt{FireFox} browser,  and the \texttt{Jinshan} browser. The \texttt{Opera Mini} and the \texttt{Baidu} browser are more likely to be adopted by the low-end users. One reason leading to the diversity is that an app can provide specific features beyond common functionalities that its competitors also provide, so as to meet requirements of specific user groups. For example, when examining the textual profile of the ten browsers, we find that the \texttt{Opera Mini} is said to save traffic by offloading computation onto cloud. As a result, more than 77\% of its users are those who hold medium-end and low-end devices. In summary, \textbf{such a finding implies that the choice of device model has impacts on competing-app selection and usage, and probably reflects the different user interests and needs}. \\

\noindent \fbox{%
\parbox{0.485\textwidth}{
     \textbf{Finding (F12)}: Users holding different device models could have various needs of specific apps or preferences against some ``competing" apps. For example, lower-end device users prefer the \texttt{Opera Mini} browser as it is said to save traffic.} }
\section{Implications and Suggestions}\label{implication}

%
So far we have investigated the user behaviors from a large-scale dataset and inferred some patterns. Besides confirming and validating some findings  that have been reported in previous studies based on a relatively small scale of users, our study results can further imply some new, open challenges and opportunities of the development, maintenance, and management of apps. In this section, we discuss some implications and suggestions that can be taken away by relevant stakeholders in the mobile-app ecosystem, including app-store operators, app developers, end-users, and network-service carriers. For ease of presentation, we denote the preceding findings as \textbf{F1, F2, ..., F12}, and discuss the problems and opportunities resulted from these findings. 
\subsection{Efficient App-Store Management}
App stores play as the core in the whole app ecosystem to connect all related stakeholders. We provide some implications to improve the recommendation quality and performance of an app store. In particular, we intend to point out some gaps that may not be well addressed by app-store operators.

\subsubsection{Improving Workloads of App Stores}
From the preceding macro-level finding of popularity distribution (\textbf{F1}), we can conclude that the app popularity generally follows the ``Pareto-like" principle. Such a finding confirms the results reported in previous work~\cite{MoreleyMao:IMC11}. In addition, we also validate that the numbers of downloads, updates, and unique users per app can follow the power law.

Indeed, the distribution of user requests to servers can help optimize the workloads of servers for better content delivery. However, to the best of our knowledge, so far very few published papers have comprehensively discussed the current practice of managing an app store's workload~\cite{Carla:CUSR16}. One reason can be the lack of actual user-request traces of app stores. A recent study~\cite{Petsas:IMC13} assumed that the time submitting a user review can be an immediately subsequent behavior of downloading an app. Based on such an assumption, this effort clustered the apps whose comments were posted in a given time interval, and simulated the possible workload for predicting which apps will be downloaded. However, such an assumption may not be always reliable. It is reported that users may not write reviews when/after downloading an app~\cite{lim2014investigating}. Additionally, users may not immediately submit their reviews when downloading an app, but submit their reviews after they try or use the app sometime later.     

Although we do not have knowledge of how app stores other than Wandoujia design their workload models, the traces of app download and update activities can help us improve the current design of the Wandoujia app store's workload, e.g., placing the \texttt{.apk} files of most popular apps in the server-side fast memory or local network cache. In practice, Wandoujia currently relies on the derived power law to place the \texttt{.apk} files on its Content Delivery Network (CDN) servers. We then simulate the workload performance from the server's point of view, by synthesizing power law and co-installation patterns. 
\begin{table*}[!t]
\centering
\small
\caption{Synthetic cache performance}
\label{cache}
\begin{tabular}{|c|rr|rr|rr|rr|rr|}
\hline
&

\multicolumn{2}{c|}{\textbf{Static Finite Cache}}                              & 
\multicolumn{2}{c|}{\textbf{Dynamic Infinite Cache}}                     & 
\multicolumn{2}{c|}{\textbf{Co-Installed and LRU-based Cache}}  
& 
\multicolumn{2}{c|}{\textbf{Adaptive Hybrid Cache}}                                 
 \\
\hline

\textbf{Date} &\textbf{\tabincell{r}{Cache Size}} & \textbf{\tabincell{r}{Hit Ratio}} 
& \textbf{\tabincell{r}{Cache Size}} & \textbf{\tabincell{r}{Hit Ratio}} 
& \textbf{\tabincell{r}{Cache Size}} & \textbf{\tabincell{r}{Hit Ratio}}
& \textbf{\tabincell{r}{Cache Size}} & \textbf{\tabincell{r}{Hit Ratio}}\\
\hline
\textbf{Day 0}& 8,260 & 95.1\% & 579,108 & 99.9\% & 3,787 & 81.2\% & 8,260 & 95.1\% \\
\textbf{Day 1} &8,260 & 95.7\%  & 580,335 & 99.9\% & 6,612 & 80.9\%  & 12,635  & 97.5\% \\
\textbf{Day 2} & 8,260 & 94.7\% & 581,562& 99.9\% & 6,781 & 79.9\% & 14,981 &97.7\% \\
\textbf{Day 3} & 8,260 & 94.4\% & 582,750  & 99.9\% & 6,873 & 79.6\% & 12,631  & 97.7\% \\
\textbf{Day 4} &8,260 & 94.9\% & 584,200 & 99.9\% & 7.027 & 80.1\% & 13,132 &98.2\% \\
\textbf{Day 5} & 8,260 & 95.0\% & 585,516& 99.9\%  & 7,322 & 79.9\% & 13,165  & 97.1\% \\
\textbf{Day 6} & 8,260  & 94.9\% & 586,585 & 99.9\%  & 6,989 & 80.4\% & 12,343  & 97.4\% \\
\hline                 
\end{tabular}
\end{table*}

For simplicity, we assume that the size of each app is the same, and assign the size of cache as the number of apps. Certainly, the size of every single app can vary in practice. We consider the following three conventional cache schemes, and simulate these schemes by replaying a successive one-week request traces of app download and update. We use request traces of the final week (September 24-September 30) in our dataset. 
\begin{itemize}
\item The first mechanism is the \textit{\textbf{power-law-driven static finite}} cache. At day zero, we fill the cache with the \texttt{.apk} copies of the most popular apps. We populate the static cache with the apps accounting for the most downloads and updates in Figure~\ref{fig:percentage_ndownload}. We use the power-law exponent (1.69) for overall app ranking against the number of downloads and updates, and obtain 8,260 apps accounting for 95\% of downloads and updates. The static cache is not changed during the one-week trace period. Such a cache is rather similar to the current design of Wandoujia, i.e., only the \texttt{.apk} files of the most popular apps are put in the cache, and the cache size is not tuned very frequently. 
\item The second mechanism is the \textit{\textbf{dynamic infinite}} cache. At day zero, the cache is filled with all apps that are ever requested before day zero (not limited to the 0.28 million apps in our dataset but those who had ever existed on Wandoujia server), and thereafter accumulatively stores any other apps requested during the trace period. 
\item The third mechanism is the \textit{\textbf{co-installed and LRU-based finite}} cache. At day zero, the cache is populated with the top pairs of co-installed apps (before day zero), whose Jaccard co-efficient exceeds 0.01. Suppose that two apps $\mathbb{A}$ and $\mathbb{B}$ both have 2,000 unique users, and such a value indicates that they have 20 shared users. As the co-installation pattern essentially implies the locality of accessing two apps, it indicates that these two apps are accessed subsequently for 20 times. After day zero, we compute the daily top co-installed apps whose Jaccard coefficient is over 0.01, and replace those who are ``\textit{Least Recently Used}" at the previous day.
\end{itemize}

In Table~\ref{cache}, we report the cache size (the number of apps placed in cache) and hit ratio of each cache design. Note that the ``hit ratio" refers to how many requests are exactly matched instead of the matched apps. Undoubtedly, the \textbf{dynamic infinite cache} reaches the highest hit ratio, but the cache size is tremendously large as well. In contrast, we can see that the simple \textbf{static finite cache} with the top popular apps can reach a very high hit ratio (around 95\%). In contrast, we see a significantly lower hit ratio (around 80\%) for the cache built upon the co-installation with LRU. It indicates that, although the typical LRU cache can exploit locality of user downloads and updates to improve the performance, it does not work very well with the co-installation patterns. Some new replacement mechanisms should be exploited to fit the behaviors of co-installation.

To balance the hit ratio and the cache size, we then aim to improve the LRU mechanism for app-store workloads by synthesizing the knowledge of power law and co-installation. We design a cache mechanism called \textit{\textbf{adaptive hybrid finite}}. At day zero, the cache is first populated with the top popular apps just like the static cache, and such a space will be tunable during the trace period. At day one, the cache reserves space for the top popular apps, and allocates some extra space for two kinds of apps: (1) the apps from the daily popular apps accounting for 95\% of downloads and updates; (2) the apps from the daily app pairs whose Jaccard co-efficient is over 0.01. If an app appears in either of the preceding cases, we assign its cache for only once. After day one, the cache daily reallocates the cache by replacing the apps that were ``\textit{Least Recently Used}" at the previous day. From Table~\ref{cache}, we can find that this mechanism requires the cache size up to around 15,000, which is nearly doubled compared to the \textbf{static cache} and \textbf{co-installated LRU-based cache}. However, the hit ratio can be significantly improved to over 97\%. Such a hybrid cache is efficient, as it compensates the limitations of co-installation patterns that can miss some very popular apps but preserves the benefits of locality. Suppose that the average size of \texttt{.apk} file is about 10 MB, and such a cache usually requires about 150 GB space for one CDN  node. As there can be various concurrent requests for the same app, we need to assign more copies for the most frequently accessed apps, and the actual cache size should be a bit larger. 

Indeed, the preceding simulation can be further explored by leveraging the diurnal patterns reported in \textbf{F2}. The cache can be more dynamically adapted when more concurrent requests arrive at the server at a fixed point, e.g., 9:00 pm in our preceding findings. In practice, app-store operators can optimize the cache adaptation on CDN servers for specific areas. In addition, we plan to further explore the timestamp of user download/update sequences for possible temporal patterns of accessing the server. In this way, the popular apps or the most probably ``co-installed'' apps that have not been downloaded by the user can be \textit{prefetched} to a nearby place so as to improve user experience and app-delivery performance. We plan to test these workload mechanisms on Wandoujia and evaluate the efficiency in real-world practices~\cite{Wang:WangSC15}.


\subsubsection{Improving App Recommendation}
The fundamental responsibility of an app store is to recommend proper and high-quality apps to end-users. A recommendation system can benefit apps and developers by identifying the apps that need to be recommended to increase their popularity, and by identifying users' interests in the respective app category. If an app is downloaded by most users in a group, then it is likely to be of interest for another user (in the same group) who has not yet downloaded it. In this way, finding \textbf{F3} can suggest the most ``co-installed'' pairs of apps that can be pushed to users. Additionally, finding \textbf{F3} also suggests that some frequently co-installed apps may come from the same category and vendor. Although such a pattern could be reasonable from the vendor's aspect and thus increase downloads, it may not be always desirable or necessary for end-users, as app vendors can purposely ``induce" the bundling downloads of some apps.

Another useful implication for app recommendation can be derived from \textbf{F9}, i.e., users with different device models can have quite various preferences towards apps with similar functionalities. For example,  lower-end users prefer the ``\texttt{Opera Mini}" browser while higher-end users prefer the ``\texttt{Chrome}" browser. However, based on our investigations of the most popular app stores such as Wandoujia, Google Play, and Tencent, device-specific recommendation is not well explored. As most app-store operators can gather information of device models, our finding can help achieve more accurate recommendation of apps for specific users. 

From \textbf{F1}, we can also detect that some possibly ``fake" downloads can exist by the metric of ``\textit{average download/update per user}.'' For example, an app has only 18 unique users but receives 3,581 downloads \& updates, and 3,563 downloads come from only one user. This metric can be  generally applied for any app stores that collect similar information. The apps that are downloaded by a limited number of unique devices in a short interval could be alerted to app-store operators and end-users. 
\subsubsection{Predicting the App Ranking}
An immediate result can be taken away from our study is the ``I/U'' ratio and the lifecycle of an app presented in \textbf{F4}. Indeed, we do not have the knowledge of other app stores except Wandoujia. Previously, Wandoujia used to  rank an app according to the number of its downloads. However, only the download count can possibly prioritize the ranking of an app with ``bursting'' downloads in a short time interval. When adding the uninstallation count, Wanoudjia now leverages the ``I/U" ratio and the lifecycle to more accurately evaluate an app. Indeed, other app stores recording uninstallation information can extrapolate the ranking accordingly.

Generally, the sentiment of an app's user ratings can reflect its users' attitudes towards the app, and may provide useful information for app ranking. As a result, the rating of an app can influence the user downloads and thus is quite significant to the revenue of the app developers and the ranking strategy of app-store operators. In a previous study~\cite{Harman:MSR2012}, it was observed that an app can probably gain higher ranking if it receives a number of ``good" ratings. However, from \textbf{F5}, we are surprised to find that such an observation may not always hold for large-scale users, at least in China. Additionally, we find that the abandonment of apps does not have strong correlation with the number of ``bad" (negative) ratings.

Such findings enable us to design new indicators for predicting an app's popularity. An important understanding about user activities is that the activities are not independent but always appear as sequences of events. Indeed, when search engines utilize user activities as the implicit feedback about the relevance of documents, the sequences of actions are usually more indicative than single clicks. Analogically, we anticipate that some sequential patterns of the app-management activities may be better indicators of app quality than downloading actions alone. In practice, our recent work~\cite{Li:WWW2016} made the first step to mine some sequential patterns of app-management activities that are actually correlated with online ratings of apps. Multiple time-aware management sequences were combined with several machine-learning algorithms such as Lasso Regression (Lasso)~\cite{Tibshirani:1996hh}, Random Forrest Regression (RF)~\cite{Breiman:2001wf}, and Gradient Boosted Regression Tree (GBRT)~\cite{Friedman:2002um}. From the activities, we derived some patterns that are surprisingly accurate to be used to produce the general rankings of an app, and may be used to effectively predict those new and high-quality apps. Given that the user reviews are sparse or even do not exist in some apps, such an effort can be useful for both app-store operators to rank an app, and for app developers to estimate the potential revenues. Due to page limit, we do not discuss more details in this article. However, this experience shows that ``mandatory" app-management activities can be a promising metric to reflect user attitudes and intensions. It would be interesting to further apply recent deep learning techniques over our user-behavioral data. 

\subsection{Avoiding Unexpected Cost}
From the finding \textbf{F7}, it is observed that the background data traffic pervasively exists, and may not be always reasonable with respect to the app's claimed functionalities. More seriously, \textbf{F6} suggests that most apps can keep ``long-live" TCP connections at background. This finding indicates that currently most Android apps could stay silently in memory even when they are switched to background. The background network connection can be useful, as the app can be quickly ``awaken'' and the network connection can be fast restored. However, these apps can also lead to memory occupation and have side effect of system performance. In fact, it is reported that many Android users complain that their devices become slower and slower to respond to user interaction~\cite{Charlie:mobicom15}. It would be interesting to explore whether such ``long-live'' network connection could be a factor. In addition, our recent work~\cite{Xu:WWW2017}  made an empirical study of 1,000 apps from three third-party app stores (including Wandoujia, Baidu, and Tencent), and surprisingly found that some apps can have ``collusion'' behavior, i.e., they can awake one another at background but users are never aware of such behavior. The hidden cost of collusion behaviors, including computation resources and energy drain, can be too significant to be ignored.  
Hence, such finding reminds that end-users should employ some tools to periodically ``clean up" their devices or terminate threads of unused apps running at background. 

The findings also imply that these apps can bring hidden and unexpected cost for users, such as the loss of data-plan traffic and energy. The findings call for efficient solutions to determine whether the undesirable cost is really necessary for an app. Preliminary efforts revealed that the undesirable cost can originate from the improper granting of permissions~\cite{Xie:USENIXSecurity13}, the use of third-party ad-network libraries~\cite{Gui:ICSE2015}, or the unreasonable API usage~\cite{Zeller:ICSE14}, etc.  However, only static analysis is not sufficient~\cite{Fang:FangSC17}. Determining whether the additional cost is ``\textbf{really}" unreasonable or malicious is quite difficult. For example, the additional data drain may not be always ``purposely" malicious, as the developers may want to collect some information such as location in order to push context-aware ads. A possible way can be combining the user-behavioral data with other analytic techniques such as natural language processing, static code analysis, library-dependency analysis, and network-trace analysis, to deeply understand the app semantics and comprehensively evaluate the cost against the functionality of apps. Additionally, it would be also appreciated to provide lightweight system services that can identify the cost of the necessary functional features from other features, respectively. Hence, we can display such information to end-users who launch the app, and let themselves decide whether to prevent some unnecessary cost. However, such a ``\textit{separation-of-concern}'' solution is not easy and must be performed very carefully, because  preventing some features may also affect the normal functionality.  Our findings can provide a genre of apps that have substantial background traffic drain, and researchers can focus on these apps for further study. 


In practice, the findings have already motivated us to apply the metric of \textbf{CBD} in the current Wandoujia app store. We can help identify some ``suspicious'' apps to which both app-store operators and end-users need to pay attention, because users may suffer from a lot of unnecessary loss of data-plan traffic and the possible overhead leading to unexpected CPU and energy cost~\cite{Charlie:mobicom15}. For example, our previous study~\cite{Li:IMC15} found that an alarm clock app daily consumes about 13 MB \textbf{CBD} and an LED flashlight app daily consumes about 7 MB \textbf{CBD}. These apps were put onto the watchlist of the Wandoujia app store, and then evaluated more comprehensively and rigorously. In fact, some of these apps have been forced off the shelf accordingly. 

Additionally, the undesirable cost of apps suggests that developers should optimize their apps, because users can give low rating and even abandon these apps when perceiving the undesirable cost. Developers need to configure the use of network connection at a finer granularity, e.g., disabling background data transfer under cellular. Developers also need to explicitly make end-users aware of the potential cost, e.g., popping up alert information at the installation wizard or displaying on the app's information webpage. 

\subsection{Addressing Device-Specific Features}
From \textbf{F10}, we can find that some apps are more frequently uninstalled on some device models. Such a finding implies that there may be some problems such as compliance with hardware, device-specific drivers, or API evolution. As reported on the StackOverflow~\cite{Web:bugs1}, some camera-related bugs have been found on \texttt{Samsung Galaxy Note2}. Indeed, the finding currently cannot comprehensively trace the root causes of these problematic issues; however, we can at least locate a genre of device models and help developers explore possible device-specific problems or bugs, especially for apps that have sparse user reviews and cannot be well amenable to existing techniques~\cite{Lorenzo:ICSE2016}.

Other than knowing only the device-specific problems, app developers have to face the challenge introduced by Android fragmentation. According to our finding \textbf{F8}, there are more than 19,000 device models in our dataset. Such a fragmentation brings significant challenges to software engineering practices for mobile apps, such as app design, development, maintenance, quality assurance, and revenue generation~\cite{han2012understanding, Zhou:SP2014, khalid:FSE2015,park2013fragmentation}. A recent study~\cite{Web:Fragmentation} (in 2013) showed that 94\% of mobile-app developers who avoid the Android platform cite fragmentation as the main reason. Android app developers need to identify the major device models out of the wide selection space to validate an app's functionality, usability (such as the GUI effects), and even revenues. 

As a result, prioritizing device models can be an important topic, especially before the release of an app. Developers need to buy some device models or use some cloud-based emulators such as Appthwack\footnote{\url{https://appthwack.com/}} and Testin\footnote{Testin. \url{http://www.testin.cn}}. In fact, due to the lack of usage data, most developers currently rely on popular device models from public market-share reports such as AppBrain~\cite{Web:AppBrain}. However, such reports are too coarse-grained to be precise enough, and the major device models for different individual apps can be quite different. A recent study~\cite{khalid:FSE2015} showed that major device models from which the users may post positive/negative reviews of a specific app can vary a lot. Given very limited resources to buy device models or cloud services, developers have strong desire to more accurately locate and invest those major device models. 

As our dataset contains detailed usage data of an app per distinct device model, we can design some prediction techniques for prioritizing device models. In practice, our recent PRADA work~\cite{Lu:ICSE2016} made a first step to prioritize Android device models for individual apps, based on mining large-scale usage data from our dataset. PRADA adapts the concept of operational profiling~\cite{musa:operational93} and includes a collaborative filtering technique to predict the usage of an app on different device models, even if the app is entirely new (without its actual usage in the market yet), based on the usage data of a large collection of apps. Compared to the coarse-grained market-share based metric, PRADA can accurately predict the device models where a new app could be used. In addition, the preceding analysis indicates that the distribution of uninstallation behaviors of an app among device models can be a more objective indicator compared to user reviews or ratings. We are currently extending the PRADA approach to prioritize device models where an app is more likely to be uninstalled.
\subsection{Addressing Various Requirements}
Our findings \textbf{F9}-\textbf{F12} demonstrate that users holding different device models may have quite various requirements. Given that a device model used by a user can imply possible preferences of the user to some extent, the diversity of user needs towards apps must be further explored.

First, the preference of selecting ``routine apps" can be diverse. For example, the users holding higher-end device models tend to more use the \textit{FINANCIAL} and \textit{NEWS\_AND\_READING} apps, while the users holding lower-end device models tend to more use the \textit{GAME} and \textit{EDUCATION} apps. The developers have to identify which users are more worth focusing on, and provide optimal services or customized features.
 
Second, the preference of selecting ``competing apps" can be diverse. Here, we refer to the ``competing apps" as the apps that have similar functionalities. For example, the users holding lower-end device models prefer the \texttt{Baidu} browser and the \texttt{Opera Mini} browser (rather than the \texttt{Chrome} browser). We inspect the textual profiles of these two browser apps to identify attractive features provided by the apps. For example, the \texttt{Opera Mini} browser claims that it can save the traffic by compressing the content and resizing the images in a front-end cloud before the page is loaded on the user side. Such a diversity can motivate the developers to address more personalized features in their release planning so as to retain the user base.

Note that \textbf{F9} to \textbf{F12} are derived from only the ``device-model" based categorization of users. In practice, the users can be categorized in different ways, e.g., by region, sex, and age. Additionally, the diversity of requirements can be more complex. As a result, addressing the various user requirements is challenging. The traditional product-oriented software delivery model often provides a large number of features that aim to meet all potential user requirements. However, mobile apps are published and delivered in a quite different ``user-oriented" model. There is a strong need of new requirements-engineering approaches to better explore the specific requirements of a user. One of the significant issues is to build up a precise, fine-grained, and extensive user profile. To this end, an important trend in requirements elicitation is to rely on the comprehensive data analytics of user reviews, bug reports, social networks,  and app stores such as our Wandoujia dataset. In addition, as the release of new versions or features of an app can be quite frequent, e.g., weekly or monthly, the gap between requirements elicitation and app development needs to be greatly shortened.

If the app developers can know that some users holding a specific device model spend more cellular time rather than Wi-Fi, the app developers can provide some customized features to these users. 
To this end, some approaches such as end-user configuration/programming and context awareness can be useful. For example, apps can be developed to enable end-users to manually ``turn-on" and pay for only the features or contents that the users need (commonly implemented via in-app purchase), or to disable the downloading of contents under cellular network. 
Also, an app can prefetch contents under Wi-Fi network for the lower-end users who may have limited cellular data plan.    
\subsection{Exploring Potential Revenues from App}
A large number of popular apps, especially Android apps, are free instead of paid ones. Mobile apps heavily rely on other revenue channels such as in-app purchase and ads, especially for specific types of apps (e.g., game and media apps). 
App developers should accurately target the users who are likely to use/buy their apps and increase potential revenue. For example, finding \textbf{F10} suggests that the users holding lower-end device models are more likely to select \textit{GAME} apps. As it was reported that a lot of \textit{GAME} apps are paid ones~\cite{Zhong:TMC13}, the \textit{GAME} app developers could provide the ``\textit{try-out-first}" feature to encourage these lower-end users to use their apps. Indeed, in practice, a lot of \textit{Game} app developers have adopted such a strategy, i.e., making the first-level play free. Besides the ``try-out-first'' strategy, our findings can further recommend app developers how to better place their in-app advertisements, which are the major revenue channel for mobile apps. From finding \textbf{F11}, the choice of device models can correlate to the network access time, so app developers can know which users more possibly spend time in their apps. From the experiences of advertisement on the Web~\cite{ducoffe:1996advertising}, longer staying time usually implies more possibility to click advertisements. In this way, developers can more precisely target potential users and increase ads-clicking opportunities. Here, one feasible way is to negotiate with the device manufacturers and make their apps preloaded on these devices. Furthermore, by learning the access-time usage among different device models from \textbf{F11}, our recent PRADA work~\cite{Lu:ICSE2016} has designed efficient machine learning techniques to predict which users are more likely to spend more time on a specific app.  

In addition, although we cannot make very strict hypothesis testing, it might be a common sense that the choice of device models can possibly reflect the economic status or other background of users. For example, users with a better economic status are more likely to use higher-end device models. In this way, developers can further provide more ``\textit{personalized}'' advertisements to fit the users' interests. Similarly, ad network providers can also leverage our findings to know which apps a specific group of users are more likely to spend time on, i.e., users with higher-end device models tend to use the   \texttt{NEWS\_AND\_READING} and \texttt{TRAVEL} apps. Therefore, ad network providers can negotiate with these developers to import their ad-network libraries. 

We can also observe that the users holding lower-end device models are more likely to pay more traffic on the \textit{MUSIC} apps under the cellular network. Given that these users may have relatively low economic background, their data plan could be a bit limited. However, the network service providers can leverage this finding and provide ``\textbf{app-specific}" data plans. For example, some carriers in China make a special data-plan contract with the \textit{MUSIC} apps (such as Baidu Music \texttt{com.ting.mp3.oemc.android} and Kuwo Music \texttt{cn.kuwo.player}) and \textit{video} apps (such as Youku \texttt{com.cibn.tv}), and users can pay for this data plan independently and enjoy unlimited cellular data-traffic to download video/audio files. Indeed, such a new ``app-specific" data plan requires supports such as independent traffic accounting.

\section{Limitations and Discussions}\label{threats}


As an empirical measurement study, considerable care and attention should be given to ensure the rigor. However, as with any chosen research methodology, it is hardly without limitations. In this section, we discuss about the major limitations and threats to validate and generalize the results of our study.\\

\noindent \textbf{Single Dataset}. The first limitation is that our dataset is collected from only a single app store. Such a limitation may have introduced some selection biases caused by the app store's specific policies. In this way, some of our results, such as the popularity distribution of apps, may not always hold in other app stores. For example, the same app can be ranked to be quite popular on the Wandoujia app store, but may be unavailable on other app stores such as Google Play. In addition, the features provided on different OSes can be various. One feature of an Android app may not be available in the corresponding iOS and Windows version of this app. As a result, care should be given to generalize our findings to other app stores or platforms. To address this issue, we plan to validate some results by leveraging the public data such as the number of downloads, user ratings, and reviews from other app stores such as Google Play and Apple AppStore. For example, for a specific app, we can investigate the difference of user attitudes towards the same app. However, the information of network traffic, online time, and device model, cannot be easily captured on other app stores, and thus the limitation caused by a single app store cannot be completely solved. In a sense, due to the uniqueness of our dataset, some potentially useful results can be further leveraged by the research community.\\

\noindent \textbf{Demographical Differences}. Another major limitation is the demographical differences. The users under study are mainly from China, so the regional differences should be considered. For example, it is reported that users from different countries can perform variously in giving reviews against apps~\cite{lim2014investigating}. However, the same limitation also exists in most of previous studies that were conducted over users from a specific region, e.g., the study conducted on some states in US~\cite{MoreleyMao:IMC11}. In practice, collecting multi-dimensional usage data of large-scale users from various regions can be quite difficult. As our study was made over millions of users, the derived findings can still be useful. In addition, we can find that the usage patterns from only Chinese users can share commonalities with the users from other countries. For example, the power law distribution of app popularity and the user interests of co-installed apps, could also exist in other app stores, and can be further generalized to improve the design of app/content delivery. \\

\noindent \textbf{Time and Versions}. Another mentionable limitation is the time sensitivity. Because the analysis of our conference version~\cite{Li:IMC15} is done on only one-month usage data, some of our findings may not be promisingly generalized to the latest released versions of some apps. It is well-known that the mobile apps are updated very frequently, and some potential bugs of some apps, e.g., misuse or over-granting of network permissions, might be already fixed in the up-to-date app versions. Indeed, we realize that the one-month data is not sufficient to comprehensively capture app quality and user behaviors. As a result, this article employs a new extensive five-month dataset to reproduce the research questions, and most of results keep consistent. In this way, the time limitation is somewhat alleviated. However, some limitations need to be further addressed, e.g., the user ratings are not identified to a specific version of an app. It is possible to align the time when a rating is committed to an app with the release time of this app, so that one can determine whether the ratings are given to a specific version. However, in fact, on some app stores, one can arbitrarily commit ratings to an app, even though he/she doesn't install the latest version of this app. In the future, we plan to consider adding other data sources such as the bug reports and textual user reviews to judge the user attitudes of a specific app version. \\

\noindent \textbf{User Classification}. We categorize users according to the device models they hold, or more specifically, the price of device models. Indeed, such an indicator cannot be always reliable for some users. It is true that we cannot well address some limitations, including the users holding second-hand or multiple device models and the inaccurate estimation of device model's prices regarding the release date. To alleviate the threats, we employ various sources to segment the device models into coarse-grained ranges. Given the large-scale of users involved, such a classification can possibly reflect the diversity of usage patterns. We plan to explore other classifiers such as screen size estate or resolution level, by which the derived knowledge could be particularly useful to GUI designs of an app.\\

\noindent \textbf{Free Apps and Paid Apps}. As mentioned previously, the apps on Wandoujia are all free. Certainly, it would be possible that user behaviors on paid apps are a bit different~\cite{Nayebi:SANER2016}. Unfortunately, our current dataset inherently cannot address such a limitation.\\

\noindent \textbf{Correlation v.s. Causation}. We made various correlation-analysis studies, such as ratings and the number of downloads/users, the user ratings, the network usage, and the choice of device models. It should be noted that not all of the analysis results can be fully interpreted. In other words, these analyses have only \textit{correlation} instead of \textit{causation}. Indeed, comprehensively interpreting causation is often rather difficult for most empirical research studies. However, correlation is the first step ahead of causation, and is very  meaningful to cause the focus from relevant stakeholders. As presented previously, one goal of this article is to motivate the related researchers in exploring more opportunities, such as finding the underlying ``causation'' and even proposing new solutions.

\section{Related Work}\label{related}

The prevalence of mobile apps significantly changes software development, deployment, delivery, maintenance, and evolution. Supporting mobility becomes a promising trend in software engineering research~\cite{Picco:FOSE14}. In the past years, various efforts have been made, covering almost all lifecycles of apps, such as requirement analysis~\cite{lim2014investigating}\cite{Falaki:IMC10}\cite{Li:IMC15}, code/library analysis~\cite{Nicolas:SigMetrics14}\cite{Xie:USENIXSecurity13}\cite{Huang:ICSE2014}\cite{Zeller:ICSE14}\cite{Thomas:ICSE14}\cite{Gui:ICSE2015}, version evolution~\cite{Geoffrey:ASE2015}, and a number of system-level supports~\cite{Chen:MobiCom2015}\cite{Fukuda:IMC2015}\cite{Crussell:MobiSys2014}. 

In a sense, empirical studies of user behaviors can be quite useful to the software engineering research of apps. Understanding user behaviors of mobile apps establishes a foundation for different stakeholders in the research community of mobile computing, e.g., app developers, network providers, app-store operators, and OS vendors. A plethora of empirical studies have been made from different perspectives.  

\noindent \textbf{Understanding User Behaviors by Field Studies.} Given that collecting large-scale user data is hardly feasible for most studies, learning user behaviors by field studies is always a straightforward way. A lot of studies were performed over specific user groups. Rahmati \textit{et al.}~\cite{Zhong:TMC13, Rahmati:MobileHCI12} performed a four-month field study of the adoption and usage of smartphone-based services by 14 novice teenage users. Tossell \textit{et al.}~\cite{Zhong:CHI2012} applied a naturalistic and longitudinal log-based approach to collect real usage data from 24 iPhone users in the wild. Sani \textit{et al.}~\cite{Zhong:SIGMobileComm13}  collected data from 387 Android users in India, where users pay for cellular data consumed, with little prevalence of unlimited data plans. Falaki \textit{et al.}~\cite{Falaki:IMC10} found that web browsing contributed over half of the traffic at that time (2010), but currently users can enjoy more Web services via apps. Using detailed traces from 255 volunteer users, Falaki \textit{et al.}~\cite{Falaki:MobiSys10} conducted a comprehensive study of smartphone use and found immense diversity of users, by characterizing intentional user activities. Lim \textit{et al.}~\cite{lim2014investigating} made a questionnaire-based study to discover the diverse usages from about 4,800 users across 15 top GDP countries. Yan \textit{et al.}~\cite{Yan:Mobisys11} developed and deployed an app to collect usage logs from over 4,600 users to find their similar interests and explore recommendation systems for smartphone apps. For a study closely related to ours, Xu \textit{et al.}~\cite{MoreleyMao:IMC11} presented usage patterns by analyzing IP-level traces of thousands of users from a tier-1 cellular carrier in U.S. They identified traffic from distinct apps based on HTTP signatures and present aggregate results on their spatial and temporal prevalence, locality, and correlation.

Some field studies were made on specific apps. B{\"{o}}hmer \textit{et al.}~\cite{Matthias:MobileHCI11, Matthias:CHI13} made a field study over three popular apps such as Angry Bird, Facebook, and Kindle. Patro \textit{et al.}~\cite{Patro:CoNEXT13} deployed a multiplayer RPG app game and an education app, respectively, and collected diverse information to understand various factors affecting app revenues.\\

\noindent \textbf{Mining App Store Data.} Some types of app related data like user reviews, star ratings, and like/dislike voting are publicly accessible. Chen \textit{et al.}~\cite{Liu:ICSE14} presented AR-Miner to extract informative user reviews and group them using topic modeling. Fu \textit{et al.}~\cite{Liu:KDD13} presented WisCom, a system that can analyze tens of millions user ratings and comments in mobile app markets. Petsas \textit{et al.}~\cite{Petsas:IMC13} monitored and mined four popular third-party Android app stores and showed that the app-popularity distribution deviates from a commonly observed Zipf-like model. User reviews are significant assets in software engineering research. Lorenzo \textit{et al.}~\cite{Lorenzo:ICSE2016} presented CLAP to make release planning based on clustering the meaningful topics from user reviews and aligning these topics to developers' bug reports. \\

\noindent \textbf{Predicting Apps Usage.}  Baeza \textit{et al.} \cite{Baeza:WSDM2015} made field studies on the sequence of launching apps, and provided a solution to predict the ``next-to-be-used'' apps. Shin \textit{et al.}~\cite{Shin:UbiComp12, Shin:UbiComp13} collected a wide range of smartphone information from 23 users, extracted and analyzed features related to app prediction. Liao \textit{et al.}~\cite{Liao:CIKM13, Liao:ICDM13} proposed a temporal-based app predictor to dynamically predict the apps that are most likely to be used. Montoliu \textit{et al.}~\cite{Perez:MultimediaTool13} presented a framework to discover places-of-interest from multimodal mobile phone sensory data.  Do \textit{et al.}~\cite{Minh:Percom14} presented a framework for predicting where users will go and which app they are to use in the next ten minutes from the contextual information collected by smartphone sensors. \\

Compared to these studies, the major differences of our study include the unique dataset covering millions of users, some unique information such as app installation, uninstallation, and diverse network usage. Although Chinese users take up majority of all users in our dataset, we believe that behavior patterns inferred from millions of users under study should be more  comprehensive than those from volunteers. With our dataset, we also validate some results that were reported over a smaller scale of users. For example, a small set of apps account for substantial portion of downloads~\cite{Petsas:IMC13} and unique users~\cite{MoreleyMao:IMC11}, some apps are more likely to be installed together~\cite{MoreleyMao:IMC11, Falaki:IMC10, Falaki:MobiSys10}, and some functionality-similar apps may vary in terms of performance~\cite{Zhong:SIGMobileComm13}. However, besides using a different dataset collected from millions of users, our study explores uniquely new findings that were unaddressed previously, but useful for software engineering towards mobility:
\begin{itemize}
\item First, we make multi-dimensional measurement of app popularity from various aspects including downloads, users, and diverse network activities. Such a measurement can present a comprehensive ranking of apps rather than download times only, and can help improve the quality of app recommendation.
\item Second, we explore which apps are likely to be uninstalled and the lifecycle of the abandoned apps. In particular, we report the inconsistency of user ratings and the management activities of apps. The results can help improve the quality assurance of apps.  
\item Third, beyond reporting the co-installation of apps, we further explore the possible reasons why these apps are selected together. The results can be helpful to improve recommendation quality and explore new value-added apps. 
\item Fourth, we make a fine-granularity analysis of network activities to identify the ``network-intensive" apps and ``problematic" apps that consume traffic at background. Such results can be very significant to identify possible bugs or problems while reducing the threats to end-users. 
\item Finally, we study the impact by the price of device models, and explore how it impacts on user behaviors on apps usage. Such results can help address various requirements of users, and increase the potential revenue of apps.

\end{itemize}

\section{Conclusion}\label{conclusion}

We have conducted a systematic descriptive analysis of a large collection of app-store service profile from millions of Android users. Diverse usage patterns are with respect to the aspects such as app popularity, app management, app selection, app abandonment, network usage, and device-specific preferences. Our findings provide implications for various stakeholders in the mobile app ecosystem, and cover a number of issues for app development, deployment, delivery, revenue, evolution, etc.

Indeed, this article mainly focuses on the descriptive analysis of the data. However, we believe that this article can make the initiative step for the research on data-driven software engineering of mobile apps. Many findings of the analysis lead to potential research questions or opportunities. In fact, some research topics such as optimizing an app store's performance, predicting an app's quality and popularity~\cite{Li:WWW2016,Liu:TOIS17}, and prioritizing the fragmented Android devices for specific apps~\cite{Lu:ICSE2016}, have already been explored based on our dataset. 

Along with opening our dataset in this article, we expect that we can contribute a valuable resource for the research community, and promote the development of new research topics to benefit researchers, practitioners, and users.

\appendices

\ifCLASSOPTIONcompsoc
  
  \section*{Acknowledgment} 
 \small
This work was supported by the National Basic Research Program (973) of China under Grant No. 2014CB347701, the Natural Science Foundation of China (Grant No. 61370020, 61421091, 61572051, 61528201, 61529201). Tao Xie's work was supported in part by National Science Foundation under grants no. CCF-1409423, CNS-1434582, CNS-1513939, CNS-1564274. Qiaozhu Mei's work was supported in part by the National Science Foundation under grant no. IIS-1054199 and an MCubed grant at the University of Michigan. Xuanzhe~Liu is the corresponding author of this work.

\ifCLASSOPTIONcaptionsoff
  \newpage
\fi



%

\bibliographystyle{IEEEtran}
\bibliography{tse-appstudy}

\begin{IEEEbiography}[{\includegraphics[width=1in,height=1.25in,keepaspectratio]{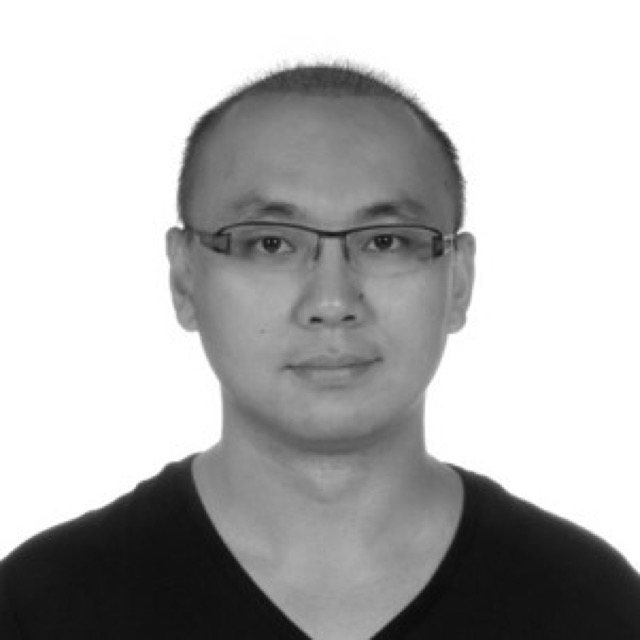}}]{Xuanzhe~Liu} is an associate professor in the School of Electronics Engineering
and Computer Science, Peking University, Beijing, China. His research interests
are in the area of services computing, mobile computing, web-based systems, and big data analytics.
\end{IEEEbiography}
\begin{IEEEbiography}[{\includegraphics[width=1in,height=1.5in,keepaspectratio]{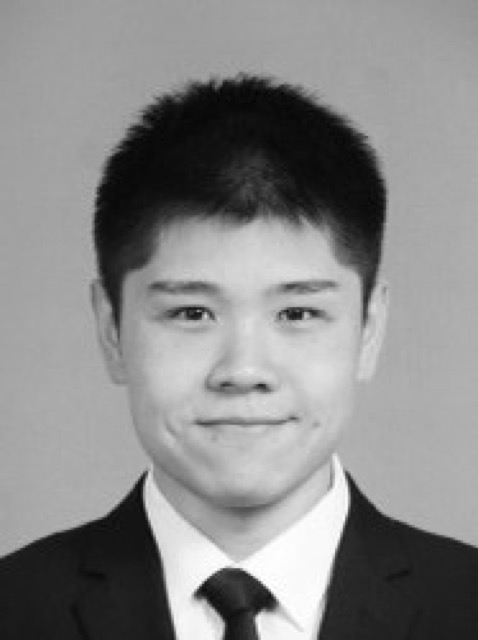}}]{Huoran~Li}
is now a Ph.D. student in the School of Electronics Engineering and Computer Science of Peking University, Beijing, China. His research interests include mobile computing, software engineering, and human computer interaction.
\end{IEEEbiography}
\begin{IEEEbiography}[{\includegraphics[width=1in,height=1.25in,clip,keepaspectratio]{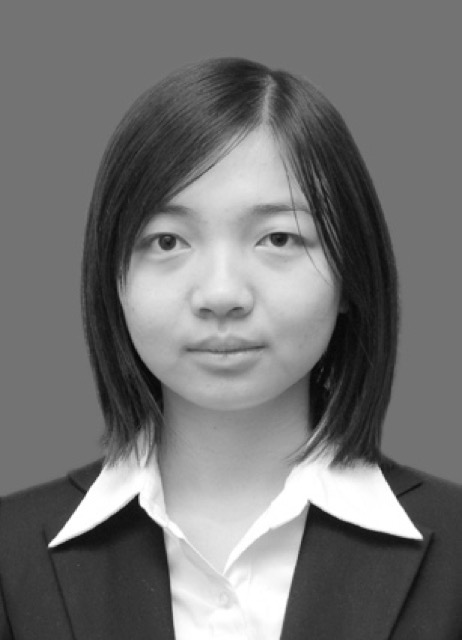}}]{Xuan~Lu}
is now a Ph.D. student in the School of Electronics Engineering and Computer Science of Peking University, Beijing, China. Her research interests include mobile computing and software analytics. 
\end{IEEEbiography}
\begin{IEEEbiography}[{\includegraphics[width=1.2in,height=1.25in,keepaspectratio]{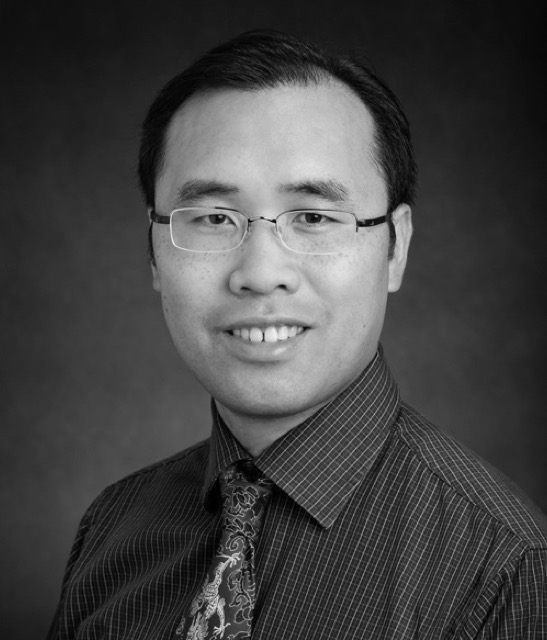}}]{Tao~Xie} is an associate professor and Willett Faculty Scholar in the Department of Computer Science at the University of Illinois at Urbana-Champaign, USA. His research interests are software testing, program analysis, software analytics, software security, and educational software engineering. He is a senior member of the IEEE.
\end{IEEEbiography}
\begin{IEEEbiography}[{\includegraphics[width=1in,height=1.25in,keepaspectratio]{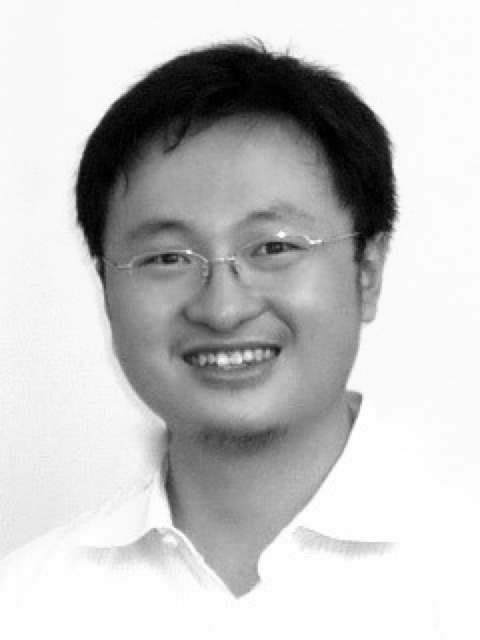}}]{Qiaozhu~Mei} is an associate professor at the University of Michigan School of Information. His major research interests include data mining and information retrieval.
\end{IEEEbiography}

\begin{IEEEbiography}[{\includegraphics[width=1in,height=1.25in,keepaspectratio]{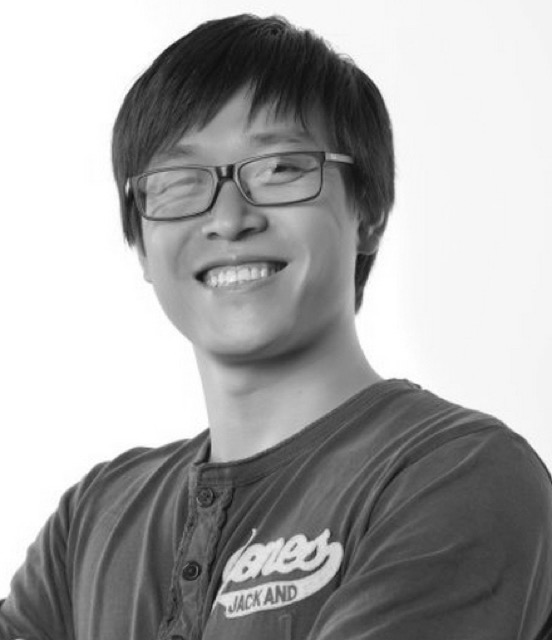}}]{Feng~Feng} is a co-founder of Wandoujia. He is the head architect of the Wandoujia marketplace and management app. He leads the engineering department and has developed dozens of products around Wandoujia.
\end{IEEEbiography}
\begin{IEEEbiography}[{\includegraphics[width=1in,height=1.25in,clip,keepaspectratio]{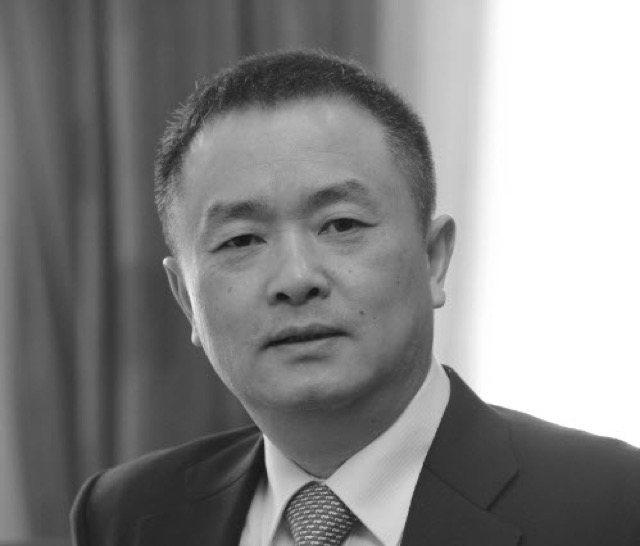}}]{Hong~Mei}
is a full professor of Beijing Institute of Technology and an adjunct professor of Peking University, Beijing, China. His current research interests are in the area of software engineering and operating systems. He is a Member of Chinese Academy of Sciences, and a Fellow of China Computer Federation (CCF). He is a Fellow of the IEEE.
\end{IEEEbiography}

\balance
%

%
%
%




\end{document}